\documentclass[amssymb,twocolumn,aps]{revtex4-2}
\usepackage{graphicx}
\usepackage{enumerate}
\usepackage{amssymb}
\usepackage{amsmath}
\usepackage{amsfonts}
\usepackage{color}
\usepackage{mathrsfs}

\begin{document}

\title{Quantum optical scattering by macroscopic lossy objects: A general approach}

\author{A. Ciattoni$^1$}
\email{alessandro.ciattoni@spin.cnr.it}
\affiliation{$^1$CNR-SPIN, c/o Dip.to di Scienze Fisiche e Chimiche, Via Vetoio, 67100 Coppito (L'Aquila), Italy}

\date{\today}

\begin{abstract}
We develop a general approach to describe the scattering of quantum light by a lossy macroscopic object placed in vacuum with no restrictions on both its dispersive optical response and its spatially inhomogeneous composition. Our analysis is based on the modified Langevin noise formalism, a recently introduced version of macroscopic quantum electrodynamics where scattering $(s)$ modes are explicitly separated from electric $(e)$ and magnetic $(m)$ medium excitations; accordingly  the formalism involves three kinds of non-interacting boson polaritons such that, in the lossless limit, $s$-polaritons reduce to standard photons whereas $e$- and $m$-polaritons disappear. We analytically derive the input-output unitary relation joining the boson operators of the ingoing and outgoing polaritons, a nontrivial result hinging upon original relations which comprehensively describe the transmission-emission-absorption interplay pertaining the classical radiation scattering, relations we here deduce by resorting to the dyadic Green's function properties. Besides we exploit the input-output relation to connect the output state of the field to the input one, this unveiling the role played by various classical electromagnetic dyadics in quantum optical scattering. We specialize the discussion to the most common situation where the object is initially not electromagnetically excited, with the ingoing electromagnetic state only containing $s$-polaritons, and we analyze the impact of the classical transmission and absorption dyadics on the transitions from ingoing to outgoing $s$-polariton and on the creation of outgoing $e$- and $m$-polaritons, respectively. Since the scattered radiation is collected in the far-field  and the object is usually left unmeasured, we analytically derive the reduced density operator of the outgoing $s$-polaritons revealing, in particular, how object losses both produce quantum decoherence of the scattered radiation and enable to exploit the possible entaglement among ingoing $s$-polaritons to dramatically manipulate the  statistical mixture of the quantum states pertaining the scattered radiation.
\end{abstract}
\maketitle

\section{Introduction}
Manipulating the quantum state of photons is one of the main subjects in quantum optics both for its conceptual value and for its applications in the prominent and radidly emerging field of quantum computing \cite{Nielse}. Common optical devices as beam splitters, mirrors, wave plates and polarizers are able to encode and decode the information carried by photons into the path, polarization, and time bin, intriguing abilities that have triggered an enormous research interest in the last decades. Such devices essentially couple the input light modes into the output modes as it was early realized in the investigation of quantum noise arising in interferometers \cite{Caves1} and produced by linear amplifiers \cite{Caves2,Loudo1,Mande1,Leyyy1,Leyyy2}. Due to its ubiquity in quantum optical setups, the lossless beam splitter has been thoroughly investigated \cite{Prasa1,Ouuuu1,Fearn1,Luiss1} and shown to support relevant quantum optical effect \cite{Makar1} as manipulation of the photon statistics \cite{Hutte1,Campo1} two-photon interference \cite{Hongg1,Fearn2,Bouch1} and generation of entangled output states out of nonclassical input states \cite{Kimmm1,Xiang1}. The theoretical description of the lossless beam splitter has also been extended to more general passive and lossless optical systems \cite{Knoll1} with particular emphasis on planar devices \cite{Leyyy3,Savas1,Khanb0,Atama1}. Basically, a lossless device is characterized by a unitary relation joining the boson operators of its input and output photonic modes, a simple description not holding for lossy devices \cite{Klysh1,Tisch1} where radiation-to-matter energy transfer exclusively 
allows for a unitary relation involving the operators of both photonic and matter modes. Accordingly suitable Langevin noise operators are introduced to model the matter degrees of freedom and their inclusion in the input-output relation restores bosonic consistency \cite{Barne1}. As a consequence, the theory of lossy four-port devices has been developed \cite{Grune1,Barne2,Knoll2} and the impact of losses on a variety of quantum optical effects has been investigated spanning from photon interference \cite{Jeffe1}, entanglement transformation \cite{Schee1,Schee2,Chizh1}, anti-coalescence of bosons \cite{Wangg1} and absorption of squeezed light \cite{Harda1}. Recently, the quantum input-output formalism has been exploited to describe the impact of metasurfaces on the quantum state of photons \cite{Jhaaa1,Wangg2,Georg1,Liiii1,Santi1} with the consequent establishment of the novel research field of quantum metaphotonic \cite{Solnt1,Dingg1,Zhang1}.

Modeling quantum functionalities of optical devices necessarily requires an underpinning theoretical scheme able to capture the interaction of the quantized electromagnetic field with macroscopic matter, in turn regarded as a continuum medium whose optical response is characterized by its electric permittivity and magnetic permeability. In the situations where losses can be neglected, such scheme is readily available since electromagnetic modes are still orthogonal and they provide a basis for field quantization in terms of photons \cite{Glaub1,Dalto1}. Converserly, when optical absorption and dispersion are relevant, the framework is both conceptually deeper due to causality and considerably more involved due to the unavoidable inclusion of matter degrees of freedom. In the early proposed schemes accounting for damping, matter polarization was modelled as an harmonic oscillator coupled to a continuum of reservoir oscillator fields 
\cite{Fanooo,Hopfie,Huttn1,Huttn2,Sutto1,Sutto2,Kheir1} and the demand to extend the treatment to arbitrary macroscopic media has lead to remove the matter oscillator from the model and to encode the medium electromagnetic response in the field-reservoir coupling term \cite{Matloo,Kheir2,Bhattt,Amoos1,Sutto3,Amoos2}. A further generalization was achieved by macroscopic quantum electrodynamics or Langevin noise formalism (LNF) \cite{Grune2,Schee3,Dungg1,Schee4,Buhma1}, a phenomenological scheme where the field is assumed to be generated by dipolar fluctuating sources within the medium  through the classical dyadic Green's function and quantization is non-canonically performed by imposing the standard field commutation relations of vacuum quantum electrodynamics. Remarkably, the lack of a canonical Lagragian formulation has been filled in Ref.\cite{Philbi} where LNF arises as a consequence of an approach where macroscopic electromagnetism is canonically quantized. Nowadays, LNF is probably the most adopted theoretical scheme for modeling the quantized light field interacting with macroscopic matter, as shown by the considerable number of applications it has succesfully outlined as dispersion forces \cite{Buhma2,Buhma3,Buhma4}, quantum emitters decay \cite{Schee5,Dungg2,Rivera,Hemmer}, cavity QED \cite{Khanb1,Khanb2,Dzsot2}, quantum nanophotonics \cite{Hakam1,Kurman,Feistt} and fast electrons scattering \cite{DiGiu1,Hayun1,Kfirr1,Ciatt1,Ciatt2}. One of the main working-assumption of LNF is that the  medium displaying absorption is spatially unbounded, i.e. its permittivity and permeabilities are complex everywhere, so that LNF does provide a description of finite-size lossy objects placed in vacuum only as a limiting situation where the permittivity and permeability limits ${\rm Im} ( \varepsilon) \rightarrow 0^+$ and ${\rm Im} ( \mu) \rightarrow 0^+$ pertaining the regions filled by vacuum is taken at the end of the calculations \cite{Hanso1}. Physically, this is due to the fact that in unbounded lossy media any field exponentially vanishes far from the sources and no asymptotically plane-wave like mode exists, so that LNF correctly retains the field contribution of the medium fluctuating sources while discarding the scattering modes contribution. Recently, the potentials of explicitly retaining the scattering modes have been mentioned in Ref.\cite{Naaaa1} in the form of a modified Langevin noise formalism (MLNF) which has been theoretically substantiated in Ref.\cite{Ciatt3} where MLNF is analytically derived from the canonical quantization scheme of macroscopic electromagnetism discussed in Ref.\cite{Philbi}. As a result, MLNF embodies a version of macroscopic quantum electrodynamics able to deal with lossy objects of finite size in vacuum by directly setting ${\rm Im} (\varepsilon) = 0$ and ${\rm Im} (\mu) = 0$ out of their volume. The main difference between the two formalisms is that LNF description hinges upon two kinds of non-interacting bosons, the $e$- and $m$-polaritons associated to the quantized electric and magnetic dipolar sources in the medium, while MLNF additionally resorts to a third kind of bosons, the $s$-polaritons related to the quantized excitation of the scattering modes. It is worth incidentally pointing out that, in the lossless limit, the MLNF coincides with the Glauber's approach to quantum optics of dielectric media of Ref.\cite{Glaub1} since $s$-polaritons reduces to standard photons while $e$- and $m$-polaritons disappear.

Quantum radiation scattering by macroscopic objects can be regarded as the utmost generalization of optical devices functionality of coupling their input photonic modes with output modes. The striking difference is that quantum optical scattering generally deals with a continuum of input and output modes associated to any possible far field spatial direction and polarization whereas the hitherto considered optical devices only couples a finite number of modes generally referred to as ports. Even though quantum optical scattering by macroscopic lossless objects is currently a textbook subject (see, e.g., Chapter 8 of Ref.\cite{Garri1}), to the best of our knowledge a general theory of the scattering of quantum light by arbitray lossy macroscopic objects is not yet available. In this paper we fill this gap by introducing a general approach to quantum optical scattering able to deal with macroscopic lossy objects of finite size with arbitrary linear causal optical response and spatially inhomogeneous composition. We derive such an approach from the MLNF, it being the most suitable version of macroscopic quantum electrodynamics for dealing with scattering problems since the involved far-field ingoing and outgoing quantum light modes are respectively excited and detected in vacuum. We start our analysis by deducing the far-field expression of the electric field operator in the Heisenberg picture and the inspection of its large-time behavior yields the ingoing ($t\rightarrow - \infty$) and outgoing ($t\rightarrow + \infty$) field operators, in turn enabling the physical identifaction of ingoing and outgoing $s$-polaritons, respectively. In order to prove that the outgoing excitations are effectively bosonic $s$-polaritons, we accompany them with additional $e$- and $m$-polariton operators and  we deduce the unitary relation connecting the annihilation operators of the outgoing polaritons to those of the ingoing ones. By virtue of its generality, such result is very remarkable and its achievement was made possible by original relations involving the transmission, emission and absorption dyadics (see below), relations which we here fully derive from the fundamental integral relation of the dyadic Green's function and which, to the best of our knowledge, have not been discussed in literature yet. The physical interpretation of such relations stems from the overall energy conservation of the radiation-matter system accompanying any classical electromagnetic scattering experiment, they detailing the interplay between scattering, absorption and possible emission of radiation when the object is initially excited. Since ingoing and outgoing polariton operators induce two different representations in the Fock space of the quantum field, we evaluate the unitary matrix of the representation change by using the input-output relation and this allows us to express the Heisenberg outgoing state of the field in terms of the ingoing one, without any restriction on the number, kind and possible entanglement of the ingoing polaritons. Such general expression reveals that in any scattering process the polariton number is conserved, an evident consequence of the linearity of the object optical response, and besides it vividly shows how classical electromagnetic dyadics trigger any possible quantum transition from the $(s,e,m)$ ingoing  polaritons to the $(s,e,m)$ outgoing ones. Of particular interest is the case where the ingoing state solely contains $s$-polaritons since it corresponds to ordinary quantum optical scattering situation where radiation is launched toward an initially inert object not displaying electromagnetic excitation. 
In this situation, we argue that quantum optical scattering is solely driven by the classical transmission and absorption dyadics, the first ruling the ingoing to outgoing $s$-polariton transitions (radiation scattering) and the second controlling the creation of outoing $e$- and $m$-polaritons (radiation absorption). Since in standard quantum optical experiments only radiation is usually detected in the far field with the object left unmeasured, we evaluate  the reduced density operator of outgoing $s$-polaritons by taking the partial trace of the full density operator over the combined outgoing $e$- and $m$-polariton subsystems. Such result enables to quantitatively assess the quantum decoherence experienced by radiation upon scattering by a macroscopic lossy object and, in addition, it unveils the impact of the possible entanglement among ingoing polaritons on the statistical mixture of states representing the scattered radiation. Both for application purposes and for their intrinsic relevance, we conclude the paper with the discussion of one and two $s$-polariton scattering, they enabling the detailed analysis of all the above presented general results.

\section{Modified Langevin noise formalism}
We start by reviewing the MLNF detailed in Ref.\cite{Ciatt3} since it is the theoretical framework on which is based our investigation of quantum optical scattering . The macroscopic scatterer is an arbitrary finite-size object filling the spatial region $V$ enclosed by the boundary surface $\partial V$ (see Fig.1), and its inhomogeneous isotropic magnetodielectric optical response is characterized, in the frequency domain ($e^{-i \omega t}$), by the dielectric permittivity $\varepsilon _\omega ^{V} \left( {\bf{r}} \right)$ and the magnetic permeability $\mu _\omega ^{V} \left( {\bf{r}} \right)$ whose only constraint is to be holomorphic functions of the complex frequency $\omega$ in the upper half-plane ${\mathop{\rm Im}\nolimits} \left( \omega  \right) > 0$ as required by causality. To emphasize that the object is placed in vacuum it is convenient to define the global permittivity ${\varepsilon _\omega  \left( {\bf{r}} \right)}$ and permeability ${\mu _\omega  \left( {\bf{r}} \right)}$ according to
\begin{equation} 
\varepsilon _\omega  \left( {\bf{r}} \right),\mu _\omega  \left( {\bf{r}} \right) = \left\{ {\begin{array}{*{20}l}
   {\varepsilon _\omega ^V \left( {\bf{r}} \right),\mu _\omega ^V \left( {\bf{r}} \right),} & {{\bf{r}} \in V,}  \\
   {1,1,} & {{\bf{r}} \notin V.}  \\
\end{array}} \right.
\end{equation}
The MLNF relies on the dyadic Green's function ${\cal G}_\omega  \left( {\left. {\bf{r}} \right|{\bf{r}}'} \right)$ \cite{CheTai,Chewww}, one of the main classical theoretical tool to characterize  
electrodynamics in the presence of the object, defined by the boundary value problem
\begin{eqnarray} \label{GreBouVal}
&& \left[ {\nabla _{\bf{r}}  \times \frac{1}{{\mu _\omega  \left( {\bf{r}} \right)}}\nabla _{\bf{r}}  \times  - k_\omega ^2 \varepsilon _\omega  \left( {\bf{r}} \right)} \right]{\cal G}_\omega  \left( {\left. {\bf{r}} \right|{\bf{r}}'} \right) = \delta \left( {{\bf{r}} - {\bf{r}}'} \right){\cal I}, \nonumber \\ 
&& {\bf{u}} \times \left[ {{\cal G}_\omega  \left( {\left. {\bf{r}} \right|{\bf{r}}'} \right)} \right]_{\partial V} = 0, \quad {\bf{u}} \times \left[ {\frac{{\nabla _{\bf{r}}  \times {\cal G}_\omega  \left( {\left. {\bf{r}} \right|{\bf{r}}'} \right)}}{{\mu _\omega  \left( {\bf{r}} \right)}}} \right]_{\partial V} = 0,\nonumber \\ 
&& {\cal G}_\omega  \left( {\left. {r{\bf{n}}} \right|{\bf{r}}'} \right) \mathop  \approx \limits_{r \to  \infty } \frac{{e^{ik_\omega  r} }}{r}{\cal W}_\omega  \left( {\left. {\bf{n}} \right|{\bf{r}}'} \right) 
\end{eqnarray}
where $k_\omega = \omega / c$, ${\cal I}$ is the identity dyadic, ${\bf u}({\bf s})$ is the outward orthogonal unit vector to $\partial V$ at the point ${\bf{s}}$, $\left[ f \right]_{\partial V}  = \lim _{\eta  \to 0^ +  } \left[ {f\left( {{\bf{s}} + \eta {\bf{u}}} \right) - f\left( {{\bf{s}} - \eta {\bf{u}}} \right)} \right]$, ${\bf n}$ is a unit vector, the symbol $\mathop  \approx \limits_{r \to \infty }$ denotes the leading order term of the asymptotic expansion for $r \rightarrow +\infty$ and ${\cal W}_\omega \left( {\left. {\bf{n}} \right|{\bf{r}}'} \right)$ is a dyadic amplitude (see Appendix A for a review of the main asymptotic, dyadic and quantum relations used in the present paper). The first of Eqs.(\ref{GreBouVal}) is the inhomogeneous Helmholtz equation elucidating that the dyadic Green's function is the kernel of the inverse Helmholtz operator. The second and the third of Eqs.(\ref{GreBouVal}) are matching conditions at the object boundary surface $\partial V$ where ${\varepsilon _\omega  \left( {\bf{r}} \right)}$ and  ${\mu _\omega  \left( {\bf{r}} \right)}$ are discontinuous and they mimic the standard electromagnetic continuity requirements of the tangential components of the electric and magnetic fields. The fourth of Eqs.(\ref{GreBouVal}) is the Sommerfeld radiation boundary condition enforcing the far-field outgoing behavior of the radiation emitted by localized sources and it introduces the dimensionless dyadic
\begin{equation} \label{Womega}
{\cal W}_\omega  \left( {\left. {\bf{n}} \right|{\bf{r}}} \right) = \mathop {\lim }\limits_{r' \to  + \infty } \left[ {r'e^{ - ik_\omega  r'} {\cal G}_\omega  \left( {\left. {r'{\bf{n}}} \right|{\bf{r}}} \right)} \right]
\end{equation}
which is the asymptotic dyadic amplitude of the dyadic Green's function, in turn left-orthogonal to the observation direction, i.e. ${\bf{n}} \cdot {\cal W}_\omega  \left( {\left. {\bf{n}} \right|{\bf{r}}} \right) = 0$. Two essential properties of the dyadic Green's function are the reciprocity relation
\begin{equation}  \label{GomRec}
{\cal G}_\omega ^T \left( {\left. {\bf{r}} \right|{\bf{r}}'} \right) = {\cal G}_\omega  \left( {\left. {{\bf{r}}'} \right|{\bf{r}}} \right)
\end{equation}
and the fundamental integral relation
\begin{eqnarray} \label{GreFunInt}
&& \int {do_{\bf{n}} } {\cal W}_\omega ^T \left( {\left. {\bf{n}} \right|{\bf{r}}} \right) \cdot {\cal W}_\omega ^* \left( {\left. {\bf{n}} \right|{\bf{r}}'} \right) \nonumber \\ 
&&  + \frac{{\pi \varepsilon _0 }}{{\hbar k_\omega ^3 }}\sum\limits_{\nu = e,m}  {\int {d^3 {\bf{s}}} \,{\cal G}_{\omega \nu } \left( {\left. {\bf{r}} \right|{\bf{s}}} \right) \cdot {\cal G}_{\omega \nu }^{T*} \left( {\left. {{\bf{r}}'} \right|{\bf{s}}} \right)}  \nonumber \\ 
&&  = \frac{1}{{k_\omega  }}{\mathop{\rm Im}\nolimits} \left[ {{\cal G}_\omega  \left( {\left. {\bf{r}} \right|{\bf{r}}'} \right)} \right],  
\end{eqnarray}
where $do_{\bf{n}}  = \sin \theta _{\bf{n}} d\theta _{\bf{n}} d\varphi _{\bf{n}}$ is the differential of the solid angle around the unit vector  ${\bf{n}} = \sin \theta _{\bf{n}} \left( {\cos \varphi _{\bf{n}} {\bf{u}}_x  + \sin \varphi _{\bf{n}} {\bf{u}}_y } \right) + \cos \theta _{\bf{n}} {\bf{u}}_z$ and
\begin{eqnarray} \label{GoeGom}
&& {\cal G}_{\omega e} \left( {\left. {\bf{r}} \right|{\bf{r}}'} \right) = {\cal G}_\omega  \left( {\left. {\bf{r}} \right|{\bf{r}}'} \right)\left[ {i\sqrt {\frac{{\hbar k_\omega ^4 }}{{\pi \varepsilon _0 }}{\mathop{\rm Im}\nolimits} \left[ {\varepsilon _\omega  \left( {{\bf{r}}'} \right)} \right]} } \right], \nonumber \\ 
&& {\cal G}_{\omega m} \left( {\left. {\bf{r}} \right|{\bf{r}}'} \right) = {\cal G}_\omega  \left( {\left. {\bf{r}} \right|{\bf{r}}'} \right) \times \mathord{\buildrel{\lower3pt\hbox{$\scriptscriptstyle\leftarrow$}} 
\over \nabla } _{{\bf{r}}'} \left\{ { \frac{1}{i}\sqrt {\frac{{\hbar k_\omega ^2 }}{{\pi \varepsilon _0 }}{\mathop{\rm Im}\nolimits} \left[ {\frac{{ - 1}}{{\mu _\omega  \left( {{\bf{r}}'} \right)}}} \right]} } \right\}. \nonumber \\ 
\end{eqnarray}
We incidentally note that the first term in the left hand side of Eq.(\ref{GreFunInt}) is a surface integral contribution not showing up in the LNF \cite{Hanso1} since that approach deals with unbounded lossy media where the exponential fall off of the dyadic Green's function at infinity results into the vanishing of its asymptotic amplitude, i.e. ${\cal W}_\omega=0$ \cite{Dreze2}.

\begin{figure}
\centering
\includegraphics[width = 1\linewidth]{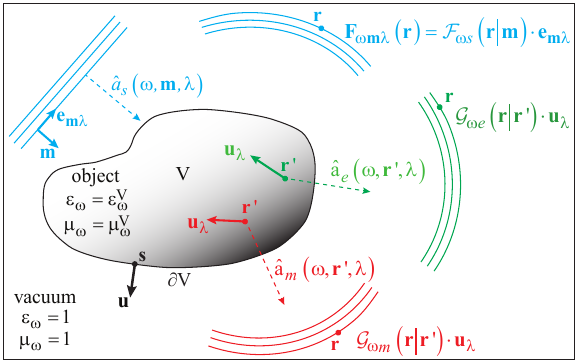}
\caption{Pictorial scketch of the MLNF portrayal of a macroscopic object quantum electrodynamics. The energy quanta of the electromagnetic field are scattering ($s$), electric ($e$) and magnetic ($m$) polaritons, here depicted in cyan, green and red, respectively. $s$-polaritons are associated to the scattering modes ${\bf{F}}_{\omega {\bf{m}}\lambda } \left( {\bf{r}} \right) = {\cal F}_{\omega s} \left( {\left. {\bf{r}} \right|{\bf{m}}} \right) \cdot {\bf{e}}_{{\bf{m}}\lambda } $ whose modal plane wave has frequency $\omega$, direction ${\bf m}$  and polarization ${\bf e}_{{\bf m}\lambda}$ whereas $e$- and $m$- polaritons pertain electric and magnetic medium dipolar sources vibrating at frequency $\omega$, located at the point ${\bf r}'$ inside the object and directed along ${\bf u}_\lambda$, in turn producing the field excitations 
${\cal G}_{\omega e} \left( {\left. {\bf{r}} \right|{\bf{r}}'} \right) \cdot {\bf{u}}_\lambda$ and ${\cal G}_{\omega m} \left( {\left. {\bf{r}} \right|{\bf{r}}'} \right) \cdot {\bf{u}}_\lambda$.}
\label{Fig1}
\end{figure}

In addition to the dyadic Green's function, MLNF also resorts to the scattering modes which are described by the modal dyadic $
{\cal F}_{\omega s} \left( {\left. {\bf{r}} \right|{\bf{m}}} \right)$ defined by the boundary value problem
\begin{eqnarray} \label{ESSBouVal}
&& \left[ {\nabla _{\bf{r}}  \times \frac{1}{{\mu _\omega  \left( {\bf{r}} \right)}}\nabla _{\bf{r}}  \times  - k_\omega ^2 \varepsilon _\omega  \left( {\bf{r}} \right)} \right]{\cal F}_{\omega s} \left( {\left. {\bf{r}} \right|{\bf{m}}} \right) = 0, \nonumber \\ 
&& {\bf{u}} \times \left[ {{\cal F}_{\omega s} \left( {\left. {\bf{r}} \right|{\bf{m}}} \right)} \right]_{\partial V}  = 0, \quad {\bf{u}} \times \left[ {\frac{{\nabla _{\bf{r}}  \times {\cal F}_{\omega s} \left( {\left. {\bf{r}} \right|{\bf{m}}} \right)}}{{\mu _\omega  \left( {\bf{r}} \right)}}} \right]_{\partial V}  = 0, \nonumber\\ 
&& {\cal F}_{\omega s} \left( {\left. {r{\bf{n}}} \right|{\bf{m}}} \right)\mathop  \approx \limits_{r \to \infty } \sqrt {\frac{{\hbar k_\omega ^3 }}{{16\pi ^3 \varepsilon _0 }}} \left[ e^{i\left( {k_\omega  r} \right){\bf{n}} \cdot {\bf{m}}} {\cal I}_{\bf{m}} \right. \nonumber \\
&& \left. + \frac{{e^{ik_\omega  r} }}{r}{\cal S}_\omega  \left( {{\bf{n}}\left| {\bf{m}} \right.} \right) \right], 
\end{eqnarray}
where, in addition to the symbols defined after Eqs.(\ref{GreBouVal}), ${\bf m}$ is a unit vector, ${\cal I}_{\bf{m}}  = {\cal I} - {\bf{mm}}$ is the dyadic projector onto the plane orthogonal to $\bf m$ and ${\cal S}_\omega  \left( {{\bf{n}}\left| {\bf{m}} \right.} \right)$ is a dyadic amplitude. The first three of Eqs.(\ref{ESSBouVal}) state that the modal dyadic ${\cal F}_{\omega s} \left( {\left. {\bf{r}} \right|{\bf{m}}} \right)$ statisfies the homogeneous Helmholtz equation and the electric field matching conditions at the discontinuity surface $\partial V$, so that it is a solution of the source-free Maxwell equations. On the other hand, the fourth of Eqs.(\ref{ESSBouVal}) is the scattering boundary condition requiring the far-field of ${\cal F}_{\omega s} \left( {\left. {\bf{r}} \right|{\bf{m}}} \right)$ to be the superposition of plane wave traveling along the direction ${\bf m}$ and an outgoing spherical wave whose amplitude is ${\cal S}_\omega  \left( {{\bf{n}}\left| {\bf{m}} \right.} \right)$ (the overall coefficient $\sqrt {\hbar k_\omega ^3 /16\pi ^3 \varepsilon _0 }$ has been introduced for later convenience). Here ${\cal S}_\omega  \left( {{\bf{n}}\left| {\bf{m}} \right.} \right)$ is the scattering dyadic \cite{Krist1}, the key quantity describing classical plane wave scattering by the object, which satisfies the two orthogonality relations ${\bf{n}} \cdot {\cal S}_\omega  \left( {{\bf{n}}\left| {\bf{m}} \right.} \right) = 0$, ${\cal S}_\omega  \left( {{\bf{n}}\left| {\bf{m}} \right.} \right) \cdot {\bf{m}} = 0$  and the reciprocity relation
\begin{equation} \label{SomRec}
{\cal S}_\omega ^T \left( {{\bf{n}}\left| {\bf{m}} \right.} \right) = {\cal S}_\omega  \left( { - {\bf{m}}\left| { - {\bf{n}}} \right.} \right).
\end{equation}
Note that the field ${\cal F}_{\omega s} \left( {\left. {\bf{r}} \right|{\bf{m}}} \right) \cdot {\bf{m}}$ is a solution of the source-free Maxwell equation with vanishing far-field (due to the relation ${\cal S}_\omega  \left( {{\bf{n}}\left| {\bf{m}} \right.} \right) \cdot {\bf{m}} = 0$ and the fourth of Eqs.(\ref{ESSBouVal})) so that it vanishes everywhere, i.e. ${\cal F}_{\omega s} \left( {\left. {\bf{r}} \right|{\bf{m}}} \right) \cdot {\bf{m}} = 0$ or, in other words, the modal dyadic is right orthogonal to the plane wave direction. Accordingly, it is convenient to introduce two mutually orthogonal unit vectors ${\bf{e}}_{{\bf{m}}1}$, ${\bf{e}}_{{\bf{m}}2}$ orthognonal to ${\bf m}$ and hence spanning the plane orthogonal to it, i.e. $\sum\nolimits_{\lambda  = 1}^2 {{\bf{e}}_{{\bf{m}}\lambda } {\bf{e}}_{{\bf{m}}\lambda } }  = {\cal I}_{\bf{m}}$, so that ${\bf{F}}_{\omega {\bf{m}}\lambda } \left( {\bf{r}} \right) = {\cal F}_{\omega s} \left( {\left. {\bf{r}} \right|{\bf{m}}} \right) \cdot {\bf{e}}_{{\bf{m}}\lambda }$ is recognized to be   the field scattered by the object (scattering mode) when illuminated by the modal plane wave of frequency $\omega$, direction $\bf m$ and polarization ${\bf{e}}_{{\bf{m}}\lambda }$ since its far field behavior, from the fourth of Eqs.(\ref{ESSBouVal}), is given by 
\begin{eqnarray} \label{sModeFarfield}
&& {\bf{F}}_{\omega {\bf{m}}\lambda } \left( {r{\bf{n}}} \right)\mathop  \approx \limits_{r \to \infty } \sqrt {\frac{{\hbar k_\omega ^3 }}{{16\pi ^3 \varepsilon _0 }}} \left[ {e^{i\left( {k_\omega  r} \right){\bf{n}} \cdot {\bf{m}}} {\bf{e}}_{{\bf{m}}\lambda } } \right. \nonumber \\ 
&& \left. { + \frac{{e^{ik_\omega  r} }}{r}{\cal S}_\omega  \left( {{\bf{n}}\left| {\bf{m}} \right.} \right) \cdot {\bf{e}}_{{\bf{m}}\lambda } } \right].
\end{eqnarray}
It is worth emphasizing that in Ref.\cite{Ciatt3} it has been shown that
\begin{eqnarray} \label{FwolProp} 
&& {\cal F}_{\omega s}  \left( {\left. {\bf{r}} \right|{\bf{m}}} \right) = \sqrt {\frac{{\hbar k_\omega ^3 }}{{\pi \varepsilon _0 }}} {\cal W}_\omega ^T \left( { - {\bf{m}}\left| {\bf{r}} \right.} \right) 
\end{eqnarray}
which enstablishes an intimate connection between the modal dyadic ${\cal F}_{\omega s}$ and the asymptotic amplitude ${\cal W}_\omega$ of the dyadic Green's function, a remarkable relation which has been exploited in Ref.\cite{Ciatt3} to theoretically substatiate the MLNF and which futher reinforces the primary role played by the dyadic Green's function in providing a complete characterization of the object electrodynamics \cite{Notaa1}. As a complement to this general observation, in Appendix B we use Eq.(\ref{FwolProp}) to show that the scattering dyadic ${\cal S}_\omega$ can be directly evaluated from the dyadic Green's function ${\cal G}_\omega$ by means of the relation
\begin{eqnarray} \label{ScatDyad}
&& {\cal S}_\omega  \left( {{\bf{n}}\left| {\bf{m}} \right.} \right) = \mathop {\lim }\limits_{r \to \infty } \mathop {\lim }\limits_{r' \to \infty } \left[ { - re^{i\left( {k_\omega  r} \right)\left( {{\bf{n}} \cdot {\bf{m}} - 1} \right)} {\cal I}_{\bf{m}} } \right. \nonumber \\ 
&&  + \left. {4\pi rr'e^{ - ik_\omega  \left( {r + r'} \right)} {\cal G}_\omega  \left( {\left. {r{\bf{n}}} \right| - r'{\bf{m}}} \right)} \right],
\end{eqnarray}
where the order of the two limits can not be interchanged, a remarkable result we also exploit in Appendix B to provide an alternative proof of the reciprocity relation of Eq.(\ref{SomRec}). Although these last results are not used to develop the quantum optical scattering approach introduced in this paper, we likewise report them here since, to the best of our knowledge, they are not present in literature.

The MLNF is characterized by the Hamiltonian operator
\begin{eqnarray} \label{Hamilt}
&& \hat H = \int\limits_0^{ + \infty } {d\omega } \;\hbar \omega \left\{ {\int {do_{\bf{m}} } {\bf{\hat g}}_{\omega s}^\dag  \left( {\bf{m}} \right) \cdot {\bf{\hat g}}_{\omega s} \left( {\bf{m}} \right)} \right. \nonumber \\ 
&& \left.  + \sum\limits_{\nu  = e,m} { \int {d^3 {\bf{r}}} \, {{\bf{\hat f}}_{\omega \nu }^\dag  \left( {\bf{r}} \right) \cdot {\bf{\hat f}}_{\omega \nu } \left( {\bf{r}} \right)} } \right\} 
\end{eqnarray}
and, in the Heisenberg picture we hereafter use, by the electric field operator
\begin{eqnarray} \label{EleHei}
&& {\bf{\hat E}}\left( {{\bf{r}},t} \right) = \int\limits_0^{ + \infty } {d\omega } e^{ - i\omega t} \left\{ {\int {do_{\bf{m}} } {\cal F}_{\omega s}  \left( {\left. {\bf{r}} \right|{\bf{m}}} \right) \cdot {\bf{\hat g}}_{\omega s} \left( {\bf{m}} \right)} \right. \nonumber \\ 
&& + \left. {\sum\limits_{\nu  = e,m} \int {d^3 {\bf{r}}'\,}  {{\cal G}_{\omega \nu } \left( {\left. {\bf{r}} \right|{\bf{r}}'} \right) \cdot {\bf{\hat f}}_{\omega \nu } \left( {{\bf{r}}'} \right)} } \right\} + h.c. 
\end{eqnarray} 
where 
\begin{eqnarray} \label{Opegosfos} 
&& {\bf{\hat g}}_{\omega s} \left( {\bf{m}} \right) = \sum\limits_{\lambda  = 1}^2 {\hat a_s \left( {\omega ,{\bf{m}},\lambda } \right){\bf{e}}_{{\bf{m}}\lambda } }, \nonumber \\
&& {\bf{\hat f}}_{\omega \nu } \left( {\bf{r}} \right) = \sum\limits_{\lambda  = 1 }^3 {\hat a_\nu  \left( {\omega ,{\bf{r}},\lambda } \right){\bf{u}}_\lambda  }  
\end{eqnarray}
are vector operators (${\bf u}_\lambda$ are the cartesian unit vectors) satisfying the bosonic commutation relation
\begin{eqnarray} \label{gfComRel}
&& \left[ {{\bf{\hat g}}_{\omega s} \left( {\bf{m}} \right),{\bf{\hat g}}_{\omega 's}^\dag  \left( {{\bf{m}}'} \right)} \right] = \delta \left( {\omega  - \omega '} \right)\delta \left( {o_{\bf{m}}  - o_{{\bf{m}}'} } \right){\cal I}_{\bf{m}}, \nonumber \\ 
&& \left[ {{\bf{\hat f}}_{\omega \nu } \left( {\bf{r}} \right),{\bf{\hat f}}_{\omega '\nu '}^\dag  \left( {{\bf{r}}'} \right)} \right] = \delta \left( {\omega  - \omega '} \right)\delta _{\nu \nu '} \delta \left( {{\bf{r}} - {\bf{r}}'} \right) {\cal I}, 
\end{eqnarray}
where $\delta \left( {o_{\bf{m}}  - o_{{\bf{m}}'} } \right) = \delta \left( {\theta _{\bf{m}}  - \theta '_{\bf{m}} } \right)\delta \left( {\varphi _{\bf{m}}  - \varphi '_{\bf{m}} } \right)/\sin \theta _{\bf{m}}$, together with the vanishing of all  the other possible commutation relations. Equations (\ref{Hamilt}), (\ref{EleHei}) and (\ref{gfComRel}) lead to the MLNF portrayal of the quantum optical field as an assembly of non-interacting bosons of three kinds, the $s$-, $e$- and $m$- polaritons, whose associated electromagnetic excitations are sketched in Fig.1. Accordingly, the operator $\hat a_s^\dag  \left( {\omega ,{\bf{m}},\lambda } \right)$ creates an $s$-polariton associated to the scattering mode ${\bf{F}}_{\omega {\bf{m}}\lambda } ({\bf r})$ whose modal plane wave has frequency $\omega$, direction ${\bf m}$  and polarization ${\bf e}_{{\bf m}\lambda}$ (see Eq.(\ref{sModeFarfield})), whereas $\hat a_e^\dag  \left( {\omega ,{\bf{r}'},\lambda} \right)$ and $\hat a_m^\dag  \left( {\omega ,{\bf{r}'},\lambda} \right)$ create $e$- and $m$-polaritons pertaining electric and magnetic dipolar sources vibrating at frequency $\omega$, located at the point ${\bf r}'$ inside the object and directed along ${\bf u}_\lambda$, in turn producing the field excitations of spatial profiles ${\cal G}_{\omega e} \left( {\left. {\bf{r}} \right|{\bf{r}}'} \right) \cdot {\bf{u}}_\lambda$ and ${\cal G}_{\omega m} \left( {\left. {\bf{r}} \right|{\bf{r}}'} \right) \cdot {\bf{u}}_\lambda$. In addition, the expression of the Hamiltonian in Eq.(\ref{Hamilt}) reveals that $s$-polariton contribution (${\bf{\hat g}}_{\omega s}^ \dagger   \cdot {\bf{\hat g}}_{\omega s}$) represents the energy of the radiation field whereas the $e$- and $m$-polariton ones (${\bf{\hat f}}_{\omega e }^ +   \cdot {\bf{\hat f}}_{\omega e }$ and ${\bf{\hat f}}_{\omega m }^ +   \cdot {\bf{\hat f}}_{\omega m }$) describe the electric and magnetic energies stored in the object volume. 

We conclude this review of the MLNF with two general observations. First,  the dyadic Green's function ${\cal G}_\omega$, in addition to supplying the full characterization of the object classical electrodynamics, also provides the complete MLNF description of the object quantum electrodynamics since the electric field operator of Eq.(\ref{EleHei}) solely results from the dyadics ${{\cal F}_{\omega s} }$ and ${{\cal G}_{\omega \nu } }$ which are trivially relatated to ${\cal G}_\omega$ by Eqs.(\ref{FwolProp}), (\ref{Womega}) and (\ref{GoeGom}). Second, in the ideal transparent limit ${\mathop{\rm Im}\nolimits} \left( \varepsilon _\omega \right) = {\mathop{\rm Im}\nolimits} \left( \mu _\omega \right) = 0$, the dyadics ${\cal G}_{\omega e}$ and ${\cal G}_{\omega m}$ of Eqs.(\ref{GoeGom}) vanish so that the operators ${\bf{\hat f}}_{\omega \nu }$ do not show up in the electric field operator of Eq.(\ref{EleHei}) and this crucially entails that  $e$- and $m$- polaritons can be neither created nor destroyed, i.e. they have no physical existence. Accordingly, in the presence of a lossless object, $e$- and $m$- polaritons disappear whereas $s$-polaritons reduce to photons of the standard quantum optical description of transparent dielectric media discussed by Glauber in Ref.\cite{Glaub1}.

\section{Quantum optical scattering}
In a general scattering setup, radiation generated by very far sources is launched onto the object and the ensuing scattered radiation is collected by very far detectors. The fact that the distance between object and sources/detectors is much larger than any involved wavelength demands a close analysis of the radiation far-field whose time asymptotical behavior in the far past and in far future describe the ingoing radiation generated by the sources and the outgoing radiation scattered the object, respectively. We perform such analysis by using the electric field operator of Eq.(\ref{EleHei}) whose far-field ($r \rightarrow + \infty$) behavior  at any time $t$, as shown in Appendix C, is given by
\begin{eqnarray} \label{EopFarFie}
&& {\bf{\hat E}}\left( {r{\bf{n}},t} \right)\mathop  \approx \limits_{r \to \infty } \int\limits_0^{ + \infty } {d\omega } \sqrt {\frac{{\hbar k_\omega  }}{{4\pi \varepsilon _0 }}} \left\{ {\frac{{e^{ - ik_\omega  \left( {r + ct} \right)} }}{{ - ir}}{\bf{\hat g}}_{\omega s} \left( { - {\bf{n}}} \right)} \right. \nonumber  \\ 
&&  + \frac{{e^{ik_\omega  \left( {r - ct} \right)} }}{{ir}}\left[ {\int {do_{\bf{m}} } {\cal T}_{\omega ss}  \left( {\left. {\bf{n}} \right|{\bf{m}}} \right) \cdot {\bf{\hat g}}_{\omega s} \left( {\bf{m}} \right)} \right. \nonumber \\ 
&& \left. {\left. { + \sum\limits_{\nu  = e,m} {\int {d^3 {\bf{r}}'\,} {\cal E}_{\omega s \nu } \left( {\left. {\bf{n}} \right|{\bf{r}}'} \right) \cdot {\bf{\hat f}}_{\omega \nu } \left( {{\bf{r}}'} \right)} } \right]} \right\} + h.c., 
\end{eqnarray}
where
\begin{eqnarray} \label{TwAweAwm}
&& {\cal T}_{\omega ss} \left( {\left. {\bf{n}} \right|{\bf{m}}} \right) = \delta \left( {o_{\bf{n}}  - o_{\bf{m}} } \right){\cal I}_{\bf{m}}  + \frac{{ik_\omega  }}{{2\pi }}{\cal S}_\omega  \left( {\left. {\bf{n}} \right|{\bf{m}}} \right), \nonumber \\ 
&& {\cal E}_{\omega se} \left( {\left. {\bf{n}} \right|{\bf{r}}} \right) =  - {\cal W}_\omega  \left( {\left. {\bf{n}} \right|{\bf{r}}} \right)\sqrt {4k_\omega ^3 {\mathop{\rm Im}\nolimits} \left[ {\varepsilon _\omega  \left( {\bf{r}} \right)} \right]} , \nonumber \\ 
&& {\cal E}_{\omega sm} \left( {\left. {\bf{n}} \right|{\bf{r}}} \right) = {\cal W}_\omega  \left( {\left. {\bf{n}} \right|{\bf{r}}} \right) \times \mathord{\buildrel{\lower3pt\hbox{$\scriptscriptstyle\leftarrow$}} 
\over \nabla } _{\bf{r}} \sqrt {4k_\omega  {\mathop{\rm Im}\nolimits} \left[ {\frac{{ - 1}}{{\mu _\omega  \left( {\bf{r}} \right)}}} \right]}. \nonumber \\
\end{eqnarray}
Here ${\cal T}_{\omega s s}$ is the classical transmission dyadic \cite{Saxon1,Rothe1} which, as a consequence of  the properties of the scattering dyadic ${\cal S}_\omega$, satisfies the orthogonality relations ${\bf{n}} \cdot {\cal T}_{\omega s s}  \left( {\left. {\bf{n}} \right|{\bf{m}}} \right) = 0$, ${\cal T}_{\omega s s}  \left( {\left. {\bf{n}} \right|{\bf{m}}} \right) \cdot {\bf{m}} = 0$ and the reciprocity relation
\begin{equation} \label{TosRec}
{\cal T}_{\omega s s} ^T \left( {\left. {\bf{n}} \right|{\bf{m}}} \right) = {\cal T}_{\omega s s}  \left( {\left. { - {\bf{m}}} \right| - {\bf{n}}} \right),
\end{equation}
whereas ${\cal E}_{\omega s e}$ and ${\cal E}_{\omega s m}$ are the electric and magnetic emission dyadics (see below). Equation (\ref{EopFarFie}) reveals that the radiation far-field displays at any time $t$ both ingoing $e^{-ik_\omega  \left( {r + ct} \right)} /r$ and outgoing $e^{ik_\omega  \left( {r - ct} \right)} /r$ spherical wave contributions, the former solely stemming from $s$-polaritons while the latter resulting from the combination of all $s$-, $e$- and $m$- polaritonts through the transmission ${\cal T}_{\omega s s}$ and emission  ${\cal E}_{\omega s e}$, ${\cal E}_{\omega s m}$ dyadics, respectively. This is physically expected since $s$-polaritons are associated to the classical scattering modes ${\bf{F}}_{\omega {\bf{m}}\lambda } \left( {\bf{r}} \right)$ whose modal plane waves display both the ingoing and outgoing behaviors in the far field (since they hit the object from one side but leave it from the opposite side) whereas $e$- and $m$-polaritons pertain medium dipolar sources radiating the fields ${\cal G}_{\omega e} \left( {\left. {\bf{r}} \right|{\bf{r}}'} \right) \cdot {\bf{u}}_\lambda$ and ${\cal G}_{\omega m} \left( {\left. {\bf{r}} \right|{\bf{r}}'} \right) \cdot {\bf{u}}_\lambda$ which are strictly outgoing in the far field due to the Sommerfeld condition. What is remarkable here is that, in the far past limit $t \rightarrow -\infty$, the outgoing spherical wave contribution $e^{ik_\omega  \left( {r - ct} \right)} /r$ in Eq.(\ref{EopFarFie}) evidently vanishes due to the Riemann-Lebesgue lemma (since $r-ct$ stays strictly positive) whereas the ingoing spherical wave contribution $e^{ - ik_\omega  \left( {r + ct} \right)} /r$ does persist if we set $r=-ct$ in the limiting process, this leading to regard the operator 
\begin{eqnarray} \label{Ein}
&& {\bf{\hat E}}^{\left( {in} \right)} \left( {r{\bf{n}},t} \right) =  \int\limits_0^{ + \infty } {d\omega } \sqrt {\frac{{\hbar k_\omega  }}{{4\pi \varepsilon _0 }}} \frac{{e^{ - ik_\omega  \left( {r + ct} \right)} }}{-i r}{\bf{\hat g}}_{\omega s} \left( { - {\bf{n}}} \right) \nonumber \\
&& + h.c.,
\end{eqnarray}
with $r$ and $-t$ very large and positive, as the ingoing field operator in the Heisenberg picture. As far as the far future limit $t \rightarrow +\infty$, a similar reasoning shows that the ingoing spherical wave contribution vanishes and that it is possibile to regard the operator
\begin{eqnarray} \label{EOut}
&& {\bf{\hat E}}^{\left( {out} \right)} \left( {r{\bf{n}},t} \right) = \int\limits_0^{ + \infty } {d\omega } \sqrt {\frac{{\hbar k_\omega  }}{{4\pi \varepsilon _0 }}} \frac{{e^{ik_\omega  \left( {r - ct} \right)} }}{i r}{\bf{\hat G}}_{\omega s} \left( {\bf{n}} \right) \nonumber \\
&& + h.c.,
\end{eqnarray}
with $r$ and $t$ very large and positive, as the outgoing field operator in the Heisenberg picture. Here we have introduced the operator
\begin{eqnarray} \label{Gos}
&& {\bf{\hat G}}_{\omega s} \left( {\bf{n}} \right) = \int {do_{\bf{m}} } {\cal T}_{\omega s s}  \left( {\left. {\bf{n}} \right|{\bf{m}}} \right) \cdot {\bf{\hat g}}_{\omega s} \left( {\bf{m}} \right) \nonumber\\ 
&& + \sum\limits_{\nu  = e,m} {\int {d^3 {\bf{r}}'} {\cal E}_{\omega s \nu } \left( {\left. {\bf{n}} \right|{\bf{r}}'} \right) \cdot {\bf{\hat f}}_{\omega \nu } \left( {{\bf{r}}'} \right)},  
\end{eqnarray}
and it is manifest that it plays in the outgoing field operator of Eq.(\ref{EOut}) the same role the operator ${\bf{\hat g}}_{\omega s}$ plays in the ingoing field operator of Eq.(\ref{Ein}). Such parallelism is made even more explicit by resorting to the plane-wave far-field representations of the ingoing and outgoing modulated spherical waves of Eqs.(\ref{InOutSphWavs}) in Appendix A which, respectively inserted into Eq.(\ref{Ein}) and Eq.(\ref{EOut}), yield
\begin{eqnarray} \label{EinEout}
&& {\bf{\hat E}}^{\left( {in} \right)} \left( {r{\bf{n}},t} \right) = \int\limits_0^{ + \infty } {d\omega } \int {do_{\bf{m}} } \sqrt {\frac{{\hbar k_\omega ^3 }}{{16\pi ^3 \varepsilon _0 }}} e^{ik_\omega  \left( {r{\bf{n}} \cdot {\bf{m}} - ct} \right)} \nonumber \\ 
&& \cdot \left[ {\theta \left( {{\bf{u}}_z  \cdot {\bf{n}}} \right)\theta \left( { - {\bf{u}}_z  \cdot {\bf{m}}} \right) + \theta \left( { - {\bf{u}}_z  \cdot {\bf{n}}} \right)\theta \left( {{\bf{u}}_z  \cdot {\bf{m}}} \right)} \right]{\bf{\hat g}}_{\omega s} \left( {\bf{m}} \right) \nonumber \\ 
&& + h.c., \nonumber \\  
&& {\bf{\hat E}}^{\left( {out} \right)} \left( {r{\bf{n}},t} \right) = \int\limits_0^{ + \infty } {d\omega } \int {do_{\bf{m}} } \sqrt {\frac{{\hbar k_\omega ^3 }}{{16\pi ^3 \varepsilon _0 }}} e^{ik_\omega  \left( {r{\bf{n}} \cdot {\bf{m}} - ct} \right)} \nonumber \\ 
&& \cdot \left[ {\theta \left( {{\bf{u}}_z  \cdot {\bf{n}}} \right)\theta \left( {{\bf{u}}_z  \cdot {\bf{m}}} \right) + \theta \left( { - {\bf{u}}_z  \cdot {\bf{n}}} \right)\theta \left( { - {\bf{u}}_z  \cdot {\bf{m}}} \right)} \right]{\bf{\hat G}}_{\omega s} \left( {\bf{m}} \right)\nonumber \\  
&& + h.c..
\end{eqnarray}
The first of Eqs.(\ref{EinEout}) shows that
the $s$-polariton operators ${\bf{\hat g}}_{\omega s} \left( {\bf{m}} \right)$ contributing to ${\bf{\hat E}}^{\left( {in} \right)}$ depend on the far-field observation direction $\bf n$, since the Heaviside functions select those $s$-polaritons whose modal plane waves $e^{ik_\omega  {\bf{m}} \cdot {\bf{r}}}$ respectively travel downward ${{\bf{u}}_z  \cdot {\bf{m}}}<0$ and upward ${{\bf{u}}_z  \cdot {\bf{m}}}>0$ in the upper ${{\bf{u}}_z  \cdot {\bf{n}}} >0$ and lower ${{\bf{u}}_z  \cdot {\bf{n}}} <0$ half-spheres, respectively. In other words, the ingoing field operator ${\bf{\hat E}}^{\left( {in} \right)}$ comprises all and only the $s$-polariton operators whose modal plane waves are directed toward the object (see Fig.2), as expected. Now it is crucial noting that the outgoing field ${\bf{\hat E}}^{\left( {out} \right)}$ in the second of Eqs.(\ref{EinEout}) has the same structure of ${\bf{\hat E}}^{\left( {in} \right)}$, with only the two differences that the traveling direction ${\bf m}$ of the involved modal plane waves is reversed in the Heaviside functions and that the opertator ${\bf{\hat G}}_{\omega s}$ replaces the operator ${\bf{\hat g}}_{\omega s}$. This observation strongly suggests that the operator ${\bf{\hat G}}_{\omega s}$ should support an $s$-polaritons description, alternative to the one provided by the operators ${\bf{\hat g}}_{\omega s}$,  since, in that case, the second of Eqs.(\ref{EinEout}) would show that the outgoing field operator comprises all and only the $s$-polaritons whose modal plane waves are directed away from the object (see Fig.2), a highly sought-after property. 

In the next Section we will prove these observations by showing that the operator ${\bf{\hat G}}_{\omega s}$ can be supported by two additional operators ${\bf{\hat F}}_{\omega e}$ and ${\bf{\hat F}}_{\omega m}$, namely
\begin{figure}
\centering
\includegraphics[width = 1\linewidth]{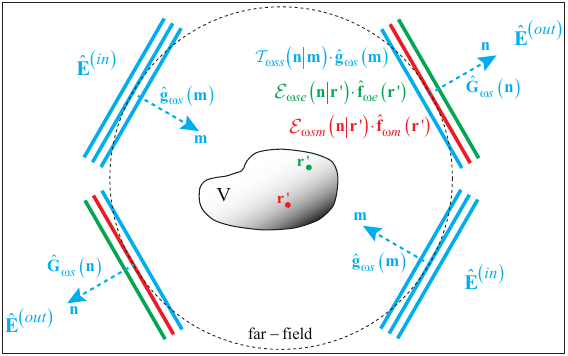}
\caption{Quantum optical scattering in the Heisenberg picture. In the far-field, here  represented by the dashed circle, the ingoing field operator ${\bf{\hat E}}^{\left( {in} \right)}$ ($t \rightarrow - \infty$) results from all and only the operators ${\bf{\hat g}}_{\omega s}$ of the ingoing $s$-polaritons whose modal plane waves are directed toward the object. Analogously, the outgoing field operator ${\bf{\hat E}}^{\left( {out} \right)}$ ($t \rightarrow + \infty$) is the superposition of all and only the operators ${\bf{\hat G}}_{\omega s}$ of the outgoing $s$-polaritons whose modal plane waves are directed away from the object. The outgoing $s$-polariton operators ${\bf{\hat G}}_{\omega s}$ are the linear mixtures of the ingoing polariton operators ${\bf{\hat g}}_{\omega s}$, ${\bf{\hat f}}_{\omega e}$ and ${\bf{\hat f}}_{\omega m}$ produced by the transmission dyadic ${\cal T}_{\omega s s}$ and the electric and magnetic emission dyadics ${\cal E}_{\omega s e}$ and ${\cal E}_{\omega s m}$, respectively.
}
\label{Fig2}
\end{figure}
\begin{eqnarray} \label{Fon}
&& {\bf{\hat F}}_{\omega \nu } \left( {\bf{r}} \right) = \int {do_{\bf{m}} } {\cal M}_{\omega \nu s } \left( {{\bf{r}}\left| {\bf{m}} \right.} \right) \cdot {\bf{\hat g}}_{\omega s} \left( {\bf{m}} \right) \nonumber\\ 
&&  + \sum\limits_{\nu ' = e,m} {\int {d^3 {\bf{r}}'} \;{\cal V}_{\omega \nu \nu '} \left( {\left. {\bf{r}} \right|{\bf{r}}'} \right) \cdot {\bf{\hat f}}_{\omega \nu '} \left( {{\bf{r}}'} \right)},
\end{eqnarray}
where ${\cal M}_{\omega \nu }$ and ${{\cal V}_{\omega \nu \nu '} }$ are dyadics to be found (compare Eq.(\ref{Fon}) with Eq.(\ref{Gos})), in such a way that the bosonic commutation relations
\begin{eqnarray} \label{GFComRel}
&& \left[ {{\bf{\hat G}}_{\omega s} \left( {\bf{n}} \right),{\bf{\hat G}}_{\omega 's}^\dag  \left( {{\bf{n}}'} \right)} \right] = \delta \left( {\omega  - \omega '} \right)\delta \left( {o_{\bf{n}}  - o_{{\bf{n}}'} } \right){\cal I}_{\bf{n}} \nonumber \\ 
&&\left[ {{\bf{\hat F}}_{\omega \nu } \left( {\bf{r}} \right),{\bf{\hat F}}_{\omega '\nu '}^\dag  \left( {{\bf{r}}'} \right)} \right] = \delta \left( {\omega  - \omega '} \right)\delta _{\nu \nu '} \delta \left( {{\bf{r}} - {\bf{r}}'} \right) {\cal I}, \nonumber \\
\end{eqnarray}
holds, together with the commutation relation
\begin{equation} \label{FGComRel}
\left[ {{\bf{\hat F}}_{\omega \nu } \left( {\bf{r}} \right),{\bf{\hat G}}_{\omega 's}^\dag  \left( {\bf{n}} \right)} \right] = 0
\end{equation}
which ensures the vanishing of all the other possible commutation relations. The comparison of Eqs.(\ref{GFComRel}) and Eqs.(\ref{gfComRel}) reveals that ${\bf{\hat G}}_{\omega s}$, ${\bf{\hat F}}_{\omega e}$ and ${\bf{\hat F}}_{\omega m} $ are bosonic operators respectively associated to outgoing $s$-, $e$- and $m$-polaritons meaning that, in the far future $t \rightarrow + \infty$, the outcome of any far-field measurement concering the scattered radiation has to be predicted by resorting to the outgoing field operator ${\bf{\hat E}}^{\left( {out} \right)}$ in Eq.(\ref{EOut}) and hence it relies on the operator ${{\bf{\hat G}}_{\omega s} }$ while, on the other hand, measurements concerning the electromagnetic excitation of the object rely on the operators ${\bf{\hat F}}_{\omega e}$ and ${\bf{\hat F}}_{\omega m}$. A similar reasoning elucidates that the operators ${\bf{\hat g}}_{\omega s}$, ${\bf{\hat f}}_{\omega e}$ and ${\bf{\hat f}}_{\omega m}$ describe, in the far past $t \rightarrow - \infty$, the ingoing $s$-, $e$- and $m$-polaritons. The quantum description of the optical scattering is therefore the following. In the Heisenberg picture we are using here, the quantum state of the electromagnetic field is time-independent and it coincides with the input (far past $t \rightarrow - \infty$) state $\left| {\Psi ^{\left( {in} \right)} } \right\rangle$ whose content of ingoing $s$-polaritons (${\bf{\hat g}}_{\omega s}$) and ingoing $e$- and $m$- polaritons (${\bf{\hat f}}_{\omega e}$,${\bf{\hat f}}_{\omega m}$) respectively accounts for the ingoing radiation and a possible initial electromagnetic excitation of the object. On the other hand, the output (far future $t \rightarrow + \infty$) state coincides with the ingoing one  $\left| {\Psi ^{\left( {out} \right)} } \right\rangle = \left| {\Psi ^{\left( {in} \right)} } \right\rangle$ but its different content of outgoing $s$-polaritons (${\bf{\hat G}}_{\omega s}$) and outgoing $e$- and $m$- polaritons (${\bf{\hat F}}_{\omega e}$, ${\bf{\hat F}}_{\omega m}$) respectively enable to predict the outcome of any measurement regarding the scattered radiation in the far field and the final electromagnetic excitation of the object.

\section{Input-output relation}
As discussed at the end of Section III, the theoretical consistency of the quantum optical scattering description we are developing in this paper is granted by the non-trivial introduction of the outgoing polariton operators in Eqs.(\ref{Gos}) and (\ref{Fon}) satisfying the bosonic commutation relations in Eqs.(\ref{GFComRel}) and (\ref{FGComRel}). To analyze this crucial point, we start by combining Eqs.(\ref{Gos}) and (\ref{Fon}) into the single input-output relation
\begin{equation} \label{InOutRel}
\begin{pmatrix} 
   {{\bf{\hat G}}_{\omega s} }  \\
   {{\bf{\hat F}}_{\omega e} }  \\
   {{\bf{\hat F}}_{\omega m} }  \\
\end{pmatrix} = 
\begin{pmatrix}
   {{\sf T}_{\omega ss} } & {{\sf E}_{\omega se} } & {{\sf E}_{\omega sm} }  \\
   {{\sf M}_{\omega es} } & {{\sf V}_{\omega ee} } & {{\sf V}_{\omega em} }  \\
   {{\sf M}_{\omega ms} } & {{\sf V}_{\omega me} } & {{\sf V}_{\omega mm} }  \\
\end{pmatrix} 
\begin{pmatrix}
   {{\bf{\hat g}}_{\omega s} }  \\
   {{\bf{\hat f}}_{\omega e} }  \\
   {{\bf{\hat f}}_{\omega m} }  \\
\end{pmatrix},
\end{equation}
where each matrix entry is an integral operator  ${\sf O}_{ \mu \mu '} :L_{\mu '}^2  \to L_\mu ^2 $ ($\mu,\mu'=s,e,m$) whose kernel we denote as $\left[\kern-0.15em\left[ {{\sf O}_{\mu \mu '} }  \right]\kern-0.15em\right]\left( {{\bf{x}}\left| {{\bf{x}}'} \right.} \right)$ is the dyadic ${\cal O}_{ \mu \mu '} \left( {{\bf{x}}\left| {{\bf{x}}'} \right.} \right)$, i.e. the operator  acts on a vector field ${{\bf{v}}_{\mu '} } \left( {{\bf{x}}'} \right)  \in L_{\mu '}^2 $ according to
\begin{equation} \label{IntOpO}
\left( {{\sf O}_{ \mu \mu '} {\bf{v}}_{ \mu '} } \right)\left( {\bf{x}} \right) = \int {d{\bf{x}}'} \,{\cal O}_{ \mu \mu '} \left( {{\bf{x}}\left| {{\bf{x}}'} \right.} \right) \cdot {\bf{v}}_{\mu '} \left( {{\bf{x}}'} \right).
\end{equation}
Note that the domain $L_{\mu '}^2$ and image $L_{\mu}^2$ spaces of the operator ${\sf O}_{ \mu \mu '}$ are generally different Hilbert spaces of square integrable vector fields. The space $L^2_s$ is the set of the vector fields ${\bf{v}}_s \left( {\bf{n}} \right)$ defined on the unit sphere $\left| {\bf{n}} \right| = 1$ and tangent to it ${\bf{n}} \cdot {\bf{v}}_s \left( {\bf{n}} \right) = 0$, such that $\left( {{\bf{v}}_s,{\bf{v}}_s} \right)  <  + \infty$ where the scalar product is $\left( {{\bf{v}}_s ,{\bf{w}}_s } \right) = \int {do_{\bf{n}} {\bf{v}}_s^* \left( {\bf{n}} \right) \cdot {\bf{w}}_s \left( {\bf{n}} \right)}$. Furthermore, the spaces $L^2_e$ and $L^2_m$ are two identical copies of the space $L^2_l$, the set of vector fields ${\bf{v}}_l \left( {\bf{r}} \right)$ defined over the whole Eucledian space and such that $\left( {{\bf{v}}_l,{\bf{v}}_l} \right) <  + \infty$, where the scalar product is $\left( {{\bf{v}}_l  ,{\bf{w}}_l  } \right) = \int {d^3 {\bf{r}}} \;{\bf{v}}_l ^* \left( {\bf{r}} \right) \cdot {\bf{w}}_l  \left( {\bf{r}} \right)$. Evidently, the space $L^2_s$ is related to $s$-polaritons and  far-field radiation whereas the spaces $L^2_e$ and $L^2_m$ are respectively tied to $e$- and $m$- polaritons and corresponding dipolar sources located within the object volume. 

Now, by enforcing the commutation relations in Eqs.(\ref{GFComRel}) and (\ref{FGComRel}), as shown in Appendix D, we get 
\begin{eqnarray} \label{MainScEq}
&& \begin{pmatrix}
   {{\sf T}_{\omega ss} } & {{\sf E}_{\omega se} } & {{\sf E}_{\omega sm} }  \\
   {{\sf M}_{\omega es} } & {{\sf V}_{\omega ee} } & {{\sf V}_{\omega em} }  \\
   {{\sf M}_{\omega ms} } & {{\sf V}_{\omega me} } & {{\sf V}_{\omega mm} } \nonumber \\
\end{pmatrix}
\begin{pmatrix}
   {{\sf T}_{\omega ss}^ +  } & {{\sf M}_{\omega es}^ +  } & {{\sf M}_{\omega ms}^ +  }  \\
   {{\sf E}_{\omega se}^ +  } & {{\sf V}_{\omega ee}^ +  } & {{\sf V}_{\omega me}^ +  }  \\
   {{\sf E}_{\omega sm}^ +  } & {{\sf V}_{\omega em}^ +  } & {{\sf V}_{\omega mm}^ +  }  \\
\end{pmatrix} \\
&& = \begin{pmatrix}
   {{\sf I}_{ss} } & 0 & 0  \\
   0 & {{\sf I}_{ee} } & 0  \\
   0 & 0 & {{\sf I}_{mm} }  \\
\end{pmatrix}
\end{eqnarray}
where  ${\sf I}_{ss}$ and ${\sf I}_{ee} = {\sf I}_{mm} = {\sf I}_{ll}$ are the identity operators in the spaces $L^2_s$ and $L^2_l$, respectively, whereas ${\sf O}_{\mu \mu '}^ +  :L_\mu ^2  \to L_{\mu '}^2$ is the Hermitian conjugate of the operator ${\sf O}_{\mu \mu '}$ (see Eq.(\ref{IntOpO})), i.e.
\begin{equation}
\left( {{\sf O}_{\mu \mu '}^ +  {\bf{v}}_{ \mu } } \right)\left( {{\bf{x}}'} \right) = \int {d{\bf{x}}} \,{\cal O}_{ \mu \mu '}^{T*} \left( {{\bf{x}}\left| {{\bf{x}}'} \right.} \right) \cdot {\bf{v}}_{\mu } \left( {\bf{x}} \right),
\end{equation}
or equivalently $[\kern-0.15em[ {{\sf O}_{\mu \mu '}^ +  } ]\kern-0.15em]\left( {{\bf{x}}'\left| {\bf{x}} \right.} \right) = {\cal O}_{\mu \mu '}^{T*} \left( {{\bf{x}}\left| {{\bf{x}}'} \right.} \right)$, where we have introduced the symbol $+$ for the Hermitian conjugation operation in $L^2_\mu$  to distinguish it from the standard dagger $\dagger$ used for the Hermitian conjugation operation in the quantum Hilbert space. Equation (\ref{MainScEq}) is the basic contraint the input-output operator matrix in Eq.(\ref{InOutRel}) has to satisfy for the correct outgoning polariton description and it deserves some discussion. As a structural observation, note that the matrix equation in Eq.(\ref{MainScEq}) amounts to seven independent operator equations (see Eqs.(\ref{COMM4}) of Appendix D) since its $(1,2)$ and $(1,3)$ matrix elements are the Hermitian conjugates of its $(2,1)$ and $(3,1)$ matrix elements, respectively. Now the transmission operator ${{\sf T}_{\omega ss} }$ together with the electric ${{\sf E}_{\omega se} }$ and magnetic ${{\sf E}_{\omega sm} }$ emission operators are explicitly provided by their dyadic kernels in Eqs.(\ref{TwAweAwm}) whereas the six operators ${\sf M}_{\omega \nu s} ,{\sf V}_{\omega \nu \nu '}$ are unknown and this requires that the equation corresponding to the matrix element ($1,1$) of Eq.(\ref{MainScEq}), i.e. 
\begin{equation} \label{TTAAAA}
{\sf T}_{\omega ss} {\sf T}_{\omega ss}^ +   + {\sf E}_{\omega se} {\sf E}_{\omega se}^ +   + {\sf E}_{\omega sm} {\sf E}_{\omega sm}^ +   = {\sf I}_{ss},
\end{equation}
has to be self-consistently satisfied while the remaining six equations have to be solved for the six unknown operators. For comparison purposes, we point out that such seven operator equations are the field-theoretic infinite-dimensional counterparts of the input-output equations used in Ref.\cite{Knoll2} to model the quantum behavior of a dispersive and absorbing four-port optical device. Unfortunately, the method proposed in Ref.\cite{Knoll2} to solve the input-output equations can not be used here since the two involved spaces $L^2_{s}$ and $L^2_l$ are evidently not isomorphic and their identity operators ${\sf I}_{ss}$ and ${\sf I}_{ll}$ are structurally different objects.

Dealing with the coupled and nonlinear integral equations in Eq.(\ref{MainScEq}) only resorting to functional analytic techniques is truly a formidable task so that we here follow a different route based on a peculiar physical analysis of the transmission-emission-absorption interplay pertaining the classical electromagnetic scattering. Such interplay is thoroughly characterized by some original relations connecting the transmission, emission and absorption dyadics to the dyadic Green's function, relations we fully derive in Appendix E by solely resorting to the fundamental integral relation in Eq.(\ref{GreFunInt}). In matrix operatorial form, such relations are 
\begin{eqnarray} \label{RelTAG}
&& \begin{pmatrix}
   {{\sf T}_{\omega ss} } & {{\sf E}_{\omega se} } & {{\sf E}_{\omega sm} }  \\
   {{\sf A}_{\omega es} } & {  {\sf Q}_{\omega ee} } & {  {\sf Q}_{\omega em} }  \\
   {{\sf A}_{\omega ms} } & {  {\sf Q}_{\omega me} } & {  {\sf Q}_{\omega mm} }  \\
\end{pmatrix}  \begin{pmatrix}
   {{\sf T}_{\omega ss}^ +  } & { {{\sf A}_{\omega es}^{ + }  }   } & { {{\sf A}_{\omega ms}^{ + }  }   }  \\
   {{\sf E}_{\omega se}^ +  } & {  {\sf Q}_{\omega ee}^ +  } & {  {\sf Q}_{\omega me}^ +  }  \\
   {{\sf E}_{\omega sm}^ +  } & {  {\sf Q}_{\omega em}^ +  } & {  {\sf Q}_{\omega mm}^ +  }  \\
\end{pmatrix} \nonumber \\
&& = \begin{pmatrix}
   {{\sf I}_{ss} } & 0 & 0  \\
   0 & {{\sf I}_{ee} } & 0  \\
   0 & 0 & {{\sf I}_{mm} }  \\
\end{pmatrix}, 
\end{eqnarray}
where 
\begin{equation} \label{AbsOper}
{\sf A}_{\omega \nu s}  = {\sf E}_{\omega s\nu }^{ + *} {\sf J}_{ss} 
\end{equation}
are the electric and magnetic absorption operators (i.e. $\left[\kern-0.15em\left[ {A_{\omega \nu s} }  \right]\kern-0.15em\right]\left( {\left. {\bf{r}} \right|{\bf{n}}} \right) = {\cal E}_{\omega s\nu }^T \left( { - {\bf{n}}\left| {\bf{r}} \right.} \right)$, see below), ${\sf J}_{ss}$ is the directional inversion operator in the space $L^2_s$, i.e. $\left( {{\sf J}_{ss} {\bf{v}}_s } \right)\left( {\bf{n}} \right) = {\bf{v}}_s \left( { - {\bf{n}}} \right)$, and the integral operators ${\sf Q}_{\omega \nu \nu '}$ are specified by their dyadic kernels ${\cal Q}_{\omega \nu \nu '} \left( {\left. {\bf{r}} \right|{\bf{r}}'} \right) = \left[\kern-0.15em\left[ {{\sf Q}_{\omega \nu \nu '} }  \right]\kern-0.15em\right]\left( {\left. {\bf{r}} \right|{\bf{r}}'} \right)$ given by
\begin{eqnarray} \label{KerQQQQ}
&& \begin{pmatrix}
   {{\cal Q}_{\omega ee} \left( {\left. {\bf{r}} \right|{\bf{r}}'} \right)} & {{\cal Q}_{\omega em} \left( {\left. {\bf{r}} \right|{\bf{r}}'} \right)}  \\
   {{\cal Q}_{\omega me} \left( {\left. {\bf{r}} \right|{\bf{r}}'} \right)} & {{\cal Q}_{\omega mm} \left( {\left. {\bf{r}} \right|{\bf{r}}'} \right)}  \\
\end{pmatrix} =  - \delta \left( {{\bf{r}} - {\bf{r}}'} \right) \begin{pmatrix}
   {\cal I} & 0  \\
   0 & {\cal I}  \\
\end{pmatrix} \nonumber \\ 
&&  - \begin{pmatrix}
   {\sqrt {\frac{{4\pi \varepsilon _0 }}{\hbar }{\mathop{\rm Im}\nolimits} \left[ {\varepsilon _\omega  \left( {\bf{r}} \right)} \right]} }   \\
   {\sqrt {\frac{{4\pi \varepsilon _0 }}{{\hbar k_\omega ^2 }}{\mathop{\rm Im}\nolimits} \left[ {\frac{{ - 1}}{{\mu _\omega  \left( {\bf{r}} \right)}}} \right]} \nabla _{\bf{r}}  \times }  \\
\end{pmatrix}\begin{pmatrix}
   {{\cal G}_{\omega e} \left( {\left. {\bf{r}} \right|{\bf{r}}'} \right)} & {{\cal G}_{\omega m} \left( {\left. {\bf{r}} \right|{\bf{r}}'} \right)}  \\
\end{pmatrix}. \nonumber \\ 
\end{eqnarray}
As a first main physical observation we note that, in the lossless limit ${\mathop{\rm Im}\nolimits} \left[ {\varepsilon _\omega  \left( {\bf{r}} \right)} \right] = {\mathop{\rm Im}\nolimits} \left[ {\mu _\omega  \left( {\bf{r}} \right)} \right] = 0$,  the  dyadics ${\cal E}_{\omega s \nu}$ manifestly vanish (see Eqs.(\ref{TwAweAwm})) whereas the other dyadics are such that ${\cal Q}_{\omega \nu \nu '}  = - \delta _{\nu \nu '} \delta \left( {{\bf{r}} - {\bf{r}}'} \right){\cal I}$ (see Eq.(\ref{KerQQQQ})) so that Eq.(\ref{RelTAG}) reduces to the single operator relation
\begin{equation}
{\sf T}_{\omega ss} {\sf T}_{\omega ss}^ +   = {\sf I}_{ss} 
\end{equation}
which states the well-known unitarity of the transmission operator of transparent objects, in turn amounting to energy conservation of radiation upon scattering \cite{Saxon1,Rothe1}. In the general case, the matrix relation in Eq.(\ref{RelTAG}) yields seven different operator relations (see Eqs.(\ref{AAQQII}) and (\ref{final0}) of Appendix E) since its $(1,2)$ and $(1,3)$ matrix elements are the Hermitian conjugates of its $(2,1)$ and $(3,1)$ matrix elements, respectively. The operator relation corresponding to the matrix element $(1,1)$ of Eq.(\ref{RelTAG}) coincide with Eq.(\ref{TTAAAA}) which is therefore proved and which physically describes the far-field energy balance between radiation transmission (transmission operator ${\sf T}_{\omega ss}$) and radiation emission by the object at expense of its initial internal energy (emission operators ${\sf E}_{\omega s\nu}$, whose name is now justified). Different versions of Eq.(\ref{TTAAAA}) are discussed in the literature concerning inverse scattering problems \cite{Colto1,Kirsc1,Kirsc2,Colto2}, where the transmission operator ${\sf T}_{\omega ss}$ is referred to as the far-field operator, but we emphasize that in such versions the emission part ${\sf E}_{\omega se} {\sf E}_{\omega se}^ +   + {\sf E}_{\omega sm} {\sf E}_{\omega sm}^ + $ is always expressed in terms of the scattering modes instead of the asymptotic amplitude ${{\cal W}_\omega  }$ of the dyadic Green's function (see Eqs.(\ref{TwAweAwm})), in agreement with the fact their intimate connection signaled by Eq.(\ref{FwolProp}) has been unveiled very recently in Ref.\cite{Ciatt3}. The four operator relations corresponding to the matrix elements $(2,2)$, $(2,3)$, $(3,2)$ and $(3,3)$ of Eq.(\ref{RelTAG}) (see Eq.(\ref{AAQQII}) of Appendix E) only involves the  operators ${\sf A}_{\omega \nu s}$ and ${\sf Q}_{\omega \nu \nu '}$ and this leads them to be physically interpreted as describing the balance between absorption of radiation energy by the object (${\sf A}_{\omega \nu s}$) and redistribution of electric-magnetic energy of the object (${\sf Q}_{\omega \nu \nu '}$ ). The remaining two operator relations corresponding to the matrix elements $(2,1)$ and $(3,1)$ of Eq.(\ref{RelTAG}) (see Eq.(\ref{final2}) of Appendix E) jointly display all the operators ${\sf T}_{\omega ss}$, ${{\sf E}_{\omega s\nu} }$, ${{\sf A}_{\omega \nu s} }$ and ${\sf Q}_{\omega \nu \nu '}$ and their structure strengthens the interpretation of the emission ${{\sf E}_{\omega s\nu} }$ and absorption ${\sf A}_{\omega \nu s}$ operators as intermediaries in the energy exchange between radiation in the far field and the object in its volume volume. A further crucial observation is that the two operator matrices in the left hand side of Eq.(\ref{RelTAG}) can be viewed as operators on the direct sum space $L_s^2 \oplus L_e^2  \oplus L_m^2$ where they are the Hermitian conjugate of each other. Now in Appendix E these matrices are also shown to satisfy the relation
\begin{eqnarray} \label{RelTAGConj}
&& \begin{pmatrix}
   {{\sf T}_{\omega ss}^ +  } & { {{\sf A}_{\omega es}^{ + }  }   } & { {{\sf A}_{\omega ms}^{ + }  }   }  \\
   {{\sf E}_{\omega se}^ +  } & {  {\sf Q}_{\omega ee}^ +  } & {  {\sf Q}_{\omega me}^ +  }  \\
   {{\sf E}_{\omega sm}^ +  } & {  {\sf Q}_{\omega em}^ +  } & {  {\sf Q}_{\omega mm}^ +  }  \\
\end{pmatrix} 
\begin{pmatrix}
   {{\sf T}_{\omega ss} } & {{\sf E}_{\omega se} } & {{\sf E}_{\omega sm} }  \\
   {{\sf A}_{\omega es} } & {  {\sf Q}_{\omega ee} } & {  {\sf Q}_{\omega em} }  \\
   {{\sf A}_{\omega ms} } & {  {\sf Q}_{\omega me} } & {  {\sf Q}_{\omega mm} }  \\
\end{pmatrix}  \nonumber \\
&& = \begin{pmatrix}
   {{\sf I}_{ss} } & 0 & 0  \\
   0 & {{\sf I}_{ee} } & 0  \\
   0 & 0 & {{\sf I}_{mm} }  \\
\end{pmatrix},
\end{eqnarray}
so that we conclude that the operator matrix is unitary, an expected property reflecting the energy conservation of the overall radiation-matter system in the scattering event. As a final observation we point out that, apart from Eq.(\ref{TTAAAA}) whose versions different from ours have been addressed in a number of scientific papers, the other six relations provided by Eq.(\ref{RelTAG}) have not been considered yet in the litterature, to the best of our knowledge, and Eq.(\ref{RelTAG}) is one of the main results of the present paper.

We are now prepared to face Eq.(\ref{MainScEq}) since its comparison with Eq.(\ref{RelTAG}) straighforwardly reveals that the six unknown operators ${\sf M}_{\omega \nu s} ,{\sf V}_{\omega \nu \nu '}$ are given by

\begin{eqnarray} \label{solut}
&& \begin{pmatrix}
   {{\sf M}_{\omega es} }  \\
   {{\sf M}_{\omega ms} }  \\
\end{pmatrix} = \begin{pmatrix}
   {{\sf U}_{ee} } & {{\sf U}_{em} }  \\
   {{\sf U}_{me} } & {{\sf U}_{mm} }  \\
\end{pmatrix}\begin{pmatrix}
   {{\sf A}_{\omega es} }  \\
   {{\sf A}_{\omega ms}}  \\
\end{pmatrix}, \nonumber \\ 
&& \begin{pmatrix}
   {{\sf V}_{\omega ee} } & {{\sf V}_{\omega em} }  \\
   {{\sf V}_{\omega me} } & {{\sf V}_{\omega mm} }  \\
\end{pmatrix} =   \begin{pmatrix}
   {{\sf U}_{ee} } & {{\sf U}_{em} }  \\
   {{\sf U}_{me} } & {{\sf U}_{mm} }  \\
\end{pmatrix}\begin{pmatrix}
   {{\sf Q}_{\omega ee} } & {{\sf Q}_{\omega em} }  \\
   {{\sf Q}_{\omega me} } & {{\sf Q}_{\omega mm} }  \\
\end{pmatrix}, \nonumber \\ 
\end{eqnarray}
where ${\sf U}_{\nu \nu'}$ are arbitrary operators satisfying the unitary requirement
\begin{equation} \label{UUU}
\begin{pmatrix}
   {{\sf U}_{ee} } & {{\sf U}_{em} }  \\
   {{\sf U}_{me} } & {{\sf U}_{mm} }  \\
\end{pmatrix}\begin{pmatrix}
   {{\sf U}_{ee}^ +  } & {{\sf U}_{me}^ +  }  \\
   {{\sf U}_{em}^ +  } & {{\sf U}_{mm}^ +  }  \\
\end{pmatrix} = \begin{pmatrix}
   {{\sf I}_{ee} } & 0  \\
   0 & {{\sf I}_{mm} }  \\
\end{pmatrix}.
\end{equation}
The arbitrariness in determining the input-output relation (also encountered in the analysis of Ref.{\cite{Knoll2}}) is here a consequence of the freedom in defining the outgoing $e$- and $m$-polariton operators since the transformation ${\bf{\hat F}}_{\omega \nu }  = {\sf U}_{\nu \nu '} {\bf{\hat F}}'_{\omega \nu '}$ evidently leaves invariant the second of Eqs.(\ref{GFComRel}). After choosing ${\sf U}_{\nu \nu '}  = \delta _{\nu \nu '} {\sf I}_{ll}$ in Eqs.(\ref{solut}) without loss of generality and inserting the resultant expressions into Eq.(\ref{InOutRel}), we get the input-output relation
\begin{equation} \label{InOutRelFinal} 
\begin{pmatrix}
   {{\bf{\hat G}}_{\omega s} }  \\
   {{\bf{\hat F}}_{\omega e} }  \\
   {{\bf{\hat F}}_{\omega m} }  \\
\end{pmatrix} = \begin{pmatrix}
   {{\sf T}_{\omega ss} } & {{\sf E}_{\omega se} } & {{\sf E}_{\omega sm} }  \\
   {{\sf A}_{\omega es} } & { {\sf Q}_{\omega ee} } & { {\sf Q}_{\omega em} }  \\
   {{\sf A}_{\omega ms} } & { {\sf Q}_{\omega me} } & { {\sf Q}_{\omega mm} }  \\
\end{pmatrix}\begin{pmatrix}
   {{\bf{\hat g}}_{\omega s} }  \\
   {{\bf{\hat f}}_{\omega e} }  \\
   {{\bf{\hat f}}_{\omega m} }  \\
\end{pmatrix}.
\end{equation}
which provides a solid foundation for the quantum optical scattering approach we are discussing in the present paper. Note that the above proved classical elecrodynamics unitarity of the operator matrix in Eq.(\ref{InOutRelFinal}) here substantiates the necessary quantum mechanical equivalence of ingoing $({\bf{\hat g}}_{\omega s} ,{\bf{\hat f}}_{\omega e} ,{\bf{\hat f}}_{\omega m})$ and outgoing $({{\bf{\hat G}}_{\omega s} ,{\bf{\hat F}}_{\omega e} ,{\bf{\hat F}}_{\omega m} })$ polariton operators so that they induce two distinct representations of the Hilbert space of the quantum electromagnetic field (see below). As a straightforward consequence of unitarity, the inverse of the input-output relation of Eq.(\ref{InOutRelFinal}) is given by
\begin{equation} \label{InversInOutRel}
\begin{pmatrix}
   {{\bf{\hat g}}_{\omega s} }  \\
   {{\bf{\hat f}}_{\omega e} }  \\
   {{\bf{\hat f}}_{\omega m} }  \\
\end{pmatrix} = \begin{pmatrix}
   {{\sf T}_{\omega ss}^ +  } & {{\sf A}_{\omega es }^ +} & {{\sf A}_{\omega ms }^ +}  \\
   {{\sf E}_{\omega se}^ +  } & { {\sf Q}_{\omega ee}^ +  } & { {\sf Q}_{\omega me}^ +  }  \\
   {{\sf E}_{\omega sm}^ +  } & { {\sf Q}_{\omega em}^ +  } & { {\sf Q}_{\omega mm}^ +  }  \\
\end{pmatrix}\begin{pmatrix}
   {{\bf{\hat G}}_{\omega s} }  \\
   {{\bf{\hat F}}_{\omega e} }  \\
   {{\bf{\hat F}}_{\omega m} }  \\
\end{pmatrix}.
\end{equation}

\section{Ingoing-outgoing quantum state relation}
As outlined at the end of Sec.III, the ingoing state $| {\Psi ^{\left( {in} \right)} } \rangle$ of the electromagnetic field encodes both impinging radiation  (ingoing $s$-polariton) and 
 object excitation (ingoing $e$- and $m$-polaritons) in the far past, whereas the outgoing state $| {\Psi ^{\left( {out} \right)} }  \rangle =  | {\Psi ^{\left( {in} \right)} }  \rangle$, once expressed in terms of outgoing polariton excitations, portrays the scattered radiation and object excitation in the far future. In other words the ingoing-outgoing quantum state relation, enabling the prediction of any quantum optical scattering outcomes, is obtained, in the Heisenberg picture, by switching from the ingoing polariton representation to the outgoing one,  the two representations being linked by the  input-output unitary relation of Eq.(\ref{InOutRelFinal}) 

The ingoing polariton representation is built up by observing that the Hilbert space of the quantum electromagnetic field is  $H = f_s  \otimes f_e  \otimes f_m$,  the tensor product of the Fock spaces $f_\mu$ with $\mu = s,e,m$,  generated by the ingoing polariton operators, so that it is spanned by the basis vectors
\begin{eqnarray} \label{IngBas}
&&\left| {\xi _s^{\left( p \right)} \xi _e^{\left( q \right)} \xi _m^{\left( r \right)} } \right\rangle  \nonumber \\
&& = \prod\limits_{i = 1}^p {\prod\limits_{j = 1}^q {\prod\limits_{k = 1}^r {\frac{{\hat a_s^\dag  \left( {\xi _s^i } \right)\hat a_e^\dag  \left( {\xi _e^j } \right)\hat a_m^\dag  \left( {\xi _m^k } \right)}}{{\sqrt {p!q!r!} }}\left| 0 \right\rangle } } } 
\end{eqnarray}
where $p$,$q$,$r$ are three integers, $\xi _\mu ^{\left( n \right)}  = \left( {\xi _\mu ^1 , \ldots ,\xi _\mu ^n } \right)$ are $n$-tuples of the polaritonic variables 
\begin{eqnarray}
&& \xi _s  = \left( {\omega ,{\bf{n}},\lambda } \right),\nonumber \\
&& \xi _e  = \xi _m  = \left( {\omega ,{\bf{r}},\lambda } \right),
\end{eqnarray} 
$\hat a_\mu  \left( {\xi _\mu  } \right)$ are the annihilation operators defined in Eqs.(\ref{Opegosfos}) and  ${\left| 0 \right\rangle }$ is the normalized vacuum state of $H$, i.e. $\hat a_\mu  \left( {\xi _\mu  } \right)\left| 0 \right\rangle  = 0$ and ${\left\langle {0} \mathrel{\left | {\vphantom {0 0}} \right. \kern-\nulldelimiterspace} {0} \right\rangle  = 1}$. The state $| {\xi _s^{\left( p \right)} \xi _e^{\left( q \right)} \xi _m^{\left( r \right)} } \rangle$ evidently contains $p$ $s$-polaritons, each labelled by its frequency $\omega^i$, direction ${\bf n}^i$ and polarization ${\bf{e}}_{{\bf{n}}^i \lambda ^i }$, together with $q$ $e$-polaritons and $r$ $m$-polaritons associated to their dipolar sources of  frequencies $\omega^i$, position ${\bf r}^i$ and cartesian direction ${\bf u}_{\lambda^i}$. A well-known result of bosonic field theory is that the commutation relations in Eqs.(\ref{gfComRel}) entail that the states in Eq.(\ref{IngBas}) satisfy the orthonormality and completeness relations
\begin{eqnarray} \label{OrtCom}
&& \left\langle {{\xi _s^{\left( p \right)} \xi _e^{\left( q \right)} \xi _m^{\left( r \right)} }}
 \mathrel{\left | {\vphantom {{\xi _s^{\left( p \right)} \xi _e^{\left( q \right)} \xi _m^{\left( r \right)} } {{\xi '}_s^{( {p'} )} {\xi '}_e^{( {q'} )} {\xi '}_m^{ {r'} } }}}
 \right. \kern-\nulldelimiterspace}
 {{{\xi '}_s^{ ({p'}) } {\xi '}_e^{( {q'} )} {\xi '}_m^{( {r'} )} }} \right\rangle \nonumber \\ 
&&  = \Delta _s \left( {\xi _s^{\left( p \right)} \left| {{\xi '}_s^{( {p'} )} } \right.} \right)\Delta _e \left( {\xi _e^{\left( q \right)} \left| {{\xi '}_e^{( {q'} )} } \right.} \right)\Delta _m \left( {\xi _m^{\left( r \right)} \left| {{\xi '}_m^{( {r'} )} } \right.} \right), \nonumber \\ 
&& \sum\limits_{p,q,r = 0}^\infty  {\int {d\xi _s^{\left( p \right)} d\xi _e^{\left( q \right)} d\xi _m^{\left( r \right)} } } \left| {\xi _s^{\left( p \right)} \xi _e^{\left( q \right)} \xi _m^{\left( r \right)} } \right\rangle \left\langle {\xi _s^{\left( p \right)} \xi _e^{\left( q \right)} \xi _m^{\left( r \right)} } \right| = \hat I, \nonumber \\ 
\end{eqnarray}
where
\begin{equation} \label{SimmDelta}
\Delta _\mu  \left( {\xi _\mu ^{\left( n \right)} \left| {{\xi '}_\mu ^{( {n'} )} } \right.} \right) = \frac{{\delta _{n,n'} }}{{n!}}\sum\limits_{\pi_n  \in S_n } {\prod\limits_{i = 1}^n {\delta_\mu \left( {\xi _\mu ^i  - {\xi '}_\mu ^{\pi_n \left( i \right)} } \right)} } 
\end{equation}
is the simmetrized delta function with $\pi_n$ spanning the $n!$ permutations of the symmetric group $S_n$ and 
\begin{eqnarray} \label{PolDeltas}
&& \delta _s \left( {\xi _s  - \xi '_s } \right) = \delta \left( {\omega  - \omega '} \right)\delta \left( {o_{\bf{n}}  - o_{{\bf{n}}'} } \right)\delta _{\lambda \lambda '}, \nonumber \\ 
&& \delta _e \left( {\xi _e  - \xi '_e } \right) = \delta _m \left( {\xi _m  - \xi '_m } \right) = \delta \left( {\omega  - \omega '} \right)\delta \left( {{\bf{r}} - {\bf{r}}'} \right)\delta _{\lambda \lambda '}, \nonumber \\ 
\end{eqnarray}
whereas the symbols
\begin{eqnarray} \label{PolInteg}
&& \int {d\xi _s^{\left( n \right)} }  = \int_0^{ + \infty } {\left( {\prod\limits_{i = 1}^n {d\omega ^i } } \right)} \int {do_{{\bf{n}}^i } } \sum\limits_{\lambda ^i  = 1}^2 , \nonumber \\ 
&& \int {d\xi _e^{\left( n \right)} }  = \int {d\xi _m^{\left( n \right)} }  = \int_0^{ + \infty } {\left( {\prod\limits_{i = 1}^n {d\omega ^i } } \right)} \int {d^3 {\bf{r}}^i } \sum\limits_{\lambda ^i  = 1}^3 \nonumber \\
\end{eqnarray}
concisely denote summations over $n$-tuples of polaritonic variables. As a consequence of Eqs.(\ref{OrtCom}) any ingoing state admits the expansion
\begin{eqnarray} \label{PsiIngo}
&& \left| {\Psi ^{\left( {in} \right)} } \right\rangle  = \sum\limits_{p,q,r = 0}^\infty  {\int {d\xi _s^{\left( p \right)} d\xi _e^{\left( q \right)} d\xi _m^{\left( r \right)} } } \nonumber \\ 
&&  \cdot \Psi _{pqr}^{\left( {in} \right)} \left( {\xi _s^{\left( p \right)} ,\xi _e^{\left( q \right)} ,\xi _m^{\left( r \right)} } \right)\left| {\xi _s^{\left( p \right)} \xi _e^{\left( q \right)} \xi _m^{\left( r \right)} } \right\rangle 
\end{eqnarray}
where the ingoing wavefunction $\Psi _{pqr}^{\left( {in} \right)} ( {\xi _s^{\left( p \right)} ,\xi _e^{\left( q \right)} ,\xi _m^{\left( r \right)} } ) =  \langle {{\xi _s^{\left( p \right)} \xi _e^{\left( q \right)} \xi _m^{\left( r \right)} }}
 \mathrel{ | {\vphantom {{\xi _s^{\left( p \right)} \xi _e^{\left( q \right)} \xi _m^{\left( r \right)} } {\Psi ^{\left( {in} \right)} }}}  \kern-\nulldelimiterspace} {{\Psi ^{\left( {in} \right)} }} \rangle $ is normalized according to
\begin{equation}
\sum\limits_{p,q,r = 0}^\infty  {\int {d\xi _s^{\left( p \right)} d\xi _e^{\left( q \right)} d\xi _m^{\left( r \right)} } } \;\left| {\Psi _{pqr}^{\left( {in} \right)} \left( {\xi _s^{\left( p \right)} ,\xi _e^{\left( q \right)} ,\xi _m^{\left( r \right)} } \right)} \right|^2  = 1,
\end{equation}
and it is separately symmetric in the three sets of polariton variables, i.e. for any three permutations $\pi _p  \in S_p$, $\pi _q  \in S_q$ and $\pi _r  \in S_r$ the relation
\begin{equation}
\Psi _{pqr}^{\left( {in} \right)} \left( {\pi _p \xi _s^{\left( p \right)} ,\pi _q \xi _e^{\left( q \right)} ,\pi _r \xi _m^{\left( r \right)} } \right) = \Psi _{pqr}^{\left( {in} \right)} \left( {\xi _s^{\left( p \right)} ,\xi _e^{\left( q \right)} ,\xi _m^{\left( r \right)} } \right),
\end{equation}
holds, where $\pi _n \xi _\mu ^{\left( n \right)}  = ( {\xi _\mu ^{\pi _n \left( 1 \right)} , \ldots ,\xi _\mu ^{\pi _n \left( n \right)} } )$. 

Due to the equivalence of the commutation realtions of Eqs.(\ref{gfComRel}) and (\ref{GFComRel}), the above construction of the ingoing polariton representation can be repeated verbatim for outgoing polaritons. To distinguish the two representations, we hereafter use capital letters to label objects in the outgoing polariton representation. Accordingly, we introduce the annihilation operators $\hat A_\mu  \left( {\Xi _\mu  } \right)$ through the expansions 
\begin{eqnarray} \label{OpeGosFos}
&& {\bf{\hat G}}_{\Omega s} \left( {\bf{N}} \right) = \sum\limits_{\Lambda  = 1}^2 {\hat A_s \left( {\Xi _s } \right){\bf{e}}_{{\bf{N}}\Lambda } } , \nonumber \\ 
&& {\bf{\hat F}}_{\Omega \nu } \left( {\bf{R}} \right) = \sum\limits_{\Lambda  = 1}^3 {\hat A_\nu  \left( {\Xi _\nu  } \right){\bf{u}}_\Lambda },  
\end{eqnarray}
where 
\begin{eqnarray}
&& \Xi _s  = \left( {\Omega ,{\bf{N}},\Lambda } \right), \nonumber \\
&& \Xi _e  = \Xi _m  = \left( {\Omega ,{\bf{R}},\Lambda } \right), 
\end{eqnarray}
and we regard the Hilbert space $H$ of the quantum electromagnetic field also as the tensor product of the Fock spaces $F_\mu$ (different from the $f_\mu$ spaces) generated by the outgoing polariton operators, i.e. $H = F_s  \otimes F_e  \otimes F_m$ in turn spanned by the basis states
\begin{eqnarray} \label{OutgBas}
&& \left| {\Xi _s^{\left( P \right)} \Xi _e^{\left( Q \right)} \Xi _m^{\left( R \right)} } \right\rangle  \nonumber \\
&& = \prod\limits_{I = 1}^P {\prod\limits_{K = 1}^R {\prod\limits_{J = 1}^Q {\frac{{\hat A_s^\dag  \left( {\Xi _s^I } \right)\hat A_e^\dag  \left( {\Xi _e^J } \right)\hat A_m^\dag  \left( {\Xi _m^K } \right)}}{{\sqrt {P!Q!R!} }}\left| 0 \right\rangle } } }.
\end{eqnarray}
Here the meaning of any capital letter symbol is precisely the same of its lowercase letter counterpart in the ingoing polariton representation and, consequently, Eqs.(\ref{OrtCom}) with substituted capital letter symbols characterize the outgoing polariton representation. Accordingly, the outgoing state can be written as 
\begin{eqnarray} \label{PsiOutgo}
&& \left| {\Psi ^{\left( {out} \right)} } \right\rangle  = \sum\limits_{P,Q,R = 0}^\infty  {\int {d\Xi _s^{\left( P \right)} d\Xi _e^{\left( Q \right)} d\Xi _m^{\left( R \right)} } } \nonumber \\ 
&&  \cdot \Psi _{PQR}^{\left( {out} \right)} \left( {\Xi _s^{\left( P \right)} ,\Xi _e^{\left( Q \right)} ,\Xi _m^{\left( R \right)} } \right)\left| {\Xi _s^{\left( P \right)} \Xi _e^{\left( Q \right)} \Xi _m^{\left( R \right)} } \right\rangle 
\end{eqnarray}
whose outgoing wavefunction $\Psi _{PQR}^{\left( {out} \right)}$ is connected to the ingoing one $\Psi _{pqr}^{\left( {in} \right)}$ by the Heisenberg equality $| {\Psi ^{\left( {out} \right)} }  \rangle =  | {\Psi ^{\left( {in} \right)} }  \rangle$ and it is easily seen to be given by
\begin{eqnarray} \label{StatIngOut}
&& \Psi _{PQR}^{\left( {out} \right)} \left( {\Xi _s^{\left( P \right)} ,\Xi _e^{\left( Q \right)} ,\Xi _m^{\left( R \right)} } \right) = \sum\limits_{p,q,r = 0}^\infty  {\int {d\xi _s^{\left( p \right)} d\xi _e^{\left( q \right)} d\xi _m^{\left( r \right)} } }  \nonumber \\ 
&&  \cdot \left\langle {{\Xi _s^{\left( P \right)} \Xi _e^{\left( Q \right)} \Xi _m^{\left( R \right)} }}
 \mathrel{\left | {\vphantom {{\Xi _s^{\left( P \right)} \Xi _e^{\left( Q \right)} \Xi _m^{\left( R \right)} } {\xi _s^{\left( p \right)} \xi _e^{\left( q \right)} \xi _m^{\left( r \right)} }}}
 \right. \kern-\nulldelimiterspace}
 {{\xi _s^{\left( p \right)} \xi _e^{\left( q \right)} \xi _m^{\left( r \right)} }} \right\rangle \Psi _{pqr}^{\left( {in} \right)} \left( {\xi _s^{\left( p \right)} ,\xi _e^{\left( q \right)} ,\xi _m^{\left( r \right)} } \right), \nonumber \\ 
\end{eqnarray}
which we hereafter label as ingoing-outgoing quantum state relation.
\begin{figure}
\centering
\includegraphics[width = 1\linewidth]{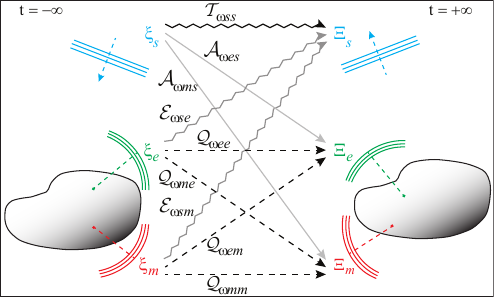}
\caption{Role played by the classical dyadics in the quantum optical scattering. Each dyadic rules the quantum transition from an ingoing polariton $\xi_\mu$ in far past $t = -\infty$ to an outgoing polariton $\Xi _ \tau$ in the far future $t = +\infty$. The involved physical energy exchange processes are: elastic scattering (black wavy arrow), radiation absorption (grey solid arrows), radiation emission (grey wavy arrows) and object energy redistribution (black dashed arrows).}
\label{Fig3}
\end{figure}
Here $\langle {{\Xi _s^{\left( P \right)} \Xi _e^{\left( Q \right)} \Xi _m^{\left( R \right)} }} \mathrel{ | {\vphantom {{\Xi _s^{\left( P \right)} \Xi _e^{\left( Q \right)} \Xi _m^{\left( R \right)} } {\xi _s^{\left( p \right)} \xi _e^{\left( q \right)} \xi _m^{\left( r \right)} }}}   \kern-\nulldelimiterspace}  {{\xi _s^{\left( p \right)} \xi _e^{\left( q \right)} \xi _m^{\left( r \right)} }} \rangle $ is the change of basis matrix which implements the switch from the ingoing to the outgoing representation and, as shown in Appendix F, it is given by
\begin{eqnarray} \label{ChaBasMat}
&&  \left\langle {{\Xi _s^{\left( P \right)} \Xi _e^{\left( Q \right)} \Xi _m^{\left( R \right)} }}
 \mathrel{\left | {\vphantom {{\Xi _s^{\left( P \right)} \Xi _e^{\left( Q \right)} \Xi _m^{\left( R \right)} } {\xi _s^{\left( p \right)} \xi _e^{\left( q \right)} \xi _m^{\left( r \right)} }}}
 \right. \kern-\nulldelimiterspace}
 {{\xi _s^{\left( p \right)} \xi _e^{\left( q \right)} \xi _m^{\left( r \right)} }} \right\rangle  = \frac{{\delta _{Nn} }}{{\sqrt {P!Q!R!} \sqrt {p!q!r!} }} \nonumber \\ 
&&  \cdot \sum\limits_{\pi _N  \in S_N } {\prod\limits_{i = 1}^N {Z_{\chi \left( {\bar \Xi ^i } \right)\chi \left( {\bar \xi ^{\pi _N \left( i \right)} } \right)} \left( {\bar \Xi ^i \left| {\bar \xi ^{\pi _N \left( i \right)} } \right.} \right)} } ,
\end{eqnarray}
where $N = P+Q+R$ and $n=p+q+r$ are the total numbers of the outgoing and ingoing polaritons, the $N$-tuple $\bar \Xi ^{\left( N \right)}$ and the $n$-tuple $\bar \xi ^{\left( n \right)}$ are
\begin{eqnarray}
&& \left( {\bar \Xi ^1  \ldots \bar \Xi ^N } \right) = \left( {\Xi _s^1 , \ldots ,\Xi _s^P ,\Xi _e^1 , \ldots ,\Xi _e^Q ,\Xi _m^1 , \ldots ,\Xi _m^R } \right), \nonumber \\ 
&& \left( {\bar \xi ^1  \ldots \bar \xi ^n } \right) = \left( {\xi _s^1 , \ldots ,\xi _s^p ,\xi _e^1 , \ldots ,\xi _e^q ,\xi _m^1 , \ldots ,\xi _m^r } \right), 
\end{eqnarray}
$\chi \left( {\xi _\mu  } \right) = \mu$ is the character $\mu = s,e,m$ of the variable $\xi_\mu$ and the $3 \times 3$ matrix  $Z_{\tau \mu } \left( {\Xi _\tau  \left| {\xi _\mu  } \right.} \right)$ is specified by its nine matrix entries
\begin{eqnarray} \label{ZmatrEntr}
&& Z_{ss} \left( {\Xi _s \left| {\xi _s } \right.} \right) = \delta \left( {\Omega  - \omega } \right){\bf{e}}_{{\bf{N}}\Lambda }  \cdot {\cal T}_{\omega ss} \left( {\left. {\bf{N}} \right|{\bf{n}}} \right) \cdot {\bf{e}}_{{\bf{n}}\lambda }  , \nonumber \\ 
&& Z_{es} \left( {\Xi _e \left| {\xi _s } \right.} \right) = \delta \left( {\Omega  - \omega } \right){\bf{u}}_\Lambda   \cdot {\cal A}_{\omega es} \left( {\left. {  {\bf{n}}} \right|{\bf{R}}} \right) \cdot {\bf{e}}_{{\bf{n}}\lambda }  , \nonumber \\ 
&& Z_{ms} \left( {\Xi _m \left| {\xi _s } \right.} \right) = \delta \left( {\Omega  - \omega } \right){\bf{u}}_\Lambda   \cdot {\cal A}_{\omega ms} \left( {\left. {  {\bf{n}}} \right|{\bf{R}}} \right) \cdot {\bf{e}}_{{\bf{n}}\lambda }  , \nonumber \\ 
&& Z_{se} \left( {\Xi _s \left| {\xi _e } \right.} \right) = \delta \left( {\Omega  - \omega } \right){\bf{e}}_{{\bf{N}}\Lambda }  \cdot {\cal E}_{\omega se} \left( {\left. {\bf{N}} \right|{\bf{r}}} \right) \cdot {\bf{u}}_\lambda   , \nonumber \\ 
&& Z_{ee} \left( {\Xi _e \left| {\xi _e } \right.} \right) = \delta \left( {\Omega  - \omega } \right){\bf{u}}_\Lambda   \cdot {\cal Q}_{\omega ee} \left( {{\bf{R}}\left| {\bf{r}} \right.} \right) \cdot {\bf{u}}_\lambda   , \nonumber \\ 
&& Z_{me} \left( {\Xi _m \left| {\xi _e } \right.} \right) = \delta \left( {\Omega  - \omega } \right){\bf{u}}_\Lambda   \cdot {\cal Q}_{\omega me} \left( {{\bf{R}}\left| {\bf{r}} \right.} \right) \cdot {\bf{u}}_\lambda   , \nonumber \\ 
&& Z_{sm} \left( {\Xi _s \left| {\xi _m } \right.} \right) = \delta \left( {\Omega  - \omega } \right){\bf{e}}_{{\bf{N}}\Lambda }  \cdot {\cal E}_{\omega sm} \left( {\left. {\bf{N}} \right|{\bf{r}}} \right) \cdot {\bf{u}}_\lambda   , \nonumber \\ 
&& Z_{em} \left( {\Xi _e \left| {\xi _m } \right.} \right) = \delta \left( {\Omega  - \omega } \right){\bf{u}}_\Lambda   \cdot {\cal Q}_{\omega em} \left( {{\bf{R}}\left| {\bf{r}} \right.} \right) \cdot {\bf{u}}_\lambda   , \nonumber \\ 
&& Z_{mm} \left( {\Xi _m \left| {\xi _m } \right.} \right) = \delta \left( {\Omega  - \omega } \right){\bf{u}}_\Lambda   \cdot {\cal Q}_{\omega mm} \left( {{\bf{R}}\left| {\bf{r}} \right.} \right) \cdot {\bf{u}}_\lambda   . \nonumber \\ 
\end{eqnarray}
We emphasize that the ingoing wavefunction $\Psi _{pqr}^{\left( {in} \right)}$ accounts for the most general quantum state pertaining the radiation-object system in the far past, with no restrictions on polariton kind, number or entanglement degree, and hence the ingoing-outgoing quantum state relation in Eq.(\ref{StatIngOut}) embodies a  comprehensive model of quantum optical scattering and it is the main result of the present paper. 

The physical content of Eq.(\ref{StatIngOut}) is that the change of basis matrix in Eq.(\ref{ChaBasMat}) represents the probability amplitude of the basic scattering process $\xi _s^{\left( p \right)} \xi _e^{\left( q \right)} \xi _m^{\left( r \right)}  \to \Xi _s^{\left( P \right)} \Xi _e^{\left( Q \right)} \Xi _m^{\left( R \right)}$ consisting in the destruction of the initial ingoing polaritons $\xi _s^{\left( p \right)}$, $\xi _e^{\left( q \right)}$, $\xi _m^{\left( r \right)} $ followed by the creation of the outgoing polaritons $\Xi _s^{\left( P \right)}$, $\Xi _e^{\left( Q  \right)}$, $\Xi _m^{\left( R \right)}$ (the proability of the process being $| {\langle {{\Xi _s^{\left( P \right)} \Xi _e^{\left( Q \right)} \Xi _m^{\left( R \right)} }} \mathrel{ | {\vphantom {{\Xi _s^{\left( P \right)} \Xi _e^{\left( Q \right)} \Xi _m^{\left( R \right)} } {\xi _s^{\left( p \right)} \xi _e^{\left( q \right)} \xi _m^{\left( r \right)} }}}   \kern-\nulldelimiterspace} {{\xi _s^{\left( p \right)} \xi _e^{\left( q \right)} \xi _m^{\left( r \right)} }} \rangle } |^2 d\Xi _s^{\left( P \right)} d\Xi _e^{\left( Q \right)} d\Xi _m^{\left( R \right)}$). What is remarkable here is that such basic quantum scattering process is fully ruled by the classical dyadic Green's function ${\cal G}_\omega$ since its probability amplitude solely results from the  dyadics ${\cal T}_{\omega ss}$, ${\cal E}_{\omega s\nu}$, ${\cal A}_{\omega \nu s}$ and $ {\cal Q}_{\omega \nu \nu '}$ through the matrix $Z_{\mu \tau }$ (see Eqs.(\ref{ZmatrEntr})). Note that this observation represents a far-reaching generalization of the well known fact that the quantum beavior of a lossy beam-splitter is characterized by its classical reflectivity and transmissivity together with appropriare Langevin operators \cite{Barne2}. An essential trait of the basic scattering process is the conservation of the total polariton number as a consequence of the $\delta_{N,n}$ factor in its probability amplitude, an expected physical feature resulting from energy conservation and ensuing unitarity discussed in Sec.IV. The particular simple form of Eq.(\ref{ChaBasMat}) unveils the physical meaning of the matrix entry $Z_{\tau \mu } \left( {\Xi _\tau  \left| {\xi _\mu  } \right.} \right)$ as the probability amplitude of the elementary process $\xi_\mu \rightarrow \Xi_\tau$ where the specific ingoing polariton $\xi _\mu$ is destroyed with subsequent creation of the specific ougoing polariton $\Xi _\tau$ and hence, due to Eqs.(\ref{ZmatrEntr}), this enable to directly shift the physical interpretation of the classical dyadics discussed in Sec.IV to the quantum domain. As a matter of fact, with reference to the pictorial sketch in Fig.3, the transmission dyadic ${\cal T}_{\omega ss}$ rules the transition from an ingoing $s$-polariton to an outgoing $s$-one, an event associated to the elastic scattering of radiation (black wavy arrow); the absorption dyadics ${\cal A}_{\omega \nu s}$ are responsible for the absorption of an ingoing $s$-polariton and the generation of outgoing $\nu$-polaritons, thus accounting for the absorption of radiation energy by the object (grey solid arrows); conversely, the emission dyadics ${\cal E}_{\omega s\nu}$ provide the destruction of ingoing $\nu$-polaritons with subsequent emission of outgoing $s$-polaritons, events describing the emission of radiation at the expense of the intial object energy (grey wavy arrows); finally, the dyadics $ {\cal Q}_{\omega \nu \nu '}$ mediate the mutual $e$ and $m$ polariton transitions, thus accounting for the redistribution in the internal energy of the object (black dashed arrows). Note that in the elementary process $\xi_\mu \rightarrow \Xi_\tau$ the polariton frequency does not chage due to the $\delta(\Omega - \omega)$ factors in Eqs.(\ref{ZmatrEntr}), as expected from the linearity of the object optical response. Besides, Eq.(\ref{ChaBasMat}) reveals that all the possible elementary processes $\bar \xi ^j \rightarrow \bar \Xi ^i$ composing the basic process $\xi _s^{\left( p \right)} \xi _e^{\left( q \right)} \xi _m^{\left( r \right)}  \to \Xi _s^{\left( P \right)} \Xi _e^{\left( Q \right)} \Xi _m^{\left( R \right)}$ are fully symmetrized to yield the overall probability amplitude, in agreement with the bosonic symmetry under polariton exchange; this is mathematically stated in Eq.(\ref{ChaBasMat}) by the permanent of the $N \times N$ matrix $\bar Z_{ij}  = Z_{\chi \left( {\bar \Xi ^i } \right)\chi \left( {\bar \xi ^j } \right)} \left( {\bar \Xi ^i \left| {\bar \xi ^j } \right.} \right)$ (see Appendix F), which is invariant under arbitrary permutations of the rows and/or columns of $\bar Z$. As a final observation on Eq.(\ref{ChaBasMat}), we point out that in the lossless limit ${\mathop{\rm Im}\nolimits} \left[ {\varepsilon _\omega  \left( {\bf{r}} \right)} \right] = {\mathop{\rm Im}\nolimits} \left[ {\mu _\omega  \left( {\bf{r}} \right)} \right] = 0$ all the classical dyadics vanish except for the transmission one $
{\cal T}_{\omega ss}$ so that, since in this case $s$-polaritons reduce to to standard photons whereas $e$- and $m$-polaritons disappear, Eq.(\ref{ChaBasMat}) reduces to
\begin{equation}
\left\langle {{\Xi _s^{\left( N \right)} }}
 \mathrel{\left | {\vphantom {{\Xi _s^{\left( N \right)} } {\xi _s^{\left( n \right)} }}}
 \right. \kern-\nulldelimiterspace}
 {{\xi _s^{\left( n \right)} }} \right\rangle  = \frac{{\delta _{Nn} }}{{N!}}\sum\limits_{\pi _N  \in S_N } {\prod\limits_{i = 1}^N {Z_{ss} \left( {\Xi _s^i \left| {\xi _s^{\pi _N \left( i \right)} } \right.} \right)} } 
\end{equation}
which coincides, in the notation adopted in this paper, with the photon scattering relation existing in literature (see, e.g., Chapter 8 of Ref.\cite{Garri1}).

\begin{figure}
\centering
\includegraphics[width = 1\linewidth]{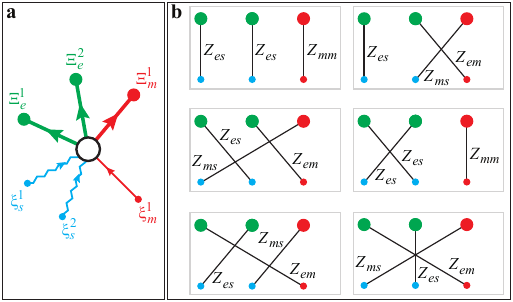}
\caption{Basic scattering process $\xi _s^{\left( 2 \right)}  \xi _m^{\left( 1 \right)}  \to \Xi _e^{\left( 2 \right)} \Xi _m^{\left( 1 \right)}$. {\bf a}. Diagrammatic representation where $s$-, $e$-, and $m$- polaritons are reported in blue, green and red, respectively (compare with Fig.1). The wavy lines label $s$-polaritons while, thin and thick lines are used for ingoing and outgoing $e$- and $m$-polaritons, respectively. {\bf b} The six contributing elementary processes. Each ingoing polariton $\xi_\mu^j$ (small solid circles) turns into an outgoing polariton $\Xi_\tau^i$ (large solid circles) with probability amplitude given by the classical scattering quantity $Z_{\tau \mu } \left( {\Xi _\tau ^i \left| {\xi _\mu ^j } \right.} \right)$ (black lines).}
\label{Fig4}
\end{figure}

An example may help to clarify the usage of such general formalism. Suppose that the ingoing radiation comprises two $s$-polaritons $\xi _s^{\left( 2 \right)}  = \left( {\xi _s^1 ,\xi _s^2 } \right)$ and the object initially hosts no $e$-polaritons (denoted with $\xi _s^{\left( 0 \right)}$) and a $m$-polariton $\xi _m^{\left( 1 \right)}  = \left( {\xi _m^1 } \right)$. We consider the basic scattering process where radiation is eventually full absorbed with no outgoing $s$-polaritons (denoted with $\Xi _s^{\left( 0 \right)}$) and the object is left with two $e$-polaritons $\Xi _e^{\left( 2 \right)}  = \left( {\Xi _e^1 ,\Xi _e^2 } \right)$ and a single $m$-polariton $\Xi _m^{\left( 1 \right)}  = \left( {\Xi _m^1 } \right)$ (see the diagrammatic representation of the process in subplot {\bf a} of Fig.4). The probability amplitude of the basic scattering process $\xi _s^{\left( 2 \right)} \xi _m^{\left( 1 \right)}  \to  \Xi _e^{\left( 2 \right)} \Xi _m^{\left( 1 \right)}$, evaluated from Eq.(\ref{ChaBasMat}), is 
\begin{eqnarray} \label{ExampProc}
&& \left\langle {{\Xi _s^{\left( 0 \right)} \Xi _e^{\left( 2 \right)} \Xi _m^{\left( 1 \right)} }}
 \mathrel{\left | {\vphantom {{\Xi _s^{\left( 0 \right)} \Xi _e^{\left( 2 \right)} \Xi _m^{\left( 1 \right)} } {\xi _s^{\left( 2 \right)} \xi _e^{\left( 0 \right)} \xi _m^{\left( 1 \right)} }}}
 \right. \kern-\nulldelimiterspace}
 {{\xi _s^{\left( 2 \right)} \xi _e^{\left( 0 \right)} \xi _m^{\left( 1 \right)} }} \right\rangle \nonumber \\ 
&&  = \frac{1}{2}\left[ {Z_{es} \left( {\Xi _e^1 \left| {\xi _s^1 } \right.} \right)Z_{es} \left( {\Xi _e^2 \left| {\xi _s^2 } \right.} \right)Z_{mm} \left( {\Xi _m^1 \left| {\xi _m^1 } \right.} \right)} \right. , \nonumber \\ 
  && + Z_{es} \left( {\Xi _e^1 \left| {\xi _s^1 } \right.} \right)Z_{ms} \left( {\Xi _m^1 \left| {\xi _s^2 } \right.} \right)Z_{em} \left( {\Xi _e^2 \left| {\xi _m^1 } \right.} \right) , \nonumber \\ 
  && + Z_{ms} \left( {\Xi _m^1 \left| {\xi _s^1 } \right.} \right)Z_{es} \left( {\Xi _e^1 \left| {\xi _s^2 } \right.} \right)Z_{em} \left( {\Xi _e^2 \left| {\xi _m^1 } \right.} \right) , \nonumber \\ 
  && + Z_{es} \left( {\Xi _e^2 \left| {\xi _s^1 } \right.} \right)Z_{es} \left( {\Xi _e^1 \left| {\xi _s^2 } \right.} \right)Z_{mm} \left( {\Xi _m^1 \left| {\xi _m^1 } \right.} \right) , \nonumber \\ 
  && + Z_{es} \left( {\Xi _e^2 \left| {\xi _s^1 } \right.} \right)Z_{ms} \left( {\Xi _m^1 \left| {\xi _s^2 } \right.} \right)Z_{em} \left( {\Xi _e^1 \left| {\xi _m^1 } \right.} \right) , \nonumber \\ 
&& + \left. {  Z_{ms} \left( {\Xi _m^1 \left| {\xi _s^1 } \right.} \right)Z_{es} \left( {\Xi _e^2 \left| {\xi _s^2 } \right.} \right)Z_{em} \left( {\Xi _e^1 \left| {\xi _m^1 } \right.} \right)} \right] ,
\end{eqnarray}
whose six contributions are associated to the six elementary processes (see panel {\bf b} of Fig.4) where each ingoing polariton $\xi_\mu^j$ turns into an outgoing polariton $\Xi_\tau^i$ with probability amplitude $Z_{\tau \mu } \left( {\Xi _\tau ^i \left| {\xi _\mu ^j } \right.} \right)$. As a paradigmatic elementary process we focus on the one corresponding to the contribution $Z_{ms} \left( {\Xi _m^1 \left| {\xi _s^1 } \right.} \right)Z_{es} \left( {\Xi _e^1 \left| {\xi _s^2 } \right.} \right)Z_{em} \left( {\Xi _e^2 \left| {\xi _m^1 } \right.} \right)$ in the right hand side of Eq.(\ref{ExampProc}), where the ingoing energy radiation carried by the $\xi_s^1$ and $\xi_s^2$ polaritons is absorbed by the object in the form of magnetic and electric energies described by the $\Xi_m^1$ and $\Xi_e^1$ polaritons, respectively, whereas the intial magnetic energy of the object represented by the $\xi^1_m$ polariton turns into final electric energy associated to the $\Xi_e^2$ polariton.

\section{Pure scattering of quantum radiation}
The approach to quantum optical scattering developed in Sec.V is completely general it enabling to deal with any initial radiation-object state. On the other hand, the most common situation encountered in actual setups is the one where the scatterer object is electromagnetically inert before radiation hits it. To investigate such practically relevant situation, describing the pure scattering and absorption of quantum radiation, we specialize the above general formalism to the case where the ingoing quantum state has no electric or magnetic polariton or, in other words, that the ingoing wavefunction is
\begin{equation} \label{IniRadStat}
\Psi _{pqr}^{\left( {in} \right)} \left( {\xi _s^{\left( p \right)} ,\xi _e^{\left( q \right)} ,\xi _m^{\left( r \right)} } \right) = \delta _{pn} \delta _{q0} \delta _{r0} \psi _n^{\left( {in} \right)} \left( {\xi _s^{\left( n \right)} } \right)
\end{equation}
where $\psi _n^{\left( {in} \right)} ( {\xi _s^{\left( n \right)} } )$ is the ingoing radiation wavefunction with $n$ $s$-polaritons, symmetric under polariton exchange and normalized as $\sum\limits_{n = 0}^\infty  {\int {d\xi _s^{\left( n \right)} } } | {\psi _n^{\left( {in} \right)} ( {\xi _s^{\left( n \right)} } )} |^2  = 1$. The ingoing-outgoing quantum state relation in this case, as shown in Appendix G, becomes 
\begin{eqnarray}  \label{IniRadOutStat}
&& \Psi _{PQR}^{\left( {out} \right)} \left( {\Xi _s^{\left( P \right)} ,\Xi _e^{\left( Q \right)} ,\Xi _m^{\left( R \right)} } \right) = \sqrt {\frac{{N!}}{{P!Q!R!}}} \nonumber  \\ 
&& \cdot \int {d\xi _s^{\left( N \right)} } \left[ {\prod\limits_{i = 1}^N {Z_{\chi \left( {\bar \Xi ^i } \right)s} \left( {\bar \Xi ^i \left| {\xi _s^i } \right.} \right)} } \right]\psi _N^{\left( {in} \right)} \left( {\xi _s^{\left( N \right)} } \right), 
\end{eqnarray}
where the exchange simmetry of the ingoing wavefunction $\psi _N^{\left( {in} \right)}$ has enabled to circumvent the summation over the permutations $\pi_N$. Note that in Eq.(\ref{IniRadOutStat}) only the ingoing wavefunction $\psi^{(in)}_N$ with $N=P+Q+R$ shows up, as a consequence of total polariton number conservation. Besides, it is worth emphasizing that $Z_{s s}$, $Z_{e s}$ and $Z_{m s}$ are the only involved entries of the matrix $Z_{\tau \mu}$ in Eq.(\ref{ZmatrEntr}) and, in consideration of the discussion of Sec.V, this is physically related to the fact that the initially inert object has no available energy to yield radiation emission or internal energy redistribution, in the far-future, so that the only elementary processes here involved are the $\xi_s^i  \rightarrow \Xi_s^j$ polariton elastic scattering (black wavy line in Fig.3) and the $\xi_s^i  \rightarrow \Xi_e^j$, $\xi_s^i  \rightarrow \Xi_m^j$ polariton absorption with increase of the object electric and magnetic internal energies (grey solid arrows in Fig.3). In the following four subsections we discuss the main properties of radiation pure scattering and we detail the two archetypal examples of one- and two-polariton scattering.

\subsection{Outgoing polariton detection probability}
The interplay of above discussed the three elementary procesesses becomes manifest when considering the probability $p_{PQR}  = \int {d\Xi _s^{\left( P \right)} d\Xi _e^{\left( Q \right)} d\Xi _m^{\left( R \right)} } | {\Psi _{PQR}^{\left( {out} \right)} ( {\Xi _s^{\left( P \right)} ,\Xi _e^{\left( Q \right)} ,\Xi _m^{\left( R \right)} } )} |^2$ of detecting in the far-future $P$, $Q$ and $R$ outgoing $s$-, $e$- and $m$- polaritons which, as show in Appendix G, is given by
\begin{eqnarray}   \label{pPQR}
&& p_{PQR}  =  \int {d\xi _s^{\left( N \right)} d{\xi '}_s^{\left( N \right)} } \psi _N^{\left( {in} \right)*} \left( {\xi _s^{\left( N \right)} } \right) \psi _N^{\left( {in} \right)} \left( {{\xi '}_s^{\left( N \right)} } \right) \nonumber\\ 
&&  \cdot \left[ \frac{{ N!}}{{P!Q!R!}}\prod\limits_{I = 1}^P {J_s \left( {\xi _s^I \left| {{\xi '}_s^I } \right.} \right)} \prod\limits_{J = 1}^Q {J_e \left( {\xi _s^{P + J} \left| {{\xi '}_s^{P + J} } \right.} \right)} \right. \nonumber \\
&& \left. \cdot \prod\limits_{K = 1}^R {J_m \left( {\xi _s^{P + Q + K} \left| {{\xi '}_s^{P + Q + K} } \right.} \right)} \right], 
\end{eqnarray}
where
\begin{equation}   \label{Jtauxx'}
J_\tau  \left( {\xi _s \left| {\xi '_s } \right.} \right) = \delta \left( {\omega  - \omega '} \right){\bf{e}}_{{\bf{n}}\lambda }  \cdot  {\left[\kern-0.15em\left[ {{\sf K}_{\omega s s }^{\tau } }  \right]\kern-0.15em\right]\left( {{\bf{n}},{\bf{n}}'} \right)}  \cdot {\bf{e}}_{{\bf{n}}'\lambda '} 
\end{equation}
with $\tau =s,e,m$ and the integral operators ${\sf K}_{\omega s s}^\tau$ are
\begin{eqnarray} \label{Komstau}
&& {\sf K}_{\omega ss}^s  =  {\sf T}_{\omega ss}^ +  {\sf T}_{\omega ss} , \nonumber  \\ 
&& {\sf K}_{\omega s s }^\nu  =  {\sf A}_{\omega \nu s}^ +  {\sf A}_{\omega \nu s} 
\end{eqnarray}
with $\nu = e,m$. The Hermitian integral operators ${\sf K}_{\omega s s}^s$ and ${\sf K}_{\omega s s }^\nu$ stem from the transmission ${\sf T}_{\omega ss}$ and absorption ${\sf A}_{\omega \nu s} $ operators, respectively, and hence they account for the basic processes of elastic scattering and electric and magnetic absortpion contributing to the probability of Eq.(\ref{pPQR}). We emphasize that the unitarity of the input-output relation discussed in Sec.IV here amounts to probability conservation since the operator relation corresponding to the matrix element $(1,1)$ of Eq.(\ref{RelTAGConj}) can be written as 
\begin{equation} \label{Kss+Kes+Kms}
{\sf K}_{\omega ss}^s  + {\sf K}_{\omega ss}^e  + {\sf K}_{\omega ss}^m  = {\sf I}_{ss}
\end{equation}
which, from Eq.(\ref{Jtauxx'}), directly yields
\begin{equation} \label{Je+Je+Jm}
J_s \left( {\xi _s \left| {\xi '_s } \right.} \right) + J_e \left( {\xi _s \left| {\xi '_s } \right.} \right) + J_m \left( {\xi _s \left| {\xi '_s } \right.} \right) = \delta _s \left( {\xi _s  - \xi '_s } \right).
\end{equation}
Trivially, if the object is transparent, $J_e =0$ and $J_m = 0$ since the absorption operators ${\sf A}_{\omega \nu s }$  vanish so that Eq.(\ref{Je+Je+Jm}) yields $J_s \left( {\xi _s \left| {\xi '_s } \right.} \right) = \delta _s \left( {\xi _s  - \xi '_s } \right)$ and, from Eq.(\ref{pPQR}), the probability that $n$ photons are detected in the far-future is 
\begin{equation}
p_{n00}  = \int {d\xi _s^{\left( n \right)} } \left| {\psi _n^{\left( {in} \right)} \left( {\xi _s^{\left( n \right)} } \right)} \right|^2 ,
\end{equation}
which coincides with the probability that $n$ photons are detected in the far-past, this restating the well known elastic nature of  photon scattering in the absence of absorption.

\subsection{Quantum decoherence of scattered radiation}
The focus of a typical scattering setup is usually concerned with the detection of the scattered radiation alone without performing any measurement on the object. Now Eq.(\ref{IniRadOutStat}) revelas that outgoing $s$-polaritons are generally entangled with outgoing $e$- and $m$-polaritons so that any measurement prediction about the scattered radiation (with the object left unmeasured) has to be performed by resorting to the reduced density operator $\hat \rho _s^{\left( {out} \right)}  = {\rm Tr}_{\Xi _e \Xi _m } \left( {\left| {\Psi ^{\left( {out} \right)} } \right\rangle \left\langle {\Psi ^{\left( {out} \right)} } \right|} \right)$ (where the partial trace ${\rm Tr}_{\Xi _e \Xi _m }$  is taken over the space $F_e  \otimes F_m$) which is an operator acting on the outgoing polariton Fock space $F_s$ , i.e.
\begin{eqnarray} \label{RedDenOpe}
&&\hat \rho _s^{\left( {out} \right)}  = \sum\limits_{P,P' = 0}^\infty  {\int {d\Xi _s^{\left( P \right)} d{\Xi '}_s^{\left( {P'} \right)} } } \rho _s^{\left( {out} \right)}\left( {\Xi _s^{\left( P \right)} \left| {{\Xi '}_s^{\left( {P'} \right)} } \right.} \right) \nonumber \\
&& \cdot \left| {\Xi _s^{\left( P \right)} } \right\rangle \left\langle {{\Xi '}_s^{\left( {P'} \right)} } \right|.
\end{eqnarray}
By taking the partial trace, as shown in Appendix H, the reduced density matrix $\rho _s^{\left( {out} \right)} ( {\Xi _s^{\left( P \right)} | {{\Xi '}_s^{\left( {P'} \right)} } } ) = \langle {\Xi _s^{\left( P \right)} } |\hat \rho _s^{\left( {out} \right)} | {{\Xi '}_s^{\left( {P'} \right)} } \rangle$ is given by
\begin{eqnarray} \label{RedDenMat}
&& \rho _s^{\left( {out} \right)} \left( {\Xi _s^{\left( P \right)} \left| {{\Xi '}_s^{\left( {P'} \right)} } \right.} \right)= \sum\limits_{S = 0}^\infty  {\int {d\xi _s^{\left( {P + S} \right)} d{\xi '}_s^{\left( {P' + S} \right)} } } \nonumber \\ 
&& \cdot \left[ \phi _S^P \left( {\Xi _s^{\left( P \right)} \left| {\xi _s^{\left( {P + S} \right)} } \right.} \right)\phi _S^{P'*} \left( {{\Xi '}_s^{\left( {P'} \right)} \left| {{\xi '}_s^{\left( {P' + S} \right)} } \right.} \right) \right. \nonumber \\
&& \cdot  \left. D_S^{PP'} \left( {\xi _s^{\left( {P + S} \right)} \left| {{\xi '}_s^{\left( {P' + S} \right)} } \right.} \right) \right],
\end{eqnarray}
where
\begin{eqnarray} \label{PhiPS}
&&  \phi _S^P \left( {\Xi _s^{\left( P \right)} \left| {\xi _s^{\left( {P + S} \right)} } \right.} \right) = \left[ {\frac{1}{{P!}}\sum\limits_{\pi _P  \in S_P } {\prod\limits_{I = 1}^P {Z_{ss} \left( {\Xi _s^{\pi _P \left( I \right)} \left| {\xi _s^I } \right.} \right)} } } \right] \nonumber \\ &&   \cdot \psi _{P+S}^{\left( {in} \right)} \left( {\xi _s^{\left( {P + S} \right)} } \right) \nonumber \\ 
\end{eqnarray}
is the ingoing wavefunction dressed by the symmetrized matrix entries $Z_{ss}$ associated to the elastic scattering and
\begin{eqnarray} \label{DecohFac}
&& D_S^{PP'} \left( {\xi _s^{\left( {P + S} \right)} \left| {{\xi '}_s^{\left( {P' + S} \right)} } \right.} \right) = \sqrt {\frac{{\left( {P + S} \right)!\left( {P' + S} \right)!}}{{P!P'!}}} \nonumber \\ 
&& \cdot \sum\limits_{Q = 0}^S {\frac{1}{{Q!\left( {S - Q} \right)!}}}  \left[ \prod\limits_{J = 1}^Q J_e^* \left( {\xi _s^{P + J} \left| {{\xi '}_s^{P' + J} } \right.} \right) \right. \nonumber \\
&&\left. \cdot \prod\limits_{K = 1}^{S - Q} {J_m^* \left( {\xi _s^{P + Q + K} \left| {{\xi '}_s^{P' + Q + K} } \right.} \right)}  \right]  
\end{eqnarray}
is the decoherence factor. The main evident feature of Eq.(\ref{RedDenMat}) is that the reduced density matrix  generally does not exhibit the factored form 
\begin{equation} \label{RhoOutFact}
\rho _s^{\left( {out} \right)} \left( {\Xi _s^{\left( P \right)} \left| {{\Xi '}_s^{\left( {P'} \right)} } \right.} \right) = \Phi \left( {\Xi _s^{\left( P \right)} } \right)\Phi ^* \left( {{\Xi '}_s^{\left( {P'} \right)} } \right)
\end{equation}
which is typical of a pure state and hence the scattered radiation is described by a statistical mixture of states physically resulting from its entanglement with the object (outgoing $e$- and $m$-polaritons). Bearing in mind that the ingoing radiation state is pure (see Eq.(\ref{IniRadStat})), we conclude that  the scattering by an absorbing object causes radiation to experience quantum decoherence, an effect here described by the decoherence factor $D_S^{PP'}$ in Eq.(\ref{DecohFac}) which, correspondingly, is entirely specified by the factors $J_e$ and $J_m$ stemming from 
electric and magnetic absorbtion. In support of this observation, we note that in the lossless limit, where $J_e=0$ and $J_e=0$, the decoherence factor survives only for $S=0$, i.e. $D_S^{PP'} = \delta_{S0}$, so that the reduced density matrix of Eq.(\ref{RedDenMat}) is easily seen to get the factored form of Eq.(\ref{RhoOutFact}) with $\Phi ( {\Xi _s^{\left( P \right)} } ) = \Psi _{P00}^{\left( {out} \right)} ( {\Xi _s^{\left( P \right)} ,\Xi _e^{\left( 0 \right)} ,\Xi _m^{\left( 0 \right)} } )$ (see Eq.(\ref{IniRadOutStat})), as expected.

\begin{figure*}
\centering
\includegraphics[width = 1\linewidth]{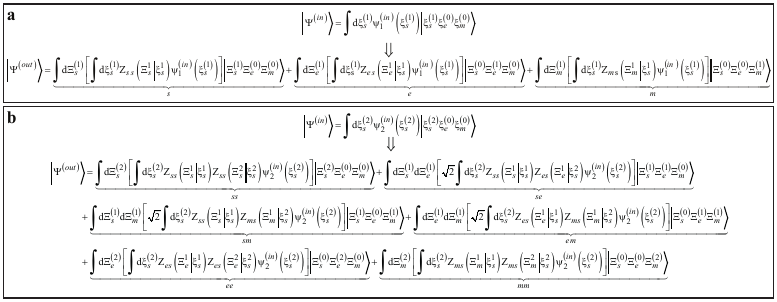}
\caption{Ingoing ${| {\Psi ^{\left( {in} \right)} } \rangle }$ and outgoing ${| {\Psi ^{\left( {out} \right)} } \rangle }$ states pertaining \textbf{a} one-polariton scattering and \textbf{b} two-polariton scattering.}
\label{Fig5}
\end{figure*}

\subsection{One-polariton scattering}
As a first example of the above general description of radiation pure scattering, we here consider the case where a single ingoing $s$-polariton is launched onto the object, i.e. the ingoing wavefunction is
\begin{equation} \label{OnePolIngWav}
\psi _n^{\left( {in} \right)} \left( {\xi _s^{\left( n \right)} } \right) = \delta _{n1} \psi _1^{\left( {in} \right)} \left( {\xi _s^{\left( 1 \right)} } \right)
\end{equation}
which is normalized as $\int {d\xi _s^{\left( 1 \right)} }  | {\psi _1^{\left( {in} \right)}  ( {\xi _s^{\left( 1 \right)} }  )}  |^2  = 1$.  The corresponding ingoing state ${| {\Psi ^{\left( {in} \right)} } \rangle }$, evaluated from Eqs.(\ref{IniRadStat}) and (\ref{PsiIngo}), is reported in Fig.5a and it is a one-polariton state comprising all the ingoing-$s$ polaritons $\xi _s^{\left( 1 \right)} = (  \xi _s^1 ) = \left( {\omega ,{\bf{n}},\lambda } \right)$ with no restrinction on the frequency, direction or polarization. The outgoing state ${| {\Psi ^{\left( {out} \right)} } \rangle }$, evaluated form Eqs.(\ref{IniRadOutStat}) and (\ref{PsiOutgo}), is reported in Fig.5a as well and it is manifestly a one-polariton state due to polariton total number conservation. The three contributions to ${| {\Psi ^{\left( {out} \right)} } \rangle }$ correspond to the processes, labelled in the underbraces as $s$, $e$ and $m$, where the ingoing $s$-polariton is elastically scattered  into an outgoing $s$-polariton or it decays into outgoing $e$- or $m$- polaritons. The probabilities of such three processes are $P_s  = p_{100}$, $P_e  = p_{010}$ and $P_m  = p_{001}$ and, from Eq.(\ref{pPQR}) they are straightforwardly given by
\begin{equation} \label{OnePolProb}
P_\tau   = \int {d\xi _s^{\left( 1 \right)} d{\xi '}_s^{\left( 1 \right)} } \psi _1^{\left( {in} \right)*} \left( {\xi _s^{\left( 1 \right)} } \right)\psi _1^{\left( {in} \right)} \left( {{\xi '}_s^{\left( 1 \right)} } \right)J_\tau  \left( {\xi _s^1 \left| {{\xi '}_s^1 } \right.} \right).
\end{equation}
Note that probability conservation 
\begin{equation} \label{OnePolProbCons}
P_s+P_e+P_m=1
\end{equation}
is manifestly assured by Eq.(\ref{Je+Je+Jm}) and the ingoing state normalization.

The reduced density operator of the scattered radiation in Eq.(\ref{RedDenOpe}), as shown in Appendix I, here becomes 
\begin{equation} \label{OnePolReDeOp}
\hat \rho _s^{\left( {out} \right)}  = P_s \left| {\Phi _{1s} } \right\rangle \left\langle {\Phi _{1s} } \right| + \left( {P_e  + P_m } \right)\left| {\Phi _{0s} } \right\rangle \left\langle {\Phi _{0s} } \right|,
\end{equation}
where 
\begin{eqnarray} \label{OnePolPh1Ph0}
&& \left| {\Phi _{1s} } \right\rangle  = \frac{1}{{\sqrt {P_s } }}\left[ {\hat I_s  \otimes \left\langle {\Phi _{0em} } \right|} \right]\left| {\Psi _{out} } \right\rangle, \nonumber  \\ 
&& \left| {\Phi _{0s} } \right\rangle  = \left| {\Xi _s^{\left( 0 \right)} } \right\rangle 
\end{eqnarray}
where ${\hat I_s }$ is the identity operator of the Fock space $F_s$ and $| {\Phi _{0em} } \rangle  = | \Xi _e ^{\left( 0 \right)} \Xi _m  ^{\left( 0 \right)}  \rangle$ is the vacuum state of the space $F_e  \otimes F_m$. Here $\left| {\Phi _{1s} } \right\rangle$ is the  state of the Fock space $F_s$ corresponding to the $s$ term in the outgoing state $| {\Psi ^{\left( {out} \right)} } \rangle$ of Fig.5a, normalized as $\left\langle {{\Phi _{1s} }}
 \mathrel{\left | {\vphantom {{\Phi _{1s} } {\Phi _{1s} }}} \right. \kern-\nulldelimiterspace} {{\Phi _{1s} }} \right\rangle  = 1$, whereas $\left| {\Phi _{0s} } \right\rangle  $ is the vacuum state of $F_s$. The particular simple form of the reduced density operator in Eq.(\ref{OnePolReDeOp}) elucidates that, in one-polariton scattering, the scattered radiation is described by the statistical mixture of the elastic scattering state $\left| {\Phi _{1s} } \right\rangle$ and the vacuum state $\left| {\Phi _{0s} } \right\rangle$ and, besides, that such mixture is fully incoherent since these two states are eigevectors of $\hat \rho _s^{\left( {out} \right)}$, Eq.(\ref{OnePolReDeOp}) exhibiting its spectral decomposition. Accordingly, the populations $P_s$ and $\left( {P_e  + P_m } \right)$  of the states $\left| {\Phi _{1s} } \right\rangle$ and $\left| {\Phi _{0s} } \right\rangle$ are the respective probabilities that the scattered radiation is fully detected (without absorption) and it is not detected (fully absorbed), with Eq.(\ref{OnePolProbCons}) assuring consistency and the necessary normalization condition ${\rm Tr} ( {\hat \rho _s^{\left( {out} \right)} } ) = 1$. A crucial remark is that, except for the peculiar case where radiation is certainly fully absorbed $P_e + P_m = 1$, Eq.(\ref{OnePolReDeOp}) shows that the reduced density opearator has two eigenvalues and this implies that the Schmidt rank \cite{Cohen1} of the outgoing state $\left| {\Psi ^{\left( {out} \right)} } \right\rangle$ is equal to $2$, that is to say, outgoing $s$-polaritons are entangled with outgoing $e$- and $m$- polaritons. In other words, in one-polariton scattering, the scattered radiation is unavoidably entangled with the object because of absorption. 

Finally, in the lossless limit where $J_e$ and $J_m$ vanish and the space $F_e  \otimes F_m$ is no longer of importance, $P_s =1$, $P_e=P_m =0$ and $\left| {\Phi _{1s} } \right\rangle  = \left| {\Psi _{out} } \right\rangle$  so that the reduced density operator $\hat \rho _s^{\left( {out} \right)}$ of Eq.(\ref{OnePolReDeOp}) becomes the full density operator $\left| {\Psi _{out} } \right\rangle \left\langle {\Psi _{out} } \right|$, as expected.

\subsection{Two-polariton scattering}
The above discussed one-polariton scattering process offers a convenient forum to examine in a simple way some relevant physical traits of quantum optical scattering as elastic scattering, absorption, radiation quantum decoherence and radiation-object entanglement. On the other hand, the one-particle nature of such process completely circumvents the possiblity to the analyze the role played by the potential entanglement among the ingoing $s$-polariton in the scattering process, a crucial physical point with an enormous relevance in quantum information and computing. In order to examine such a very important quantum phenomenon in the presence of matter absorption, i.e. by means of the quantum scattering approach introduced in the preset paper, we here consider the case where two ingoing $s$-polariton are launched onto the object, i.e. the ingoing wavefunction is
\begin{equation} \label{TwoPolIngWav}
\psi _n^{\left( {in} \right)} \left( {\xi _s^{\left( n \right)} } \right) = \delta _{n2} \psi _2^{\left( {in} \right)} \left( {\xi _s^{\left( 2 \right)} } \right),
\end{equation}
where we repeat that $\xi _s^{\left( 2 \right)}  = \left( {\xi _s^1 ,\xi _s^2 } \right)$ are the two ingoing $s$-polariton variables, which is exchange symmetric $\psi _2^{\left( {in} \right)} \left( {\left( {\xi _s^1 ,\xi _s^2 } \right)} \right) = \psi _2^{\left( {in} \right)} \left( {\left( {\xi _s^2 ,\xi _s^1 } \right)} \right)$ and  normalized as $\int {d\xi _s^{\left( 2 \right)} } | {\psi _2^{\left( {in} \right)} ( {\xi _s^{\left( 2 \right)} } )} |^2  = 1$. 
The corresponding ingoing state ${| {\Psi ^{\left( {in} \right)} } \rangle }$ is reported in Fig.5b and it is a two-polariton state involving all the ingoing-$s$ polaritons $\xi _s^1  = \left( {\omega ^1 ,{\bf{n}}^1 ,\lambda ^1 } \right)$  and $\xi _s^2  = \left( {\omega ^2 ,{\bf{n}}^2 ,\lambda ^2 } \right)$ with no restrinction on their frequencies, directions,  polarizations and entanglement. Due to bosonic indistinguishability, the two ingoing $s$-polaritons are always entangled unless  the ingoing wavefunction factor as 
\begin{equation} \label{NotEntangled}
\psi _2^{\left( {in} \right)} ( {\xi _s^{\left( 2 \right)} } ) = \varphi \left( {\xi _s^1 } \right) \varphi \left( {\xi _s^2 } \right),
\end{equation}
this corresponding to the situation where the polaritons are in the same quantum state \cite{Garri1}. The outgoing state $| {\Psi ^{\left( {out} \right)} } \rangle$, evaluated from Eqs.(\ref{IniRadOutStat}) and (\ref{PsiOutgo}), is reported in Fig.5b as well and it is evindently a two-polariton state again due to polariton number conservation. The six contributions to $| {\Psi ^{\left( {out} \right)} } \rangle$ correpond to the processes, labelled in the underbaces as $ss$, $se$, $sm$, $em$, $ee$ and $mm$, where both ingoing $s$-polaritons undergo elastic scattering (first term), one ingoing $s$-polariton is elastically scattered and the other is absorbed (second and third  terms) and both ingoing $s$-polaritons are absorbed (fourth, fifth and sixth terms). Note that, in the ougoing state $| {\Psi ^{\left( {out} \right)} } \rangle$ of Fig.5b, the indistinguishability of the two ingoing $s$-polaritons ${( {\xi _s^1 ,\xi _s^2 } ) \to ( {\xi _s^2 ,\xi _s^1 } )}$  is accounted by the exchange symmetry of the ingoing wavefunction $\psi _2^{\left( {in} \right)} ( {\xi _s^{\left( 2 \right)} } )$. The probabilities of the six processes are $P_{ss}  = p_{200}$, $P_{se}  = p_{110}$, $P_{sm}  = p_{101}$, $P_{em}  = p_{011}$, $P_{ee}  = p_{020}$ and $P_{mm}  = p_{002}$ and, from Eq.(\ref{pPQR}), they are directly seen to be given by 
\begin{eqnarray} \label{TwoPolProb}
&& P_{\tau \mu }  = \left( {2 - \delta _{\tau \mu } } \right)\int {d\xi _s^{\left( 2 \right)} d{\xi '}_s^{\left( 2 \right)} } \psi _2^{\left( {in} \right)*} \left( {\xi _s^{\left( 2 \right)} } \right)\psi _2^{\left( {in} \right)} \left( {{\xi '}_s^{\left( 2 \right)} } \right) \nonumber \\ 
&& \cdot J_\tau  \left( {\xi _s^1 \left| {{\xi '}_s^1 } \right.} \right)J_\mu  \left( {\xi _s^2 \left| {{\xi '}_s^2 } \right.} \right)  
\end{eqnarray}
where the factor ${2 - \delta _{\tau \mu } }$ reproduces the coefficient $N!/(P!Q!R!)$ for the $N=2$ case we are considering here. Probability conservation
\begin{equation} \label{TwoPolProbCons}
P_{ss}  + P_{se}  + P_{sm}  + P_{em}  + P_{ee}  + P_{mm}  = 1
\end{equation}
can here be easily verified by noting that $P_{\tau \mu} = P_{\tau \mu}$ and using Eq.(\ref{Je+Je+Jm}) written in the form
\begin{eqnarray}
&& \left[ {J_s \left( {\xi _s^1 \left| {{\xi '}_s^1 } \right.} \right) + J_e \left( {\xi _s^1 \left| {{\xi '}_s^1 } \right.} \right) + J_m \left( {\xi _s^1 \left| {{\xi '}_s^1 } \right.} \right)} \right] \nonumber \\ 
&&  \cdot \left[ {J_s \left( {\xi _s^2 \left| {{\xi '}_s^2 } \right.} \right) + J_e \left( {\xi _s^2 \left| {{\xi '}_s^2 } \right.} \right) + J_m \left( {\xi _s^2 \left| {{\xi '}_s^2 } \right.} \right)} \right]  \nonumber \\ 
&&  = \delta _s \left( {\xi _s^1  - {\xi '}_s^1 } \right)\delta _s \left( {\xi _s^2  - {\xi '}_s^2 } \right) 
\end{eqnarray}
together with the ingoing state normalization.

The reduced density operator of the scattered radiation in Eq.(\ref{RedDenOpe}), as shown in Appendix J, here becomes 

\begin{eqnarray} \label{TwoPolReDeOp}
&& \hat \rho _s^{\left( {out} \right)}  = P_{ss} \left| {\Phi _{2s} } \right\rangle \left\langle {\Phi _{2s} } \right| + \left( {P_{se}  + P_{sm} } \right) \hat \rho _{1s}  \nonumber \\
&& + \left( { P_{em}  + P_{ee}  + P_{mm} } \right)\left| {\Phi _{0s} } \right\rangle \left\langle {\Phi _{0s} } \right|
\end{eqnarray}
where
\begin{eqnarray}  \label{TwoPolDefin}
&& \left| {\Phi _{2s} } \right\rangle  = \frac{1}{{\sqrt {P_{ss} } }}\left[ {\hat I_s  \otimes \left\langle {\Phi _{0em} } \right|} \right]\left| {\Psi ^{\left( {out} \right)} } \right\rangle, \nonumber \\ 
&& \hat \rho _{1s}  = \frac{1}{{P_{se}  + P_{sm} }} \int {d\Xi _s^{\left( 1 \right)} d{\Xi '}_s^{\left( 1 \right)} } \left\{ {2\int {d\xi _s^{\left( 2 \right)} d{\xi '}_s^{\left( 2 \right)} } } \right.  \nonumber \\ 
&& \psi _2^{\left( {in} \right)} \left( {\xi _s^{\left( 2 \right)} } \right)\psi _2^{\left( {in} \right)*} \left( {{\xi '}_s^{\left( 2 \right)} } \right)Z_{ss} \left( {\Xi _s^1 \left| {\xi _s^1 } \right.} \right)Z_{ss}^* \left( {{\Xi '}_s^1 \left| {{\xi '}_s^1 } \right.} \right)  \nonumber \\ 
&&  \cdot \left. {\left[ {J_e^* \left( {\xi _s^2 \left| {{\xi '}_s^2 } \right.} \right) + J_m^* \left( {\xi _s^2 \left| {{\xi '}_s^2 } \right.} \right)} \right]} \right\}\left| {\Xi _s^{\left( 1 \right)} } \right\rangle \left\langle {{\Xi '}_s^{\left( 1 \right)} } \right|, \nonumber \\ 
&& \left| {\Phi _{0s} } \right\rangle  = \left| {\Xi _s^{\left( 0 \right)} } \right\rangle, 
\end{eqnarray}
where ${\hat I_s }$ is the identity operator of the Fock space $F_s$, $\hat \rho _{1s}$ is a one-polariton operator acting on $F_s$ normalized as ${{\rm Tr}\left( {\hat \rho _{1s} } \right) = 1}$ and $| {\Phi _{0em} } \rangle  = | \Xi _e ^{\left( 0 \right)} \Xi _m  ^{\left( 0 \right)}  \rangle$ is the vacuum state of the space $F_e  \otimes F_m$. In analogy with one-polariton scattering, $\left| {\Phi _{2s} } \right\rangle$ is the two-polariton normalized state of the Fock space $F_s$ corresponding to the $ss$ term in the outgoing state $| {\Psi ^{\left( {out} \right)} } \rangle$ of Fig.5b and $\left| {\Phi _{0s} } \right\rangle$ is the zero-polariton vacuum state of $F_s$. On the other hand, the one-polariton contribution $\rho _{1s}$  does not show up as simple spectral projector associated to a definite quantum state of $F_s$ and it has evidently  no counterpart in the reduced density operator in Eq.(\ref{OnePolReDeOp}) of the one-polariton scattering. Now $\hat \rho _{1s}$ is clearly an Hermitian operator so that it admits the spectral decomposition
\begin{equation} \label{rho1sSpecDeco}
\hat \rho _{1s}  = \sum\limits_\alpha {\rho _{1s}^\alpha \left| {\Phi _{1s}^\alpha } \right\rangle \left\langle {\Phi _{1s}^\alpha } \right|} 
\end{equation}
displaying its multiple eigenstates ${| {\Phi _{1s}^\alpha } \rangle }$ and its eigenvalues satisfying the relation $\sum\nolimits_\alpha {\rho _{1s}^\alpha }  = 1$. Consequently, the fully incoherent statistical mixture  describing the scattered radiation here comprises the states $| {\Phi _{2s} } \rangle$, $ {| {\Phi _{1s}^\alpha } \rangle }$ and $| {\Phi _{0s} } \rangle$ whose populations $P_{ss}$, $\left( {P_{se}  + P_{sm} } \right)\rho _{1s}^a$ and $\left( {P_{ee}  + P_{em}  + P_{mm} } \right)$ are the respective probabilities that the scattered radiation is fully detected (without absorption), it is detected with a single outgoing $s$-polariton found in the state $\left| {\Phi _{1s}^\alpha } \right\rangle$ (partially absorbed) and it is not detected (fully absorbed). Again, probability conservation of Eq.(\ref{TwoPolProbCons}), together with the relation $\sum\nolimits_\alpha {\rho _{1s}^\alpha }  = 1$, provides consistency and it assures the necessary normalization condition ${\rm Tr} ( {\hat \rho _s^{\left( {out} \right)} } ) = 1$. Note that, except for the peculiar case where radiation is certainly fully absorbed $P_{ee} + P_{em} + P_{mm} = 1$, ${\hat \rho _s^{\left( {out} \right)} }$ has at least two eigenvalues so that the Schmidt rank \cite{Cohen1} of the outgoing state $\left| {\Psi ^{\left( {out} \right)} } \right\rangle$ is greater/equal than $2$ and again outgoing $s$-polaritons are entangled with outgoing $e$- and $m$- polaritons. Therefore, as in the one-polariton scattering, the scattered radiation is unavoidably entangled with the object because of absorption.

The relevant point here is that the one-polariton eigenstates ${\left| {\Phi _{1s}^\alpha  } \right\rangle }$ of $\hat \rho _{1s}$ and their eigenvalues ${\rho _{1s}^\alpha  }$ crucially depend on the entanglement between the two ingoing $s$-polariton. To substantiate this assertion we first focus on the situation where the two-ingoing $s$-polaritons are not entangled so that Eq.(\ref{NotEntangled}) displays their factored ingoing wavefunction for which, as shown in Appendix K, the operator $\hat \rho _{1s}$ is given by
\begin{equation} \label{TwoPolNotEntarho1s}
{\hat \rho _{1s}  = \left| {\Phi _{1s} } \right\rangle \left\langle {\Phi _{1s} } \right|}
\end{equation}
where
\begin{equation} \label{TwoPolPh1s}
\left| {\Phi _{1s} } \right\rangle  = \frac{\displaystyle {\int {d\Xi _s^{\left( 1 \right)} } \left[ {\int {d\xi _s^{\left( 1 \right)} } Z_{ss} \left( {\Xi _s^1 \left| {\xi _s^1 } \right.} \right)\varphi \left( {\xi _s^1 } \right)} \right]\left| {\Xi _s^{\left( 1 \right)} } \right\rangle }}{{\sqrt {\int {d\xi _s^{\left( 1 \right)} d{\xi '}_s^{\left( 1 \right)} } \varphi ^* \left( {\xi _s^1 } \right)\varphi \left( {{\xi '}_s^1 } \right)J_s \left( {\xi _s^1 \left| {{\xi '}_s^1 } \right.} \right)} }}.
\end{equation}
Note that the normalized state $\left| {\Phi _{1s} } \right\rangle$ in Eq.(\ref{TwoPolPh1s}) coincides with the state $\left| {\Phi _{1s} } \right\rangle$ in the first of Eqs.(\ref{OnePolPh1Ph0}) after the replacement $\psi _1^{\left( {in} \right)}  \to \varphi$ (see Fig.5a) and therefore it describes the elastic scattering of an $s$-polariton whose ingoing state is $\left| \varphi  \right\rangle$. Therefore, Eq.(\ref{TwoPolNotEntarho1s}) states that, when the two ingoing polaritons are not entangled being in the same state $\left| \varphi  \right\rangle$, the one-polariton contribution to the statistical mixture describing the scattered radiation comprises just the single state  $\left| {\Phi _{1s} } \right\rangle$ of Eq.(\ref{TwoPolPh1s}) and this situation physically amounts to the strict elastic scattering of one of the two ingoing polaritons initially prepared in the state $\left| \varphi \right\rangle$, as if it were alone (the other polariton being actually absorbed), in agreement with the full lack of correlation between two not-entangled particles. On the other hand, the situation where the two ingoing polaritons are entangled  
is much more involved and it displays a much rich phenomenology. To explore this situation, we now assume that the two-polariton ingoing wavefunction is
\begin{equation} \label{TwoPolEntIngWav}
\psi _2^{\left( {in} \right)} \left( {\xi _s^{\left( 2 \right)} } \right) = \frac{1}{{\sqrt 2 }}\left[ {\varphi ^1 \left( {\xi _s^1 } \right)\varphi ^2 \left( {\xi _s^2 } \right) + \varphi ^1 \left( {\xi _s^1 } \right)\varphi ^2 \left( {\xi _s^2 } \right)} \right]
\end{equation}
describing two entangled polaritons with orthonormal wavefunctions $\varphi^1$ and $\varphi^2$, i.e.
\begin{equation}
\int {d\xi _s \varphi ^{i*} \left( {\xi _s } \right)\varphi ^j \left( {\xi _s } \right)}  = \delta _{ij}, 
\end{equation}
this assuring the overall normalization condition $\int {d\xi _s^{\left( 2 \right)} } | {\psi _2^{\left( {in} \right)} ( {\xi _s^{\left( 2 \right)} } )} |^2  = 1$. Using the two wavefunctions we form the $2 \times 2$  Hermitian matrices $X_s  = \{ {X_s^{ij} } \}$ and $X_{em}  = \{ {X_{em}^{ij} } \}$ whose entries are
\begin{eqnarray} \label{TwoPolEntaXsXem}
&& X_s^{ij}  = \int {d\xi _s d\xi '_s } \varphi ^{i*} \left( {\xi _s } \right)\varphi _s^j \left( {\xi '_s } \right)J_s \left( {\xi _s \left| {\xi '_s } \right.} \right), \nonumber \\ 
&& X_{em}^{ij}  = \int {d\xi _s d\xi '_s } \varphi ^{i*} \left( {\xi _s } \right)\varphi _s^j \left( {\xi '_s } \right)\left[ {J_e \left( {\xi _s \left| {\xi '_s } \right.} \right) + J_m \left( {\xi _s \left| {\xi '_s } \right.} \right)} \right]. \nonumber \\
\end{eqnarray}
For simplicity purposes, we focus on the case $\det \left( {X_s } \right) \ne 0$ where, as shown in Appendix K, the operator $\hat \rho _{1s}$ is given by
\begin{equation} \label{TwoPolEntarho1s}
\hat \rho _{1s}  = \rho _{1s}^a \left| {\Phi _{1s}^a } \right\rangle \left\langle {\Phi _{1s}^a } \right| + \rho _{1s}^a \left| {\Phi _{1s}^b } \right\rangle \left\langle {\Phi _{1s}^b } \right|
\end{equation}
where
\begin{eqnarray} \label{TwoPolPh1salpha}
&& \left| {\Phi _{1s}^\alpha  } \right\rangle  = \int {d\Xi _s^{\left( 1 \right)} } \left\{ {\int {d\xi _s^{\left( 1 \right)} Z_{ss} \left( {\Xi _s^1 \left| {\xi _s^1 } \right.} \right)} } \right. \nonumber \\ 
&& \left. { \cdot \left[ {\left( {\Phi _{1s}^\alpha  } \right)_1 \varphi ^1 \left( {\xi _s^1 } \right) + \left( {\Phi _{1s}^\alpha  } \right)_2 \varphi ^2 \left( {\xi _s^1 } \right)} \right]} \right\}\left| {\Xi _s^{\left( 1 \right)} } \right\rangle   
\end{eqnarray}
with $\alpha = a,b$ and where the real numbers $\rho _{1s}^\alpha$ and the complex vectors $\Phi _{1s}^\alpha   =\begin{pmatrix}   {\left( {\Phi _{1s}^\alpha  } \right)_1 } & {\left( {\Phi _{1s}^\alpha  } \right)_2 }  \\
\end{pmatrix}^T$ satisfy the eigenvalue problem 

\begin{equation} \label{TwoPolEntaEigProb}
{\left[ {\frac{1}{{P_{se}  + P_{sm} }}\begin{pmatrix}
   {X_{em}^{22} } & {X_{em}^{12} }  \\
   {X_{em}^{12*} } & {X_{em}^{11} }  \\
\end{pmatrix}\begin{pmatrix}
   {X_s^{11} } & {X_s^{12} }  \\
   {X_s^{12*} } & {X_s^{22} }  \\
\end{pmatrix}} \right]\Phi _{1s}  = \rho _{1s} \Phi _{1s} }.
\end{equation}
As detailed in Appendix K, the states $| {\Phi _{1s}^a } \rangle$ and $| {\Phi _{1s}^b } \rangle$ are orthonormal, i.e.
\begin{equation}
\left\langle {{\Phi _{1s}^\alpha  }}
 \mathrel{\left | {\vphantom {{\Phi _{1s}^\alpha  } {\Phi _{1s}^\beta  }}}
 \right. \kern-\nulldelimiterspace}
 {{\Phi _{1s}^\beta  }} \right\rangle  = \delta _{\alpha \beta }, 
\end{equation}
so that Eq.(\ref{TwoPolEntarho1s}) provides the spectral decomposition of the operator $\hat \rho _{1s}$ (compare with Eq.(\ref{rho1sSpecDeco})) for the specific two-polariton ingoing wavefunction of Eq.(\ref{TwoPolEntIngWav}). Equation (\ref{TwoPolEntarho1s}) manifestly shows that, in the here considered process where one ingoing polariton is scattered while the other one is absorbed, their mutual entanglement generally produces two one-polartion states, $\left| {\Phi _{1s}^a } \right\rangle$ and $\left| {\Phi _{1s}^b } \right\rangle$, contributing to the statistical mixture describing the scattered radiation, in striking difference with the above discussed situation where the ingoing polaritons are not entangled (see Eq.(\ref{TwoPolNotEntarho1s})). In order to grasp the physical mechanism ruling the process, we note that the state $\left| {\Phi _{1s}^\alpha  } \right\rangle$ in Eq.(\ref{TwoPolPh1salpha}) can be written as
\begin{equation}  \label{TwoPolPh1salphaComb}
{\left| {\Phi _{1s}^\alpha  } \right\rangle  = \left( {\Phi _{1s}^\alpha  } \right)_1 \left| {\tilde \Phi ^1 } \right\rangle  + \left( {\Phi _{1s}^\alpha  } \right)_2 \left| {\tilde \Phi ^2 } \right\rangle }
\end{equation}
where the not-normalized states
\begin{equation}
\left| {\tilde \Phi ^i } \right\rangle  = \int {d\Xi _s^{\left( 1 \right)} } \left[ {\int {d\xi _s^{\left( 1 \right)} } Z_{ss} \left( {\Xi _s^1 \left| {\xi _s^1 } \right.} \right)\varphi ^i \left( {\xi _s^1 } \right)} \right]\left| {\Xi _s^{\left( 1 \right)} } \right\rangle 
\end{equation}
nonetheless have the same structure of the state $\left| {\Phi _{1s} } \right\rangle$ in Eq.(\ref{TwoPolPh1s}) so that we conclude that $\left| {\Phi _{1s}^\alpha  } \right\rangle$ is the superposition of the elastic scattering states pertaining the two polaritons of ingoing wavefunctions $\varphi ^1$ and $\varphi ^2$ as if they were alone. This is clearly a physical consequence of the entanglement between the ingoing polaritons since their correlation strictly forbids the simple process where one polariton is scattered and the other one is absorbed, this forcing both $\varphi ^1$ and $\varphi ^2$ to contribute to the state $\left| {\Phi _{1s}^\alpha  } \right\rangle$ as in Eq.(\ref{TwoPolPh1salphaComb}). The remarkable point here is that the specific mixing of  $| {\tilde \Phi ^1 } \rangle$ and $| {\tilde \Phi ^2 } \rangle$ in Eq.(\ref{TwoPolPh1salphaComb}) is set up  by the complex vector ${\Phi _{1s}^\alpha  }$ which in turn results from the matrices $X_s$ and $X_{em}$ accounting for any scattering and absorption basic process (see Eqs.(\ref{TwoPolEntaXsXem})) and we thus conclude that the states $\left| {\Phi _{1s}^a } \right\rangle$ and $\left| {\Phi _{1s}^b } \right\rangle$ physically result from a complex interplay between ingoing polariton entanglement and object absorption. In other words, the overall quantum decoherence of the scattered radiation,  produced by the dispersive nature of the object optical response, dramatically depends on the possibile entanglement of the ingoing $s$-polaritons.

\section{Conclusion}

In summary, we have proposed a general and comprehensive theoretical framework able to describe the scattering of quantum radiation by a lossy macroscopic object of finite size in vacuum with no restrictions on both the quantum state of the impinging radiation and on the dispersive optical response of the spatially inhomogeneous object. The approach is based on the MLFN whose polaritonic description of the macroscopic quantum electrodynamics of a finite size object in vaccum has enabled the identification of ingoing and outgoing polaritons, bosons describing far-field radiation ($s$-polaritons) and medium excitation ($e$- and $m$-polaritons) in the far past and far future, respectively. One crucial result of our investigation is the input-output relation of Eq.(\ref{InOutRelFinal}) connecting the bosonic operators of the ingoing and outgoing polaritons, the utmost generalization of the input-output relations commonly used in literature to model the quantum behavior of optical devices with a finite number of ports. A remarkable point is that the input-output relation dramatically incorporates the classical electromagnetic dyadics into the quantum description whose insight is thus facilitated by our confort with classical electrodynamics. Accordingly, we have elucidated that the physical mechanisms underpinning the quantum input-output relation are provided by the energy balance between scattering, emission and absortpion  of classical radiation. In the here adopted Heisenberg picture, the prediction of the quantum optical scattering outcomes amounts to changing from the representation of the field Fock space induced by ingoing polaritons to the one induced by outgoing polaritons, this resulting in the ingoing-outgoing quantum state relation of Eq.(\ref{StatIngOut}) which is the main result of the present paper. As a matter of fact, such relation provides a comprehensive description of quantum optical scattering since it yields the outgoing state (in the far future) out of the most general ingoing state (in the far past) of the radiation-object system with no restrictions on the polariton kind, number or entanglement degree. In order to physically grasp the quantum optical scattering phenomenology, we have identified the nine fundamental quantum transitions between single ingoning and outgoing polaritons and we have found that their amplitudes basically coincide with appropriate matrix elements of the classical dyadics pertaining the transmission, emission and absorbtion of radiation and the redistribution of electric and magnetic internal  energies of the object (see Eq.(\ref{ZmatrEntr})). In addition, we have specialized the general approach to the situation where radiation is launched onto an object not displaying electromagnetic excitation in the far past which is probably the most common situation encountered in actual scattering setups. In this situation, the outgoing quantum state, as opposed to the ingoing one, generally contains polaritons of all the three kinds and it manifestly displays the entanglement between radiation (outgoing $s$-polaritons) and object (outgoing $e$- and $m-$ polaritons) produced by the lossy scattering event. Such a physical effect results into the loss of coherence experied by radiation upon scattering and we assess the ensuing quantum decoherence by evaluating the reduced density operator of the scattered radiation. In order to analyze in more details the general results of our approach and to appreciate its practical versatility, we have discussed the two relevant example situations of one and two-polariton scattering. The first example situation is simple enough to provide a clear picture of the interplay between scattering and absorption since we have found that the statistical mixture of the quantum states describing the scattered radiation only involves the elastic scattering state and the polaritonic vacuum. The second example situation is conversely more involved and it displays a much rich phenomenology associated to the possible entanglement bewteen the two ingoing polariations. As a matter of fact, we have found that reduced density operator, in addition to the two projectors pertaining elastic scattering and full absorbion of radiation, also generally displays a one-polariton contribution whose eigenstates and eigevalues crucially depend on the possible entanglement between the two ingoing polaritons. We conclude that also in the general situation involving any number of ingoing polaritons, the statistical mixture of quantum states describing the scattered radiation remarkably depend on the possible entanglement among the ingoing $s$-polaritons of the ingoing radiation. This suggests that the ingoing polariton entanglement provides a terrific potential to manipulating the quantum features of the scattered radiation since the ensuing quantum steering has a wide flexibility  supplied by the freedom in selecting the object dispersive optical response.

\newpage

\appendix
\begin{widetext}

\section{Asymptotic, dyadic and quantum relations}
The leading order term $F^{(1)}(r)$ of the asymptotic expansion of the function $F(r)$ for $r \rightarrow +\infty$ is defined by the relation
\begin{equation}
F\left( r \right) = F^{\left( 1 \right)} \left( r \right)\left[ {1 + O\left( {\frac{1}{r}} \right)} \right],\quad r \to \infty 
\end{equation}
to which we refer with the concise notation $F\left( r \right)\mathop  \approx \limits_{r \to \infty } F^{\left( 1 \right)} \left( r \right)$. 
If ${\bf n}$ and ${\bf m}$ are unit vectors and $f({\bf m})$ is a function defined on the unit sphere, the classical Jones' lemma (see Appendix XII of Ref.\cite{Bornnn}) states that
\begin{equation} \label{JonLem}
\int {do_{\bf{m}} } e^{i\left( {k_\omega  r} \right){\bf{n}} \cdot {\bf{m}}} f\left( {\bf{m}} \right) \mathop  \approx \limits_{r \to \infty }  \frac{{2\pi }}{{k_\omega  }} \left[
 \frac{{e^{ - ik_\omega  r} }}{-i r} f\left( - {\bf{n}} \right) + \frac{{e^{ik_\omega  r} }}{i r}f\left( {\bf{n}} \right)  \right].
\end{equation}
which can be written in distributional form as
\begin{equation} \label{DistJonLem}
e^{i\left( {k_\omega  r} \right){\bf{n}} \cdot {\bf{m}}} \mathop  \approx \limits_{r \to \infty } \frac{{2\pi }}{{k_\omega  }}\left[ {\frac{{e^{ - ik_\omega  r} }}{{ - ir}}\delta \left( {o_{- \bf{n}}  - o_{  {\bf{m}}} } \right) + \frac{{e^{ik_\omega  r} }}{{ir}}\delta \left( {o_{\bf{n}}  - o_{\bf{m}} } \right)} \right].
\end{equation}
Straightforward but relevant consequences of the Jones' lemma are the two asymptotic expansions
\begin{eqnarray} \label{InOutSphWavs}
&& \frac{k_\omega  }{2\pi} \int {do_{\bf{m}} } e^{i\left( {k_\omega  r} \right){\bf{n}} \cdot {\bf{m}}} \left[ {\theta \left( {{\bf{u}}_z  \cdot {\bf{n}}} \right)\theta \left( { - {\bf{u}}_z  \cdot {\bf{m}}} \right)} + {\theta \left( { - {\bf{u}}_z  \cdot {\bf{n}}} \right)\theta \left( {{\bf{u}}_z  \cdot {\bf{m}}} \right)} \right]f\left( {\bf{m}} \right)\mathop  \approx \limits_{r \to \infty }  {  \frac{{e^{ - ik_\omega  r} }}{- i r}f\left( { - {\bf{n}}} \right)}, \nonumber \\ 
&& \frac{k_\omega  }{2\pi} \int {do_{\bf{m}} } e^{i\left( {k_\omega  r} \right){\bf{n}} \cdot {\bf{m}}} \left[ {\theta \left( {{\bf{u}}_z  \cdot {\bf{n}}} \right)\theta \left( {{\bf{u}}_z  \cdot {\bf{m}}} \right)} + {\theta \left( { - {\bf{u}}_z  \cdot {\bf{n}}} \right)\theta \left( { - {\bf{u}}_z  \cdot {\bf{m}}} \right)} \right]f\left( {\bf{m}} \right)\mathop  \approx \limits_{r \to \infty } {\frac{{e^{ik_\omega  r} }}{i r}f\left( {\bf{n}} \right)},
\end{eqnarray}
where $\theta\left(\xi\right)$ is the Heaviside step-function, which provide the plane-wave representations in the far-field of the ingoing and outgoing modulated spherical waves, respectively.

We label with ${\bf{ab}}$ the dyad formed by the vectors ${\bf{a}}$ and ${\bf{b}}$. The identity dyadic is ${\cal I} = {\bf{u}}_x {\bf{u}}_x  + {\bf{u}}_y {\bf{u}}_y  + {\bf{u}}_z {\bf{u}}_z$, where ${\bf{u}}_x$, ${\bf{u}}_y$ and ${\bf{u}}_z$ are cartesian unit vectors. Any dyadic ${\cal D}$ admits the cartesian expansion ${\cal D} = D_{ij} {\bf{u}}_i {\bf{u}}_j $, where $D_{ij}  = {\bf{u}}_i  \cdot {\cal D} \cdot {\bf{u}}_j$, and its transpose is the dyadic ${\cal D}^T  = D_{ji} {\bf{u}}_i {\bf{u}}_j$ which, for any vector ${\bf V} = V_i {\bf u}_i$, satisfties the relation ${\cal D}^T  \cdot {\bf{V}} = {\bf{V}} \cdot {\cal D}$. In correspondence of the vector field ${\bf{V}}$ and the dyadic field  ${\cal D}$,  left ${\nabla  \times }$ and right ${ \times \mathord{\buildrel{\lower3pt\hbox{$\scriptscriptstyle\leftarrow$}} \over \nabla } }$ curls can be defined according to
\begin{equation}
\begin{array}{*{20}c}
   {\nabla  \times {\bf{V}} = \varepsilon _{ijk} \partial _i V_j {\bf{u}}_k ,} \hfill & \quad \quad {\nabla  \times {\cal D} = \varepsilon _{kip} \partial _k D_{ij} {\bf{u}}_p {\bf{u}}_j ,} \hfill  \\
   {{\bf{V}} \times \mathord{\buildrel{\lower3pt\hbox{$\scriptscriptstyle\leftarrow$}} 
\over \nabla }  = \varepsilon _{jik} \partial _i V_j {\bf{u}}_k ,} \hfill & \quad \quad {{\cal D} \times \mathord{\buildrel{\lower3pt\hbox{$\scriptscriptstyle\leftarrow$}} 
\over \nabla }  = \varepsilon _{jkp} \partial _k D_{ij} {\bf{u}}_i {\bf{u}}_p ,} \hfill  \\
\end{array}
\end{equation}
whose connection is established by relations
\begin{equation}
{\bf{V}} \times \mathord{\buildrel{\lower3pt\hbox{$\scriptscriptstyle\leftarrow$}} 
\over \nabla }  =  - \nabla  \times {\bf{V}},\quad \quad \quad \quad {\cal D} \times \mathord{\buildrel{\lower3pt\hbox{$\scriptscriptstyle\leftarrow$}} 
\over \nabla }  =  - \left( {\nabla  \times {\cal D}^T } \right)^T .
\end{equation}
If the vector fields ${\bf{V}}$ and ${\bf{U}}$ are piecewise continuous, the divergence theorem intended in the sense of distributions \cite{Kanwal} readily provides the relation
\begin{equation} \label{FrotF}
\int {d^3 {\bf{r}}} \;{\bf{V}} \cdot \left( {\nabla  \times {\bf{U}}} \right) = \int {d^3 {\bf{r}}} \left( {\nabla  \times {\bf{V}}} \right) \cdot {\bf{U}} - \int\limits_{S_\infty  } {dS} \;{\bf{u}}_{\bf r} \cdot \left( {{\bf{V}} \times {\bf{U}}} \right)
\end{equation}
where ${\bf{u}}_{\bf r} = {\bf r}/r$ is the radial unit vectors and the surface integral is performed over the sphere $S_\infty$ of infinite radius by means of the limiting prescription $\int_{S_\infty  } {dS} \;f = \mathop {\lim }\limits_{R \to \infty } \int_{S_R } {dS} \;f$, where $S_R$ is the surface of the sphere of radius $R$. 

The commutator of two vector operators ${\bf{\hat V}}= \hat V_i {\bf{u}}_i $ and ${\bf{\hat U}} = \hat U_j {\bf{u}}_j$ is the dyadic operator 
\begin{equation} \label{FielComm1}
\left[ {{\bf{\hat V}},{\bf{\hat U}}} \right] = \left[ {\hat V_i ,\hat U_j } \right]{\bf{u}}_i {\bf{u}}_j
\end{equation}
and, if ${\cal D}$ and ${\cal Q}$ are two dyadics, from Eq.(\ref{FielComm1}) it is easy to prove that
\begin{equation} \label{FielComm2}
\left[ {{\cal D} \cdot {\bf{\hat V}},{\cal Q} \cdot {\bf{\hat U}}} \right] = {\cal D} \cdot \left[ {{\bf{\hat V}},{\bf{\hat U}}} \right] \cdot {\cal Q}^T.
\end{equation}

\section{The scattering dyadics}
The relation between the modal dyadic ${\cal F}_{\omega s}$ and the asymptotic amplitude ${\cal W}_{\omega}$ of the dyadic Green's function in Eq.(\ref{FwolProp}) suggests an alternative way to evaluate the scattering dyadic ${\cal S}_{\omega}$. We start by noting that the fourth of Eqs.(\ref{GreBouVal}), after a suitable variables relabeling, can evidently be turned into 
\begin{equation}
{\cal W}_\omega  \left( {\left. {\bf{n}} \right|{\bf{r}}} \right) = r'e^{ - ik_\omega  r'} {\cal G}_\omega  \left( {\left. {r'{\bf{n}}} \right|{\bf{r}}} \right) + O\left( {\frac{1}{{r'}}} \right)
\end{equation}
for $r' \rightarrow +\infty$ so that Eq.(\ref{FwolProp}) readily yields
\begin{equation} \label{FosLim}
{\cal F}_{\omega s} \left( {\left. {\bf{r}} \right|{\bf{m}}} \right) = \sqrt {\frac{{\hbar k_\omega ^3 }}{{\pi \varepsilon _0 }}} \mathop {\lim }\limits_{r' \to + \infty }  {r'e^{ - ik_\omega  r'} {\cal G}_\omega  \left( {\left. {\bf{r}} \right| - r'{\bf{m}}} \right)},
\end{equation}
where the reciprocity relation of Eq.(\ref{GomRec}) has been used to take the transpose of the dyadic Green's function. Besides, the fourth of Eqs.(\ref{ESSBouVal}) can be written as
\begin{equation} \label{SomLim}
{\cal S}_\omega  \left( {{\bf{n}}\left| {\bf{m}} \right.} \right) = re^{ - ik_\omega  r} \left[ {\sqrt {\frac{{16\pi ^3 \varepsilon _0 }}{{\hbar k_\omega ^3 }}} {\cal F}_{\omega s} \left( {\left. {r{\bf{n}}} \right|{\bf{m}}} \right)}  { - e^{i\left( {k_\omega  r} \right){\bf{n}} \cdot {\bf{m}}} {\cal I}_{\bf{m}} } \right] + O\left( {\frac{1}{r}} \right)  
\end{equation}
for $r \rightarrow +\infty$ so that Eqs.(\ref{FosLim}) and (\ref{SomLim}) yields
\begin{equation} \label{SomGom}
{\cal S}_\omega  \left( {{\bf{n}}\left| {\bf{m}} \right.} \right) = \mathop {\lim }\limits_{r \to + \infty } \mathop {\lim }\limits_{r' \to + \infty } \left[ {4\pi rr'e^{ - ik_\omega  \left( {r + r'} \right)} {\cal G}_\omega  \left( {\left. {r{\bf{n}}} \right| - r'{\bf{m}}} \right) - re^{i\left( {k_\omega  r} \right)\left( {{\bf{n}} \cdot {\bf{m}} - 1} \right)} {\cal I}_{\bf{m}} } \right].  
\end{equation}
Note that each of the two terms inside the square bracket of Eq.(\ref{SomGom}) has no $r \rightarrow + \infty$ limit (as opposed to their difference) so that the order of the two limits can not be interchanged.
 
We now use such results to provide an alternative proof of the scattering dyadic reciprocity relation of Eq.(\ref{SomRec}), whose standard derivation can be found in textbooks \cite{Krist1}. Due to the singular behavior for $r \rightarrow + \infty$ of the two separate terms in Eq.(\ref{SomGom}), it is more convenient to prove that the relation
\begin{equation}
\int {do_{\bf{n}} } \left[ {{\cal S}_\omega ^T \left( {{\bf{n}}\left| {\bf{m}} \right.} \right) - {\cal S}_\omega  \left( { - {\bf{m}}\left| { - {\bf{n}}} \right.} \right)} \right] \cdot {\bf{J}}\left( {\bf{n}} \right) = 0
\end{equation}
holds for any vector field ${\bf{J}}\left( {\bf{n}} \right)$, which is evidently an equivalent way to state Eq.(\ref{SomRec}). Using Eqs.(\ref{FosLim}) and (\ref{SomLim}) together with the Green's dyadic reciprocity relation of Eq.(\ref{GomRec}) we readily get
\begin{eqnarray} \label{SoTRSoRW}
&& \int {do_{\bf{n}} } {\cal S}_\omega ^T \left( {{\bf{n}}\left| {\bf{m}} \right.} \right) \cdot {\bf{J}}\left( {\bf{n}} \right) = \mathop {\lim }\limits_{r \to \infty } re^{ - ik_\omega  r} \int {do_{\bf{n}} } \left[ { - e^{i\left( {k_\omega  r} \right){\bf{n}} \cdot {\bf{m}}} {\cal I}_{\bf{m}} }  { + \mathop {\lim }\limits_{r' \to \infty } 4\pi r'e^{ - ik_\omega  r'} {\cal G}_\omega  \left( {\left. { - r'{\bf{m}}} \right|r{\bf{n}}} \right)} \right] \cdot {\bf{J}}\left( {\bf{n}} \right), \nonumber \\ 
&& \int {do_{\bf{n}} } {\cal S}_\omega  \left( { - {\bf{m}}\left| { - {\bf{n}}} \right.} \right) \cdot {\bf{J}}\left( {\bf{n}} \right)   = \mathop {\lim }\limits_{r \to \infty } re^{ - ik_\omega  r} \int {do_{\bf{n}} } \left[ { - e^{i\left( {k_\omega  r} \right){\bf{n}} \cdot {\bf{m}}} {\cal I}_{\bf{n}} }  { + \mathop {\lim }\limits_{r' \to \infty } 4\pi r'e^{ - ik_\omega  r'} {\cal G}_\omega  \left( {\left. { - r{\bf{m}}} \right|r'{\bf{n}}} \right)} \right] \cdot {\bf{J}}\left( {\bf{n}} \right), 
\end{eqnarray}
where we stress that the $r \rightarrow +\infty$ limit is taken after integration as a consequence of the asymptotic character of Eq.(\ref{SomLim}). Taking the difference of Eqs.(\ref{SoTRSoRW}) we get
\begin{eqnarray} \label{SoTRminSoRW}
&& \int {do_{\bf{n}} } \left[ {{\cal S}_\omega ^T \left( {{\bf{n}}\left| {\bf{m}} \right.} \right) - {\cal S}_\omega  \left( { - {\bf{m}}\left| { - {\bf{n}}} \right.} \right)} \right] \cdot {\bf{J}}\left( {\bf{n}} \right)  = \mathop {\lim }\limits_{r \to \infty } re^{ - ik_\omega  r} \int {do_{\bf{n}} } e^{i\left( {k_\omega  r} \right){\bf{n}} \cdot {\bf{m}}} \left[ \left( {{\cal I}_{\bf{n}}  - {\cal I}_{\bf{m}} } \right) \cdot {\bf{J}}\left( {\bf{n}} \right) \right] \nonumber \\ 
&&  + \mathop {\lim }\limits_{r \to \infty } \mathop {\lim }\limits_{r' \to \infty } 4\pi rr'e^{ - ik_\omega  \left( {r + r'} \right)}  \cdot \int {do_{\bf{n}} } \left[ {{\cal G}_\omega  \left( {\left. { - r'{\bf{m}}} \right|r{\bf{n}}} \right) - {\cal G}_\omega  \left( {\left. { - r{\bf{m}}} \right|r'{\bf{n}}} \right)} \right] \cdot {\bf{J}}\left( {\bf{n}} \right),
\end{eqnarray}
where it has been possible to split the two $r \rightarrow + \infty$ limits since they separaterly exist. In fact, Jones' lemma of Eq.(\ref{JonLem}) readly yields
\begin{eqnarray}
&& re^{ - ik_\omega  r} \int {do_{\bf{n}} } e^{i\left( {k_\omega  r} \right){\bf{n}} \cdot {\bf{m}}} \left[ {\left( {{\cal I}_{\bf{n}}  - {\cal I}_{\bf{m}} } \right) \cdot {\bf{J}}\left( {\bf{n}} \right)} \right]  \mathop  \approx \limits_{r \to \infty } \frac{{2\pi i}}{{k_\omega  }} \left\{ e^{ - i2k_\omega  r} \left[ {\left( {{\cal I}_{\bf{n}}  - {\cal I}_{\bf{m}} } \right) \cdot {\bf{J}}\left( {\bf{n}} \right)} \right]_{{\bf{n}} =  - {\bf{m}}} - \left[ {\left( {{\cal I}_{\bf{n}}  - {\cal I}_{\bf{m}} } \right) \cdot {\bf{J}}\left( {\bf{n}} \right)} \right]_{{\bf{n}} = {\bf{m}}} \right\}= 0, \nonumber \\
\end{eqnarray}
so that the first contribution in the right hand side of Eq.(\ref{SoTRminSoRW}) vanishes. On the other, the symmetry of the roles played by $r$ and $r'$ in the second contribution in the right hand side of Eq.(\ref{SoTRminSoRW}) evidently entails its vanishing. Accordingly this proves the scattering dyadic reciprocity relation of Eq.(\ref{SomRec}) since the above argument is valid for an arbitrary field ${\bf{J}}\left( {\bf{n}} \right)$.

\section{Far-field behavior of the electric field operator}
To get the far-field asymptotic behavior at any time $t$ of the electric field operator of Eq.(\ref{EleHei}), we first consider its contributions arising from the three kinds of polaritons. By using the far-field behavior of the modal dyadic in the fourth of Eqs.(\ref{ESSBouVal}), the s-polariton contribution to the far-field can be written as
\begin{equation} \label{AsySpol1}
\int {do_{\bf{m}} } {\cal F}_{\omega s}  \left( {\left. r {\bf{n}} \right|{\bf{m}}} \right) \cdot {\bf{\hat g}}_{\omega s} \left( {\bf{m}} \right)\mathop  \approx \limits_{r \to  \infty }  \sqrt {\frac{{\hbar k_\omega ^3 }}{{16\pi ^3 \varepsilon _0 }}} \left[ {\int {do_{\bf{m}} } e^{i\left( {k_\omega  r} \right){\bf{n}} \cdot {\bf{m}}} {\bf{\hat g}}_{\omega s} \left( {\bf{m}} \right)}  { + \frac{{e^{ik_\omega  r} }}{r}\int {do_{\bf{m}} } {\cal S}_\omega  \left( {{\bf{n}}\left| {\bf{m}} \right.} \right) \cdot {\bf{\hat g}}_{\omega s} \left( {\bf{m}} \right)} \right]. 
\end{equation}
The first term in the right hand side of Eq.(\ref{AsySpol1}) is the superposition of all the plane waves pertaining  the far-field of the scattering modes and its asymptotic behavior can be readily evaluated by means of the Jones' lemma of Eq.(\ref{JonLem}) by using which Eq.(\ref{AsySpol1}) becomes
\begin{eqnarray} \label{AsySpol2}
\int {do_{\bf{m}} } {\cal F}_{\omega s} \left( {\left. {r{\bf{n}}} \right|{\bf{m}}} \right) \cdot {\bf{\hat g}}_{\omega s} \left( {\bf{m}} \right)\mathop  \approx \limits_{r \to \infty }  \sqrt {\frac{{\hbar k_\omega  }}{{4\pi \varepsilon _0 }}} \left\{ {  \frac{{e^{ - ik_\omega  r} }}{-i r}{\bf{\hat g}}_{\omega s} \left( { - {\bf{n}}} \right)} + \frac{{e^{ik_\omega  r} }}{i r}\int {do_{\bf{m}} } {\left[ {\delta \left( {o_{\bf{n}}  - o_{\bf{m}} } \right){\cal I}_{\bf{m}}  + \frac{{ik_\omega  }}{{2\pi }}{\cal S}_\omega  \left( {{\bf{n}}\left| {\bf{m}} \right.} \right)} \right] \cdot {\bf{\hat g}}_{\omega s} \left( {\bf{m}} \right)} \right\}. \nonumber \\
\end{eqnarray}
Note that in Eq.(\ref{AsySpol2}) both ingoing and outgoing spherical waves appear, this physically indicating that $s$-polaritons both supply ingoing and outgoing contributions to the radiation far-field, as expected. By using Eqs.(\ref{GoeGom}) together with the fourth of Eqs.(\ref{GreBouVal}) we get the $e$- and $m$-polaritons spectral contribution to the far-field
\begin{eqnarray} \label{AsyEMpol}
&& \sum\limits_{\nu  = e,m} {\int {d^3 {\bf{r}}'\,} {\cal G}_{\omega \nu } \left( {\left. r{\bf{n}} \right|{\bf{r}}'} \right) \cdot {\bf{\hat f}}_{\omega \nu } \left( {{\bf{r}}'} \right)} \mathop  \approx \limits_{r \to \infty } \frac{{e^{ik_\omega  r} }}{i r}\int {d^3 {\bf{r}}'\,} \left\{ {-\sqrt {\frac{{\hbar k_\omega ^4 }}{{\pi \varepsilon _0 }}{\mathop{\rm Im}\nolimits} \left[ {\varepsilon _\omega  \left( {{\bf{r}}'} \right)} \right]} {\cal W}_\omega  \left( {\left. {\bf{n}} \right|{\bf{r}}'} \right) \cdot {\bf{\hat f}}_{\omega e} \left( {{\bf{r}}'} \right)} \right.  \nonumber \\ 
&& \left. { + \sqrt {\frac{{\hbar k_\omega ^2 }}{{\pi \varepsilon _0 }}{\mathop{\rm Im}\nolimits} \left[ {\frac{{ - 1}}{{\mu _\omega  \left( {{\bf{r}}'} \right)}}} \right]} \left[ {{\cal W}_\omega  \left( {\left. {\bf{n}} \right|{\bf{r}}'} \right) \times \mathord{\buildrel{\lower3pt\hbox{$\scriptscriptstyle\leftarrow$}} 
\over \nabla } _{{\bf{r}}'} } \right] \cdot {\bf{\hat f}}_{\omega m} \left( {{\bf{r}}'} \right)} \right\},  
\end{eqnarray}
which contains only the outgoing spherical wave, in agreement with the physical fact that $e$- and $m$-polaritons are associated to medium dipole sources which  produce outgoing radiation. With the help of Eqs.(\ref{AsySpol2}) and (\ref{AsyEMpol}) we easily get from Eq.(\ref{EleHei}) the far-field behavior 
\begin{eqnarray}
&& {\bf{\hat E}}\left( {r{\bf{n}},t} \right)\mathop  \approx \limits_{r \to \infty } \int\limits_0^{ + \infty } {d\omega } \sqrt {\frac{{\hbar k_\omega  }}{{4\pi \varepsilon _0 }}} \left\{ {\frac{{e^{ - ik_\omega  \left( {r + ct} \right)} }}{{ - ir}}{\bf{\hat g}}_{\omega s} \left( { - {\bf{n}}} \right)} \right. + \frac{{e^{ik_\omega  \left( {r - ct} \right)} }}{{ir}}\left[ {\int {do_{\bf{m}} } {\cal T}_{\omega s}  \left( {\left. {\bf{n}} \right|{\bf{m}}} \right) \cdot {\bf{\hat g}}_{\omega s} \left( {\bf{m}} \right)} \right. \nonumber \\ 
&& \left. {\left. { + \sum\limits_{\nu  = e,m} {\int {d^3 {\bf{r}}'\,} {\cal E}_{\omega \nu } \left( {\left. {\bf{n}} \right|{\bf{r}}'} \right) \cdot {\bf{\hat f}}_{\omega \nu } \left( {{\bf{r}}'} \right)} } \right]} \right\} + h.c.,  
\end{eqnarray}
where the dyadics ${{\cal T}_{\omega s}  \left( {\left. {\bf{n}} \right|{\bf{m}}} \right)}$, ${\cal E}_{\omega e } \left( {\left. {\bf{n}} \right|{\bf{r}}'} \right)$ and ${\cal E}_{\omega m } \left( {\left. {\bf{n}} \right|{\bf{r}}'} \right)$ are defined in Eqs.(\ref{TwAweAwm}) of the main text.

\section{The input-output equation}
In order to derive the input-output equation in Eq.(\ref{MainScEq}), we start by evaluating the commutators in the left hand sides of Eqs.(\ref{GFComRel}) and (\ref{FGComRel}) by using the expressions of the outgoing polariton operators of Eqs.(\ref{Gos}) and (\ref{Fon}), thus getting after some algebra
\begin{eqnarray} \label{COMM1}
&& \left[ {{\bf{\hat G}}_{\omega s} \left( {\bf{n}} \right),{\bf{\hat G}}_{\omega 's}^\dag  \left( {{\bf{n}}'} \right)} \right] = \int {do_{\bf{m}} do_{{\bf{m}}'} } \left[ {{\cal T}_{\omega ss} \left( {\left. {\bf{n}} \right|{\bf{m}}} \right) \cdot {\bf{\hat g}}_{\omega s} \left( {\bf{m}} \right),{\cal T}_{\omega 'ss}^* \left( {\left. {{\bf{n}}'} \right|{\bf{m}}'} \right) \cdot {\bf{\hat g}}_{\omega 's}^\dag  \left( {{\bf{m}}'} \right)} \right] \nonumber \\ 
&&  + \sum\limits_{\alpha ,\alpha ' = e,m} {\int {d^3 {\bf{s}}d^3 {\bf{s}}'} \left[ {{\cal E}_{\omega s\alpha } \left( {\left. {\bf{n}} \right|{\bf{s}}} \right) \cdot {\bf{\hat f}}_{\omega \alpha } \left( {\bf{s}} \right),{\cal E}_{\omega 's\alpha '}^* \left( {\left. {{\bf{n}}'} \right|{\bf{s}}'} \right) \cdot {\bf{\hat f}}_{\omega '\alpha '}^\dag  \left( {{\bf{s}}'} \right)} \right],} \nonumber \\ 
&& \left[ {{\bf{\hat F}}_{\omega \nu } \left( {\bf{r}} \right),{\bf{\hat F}}_{\omega '\nu '}^\dag  \left( {{\bf{r}}'} \right)} \right] = \int {do_{\bf{m}} do_{{\bf{m}}'} } \left[ {{\cal M}_{\omega \nu s} \left( {{\bf{r}}\left| {\bf{m}} \right.} \right) \cdot {\bf{\hat g}}_{\omega s} \left( {\bf{m}} \right),{\cal M}_{\omega '\nu 's}^* \left( {{\bf{r}}'\left| {{\bf{m}}'} \right.} \right) \cdot {\bf{\hat g}}_{\omega 's}^\dag  \left( {{\bf{m}}'} \right)} \right] \nonumber \\ 
&&  + \sum\limits_{\alpha ,\alpha ' = e,m} {\int {d^3 {\bf{s}}d^3 {\bf{s}}'} } \left[ {{\cal V}_{\omega \nu \alpha } \left( {\left. {\bf{r}} \right|{\bf{s}}} \right) \cdot {\bf{\hat f}}_{\omega \alpha } \left( {\bf{s}} \right),{\cal V}_{\omega '\nu '\alpha '}^* \left( {\left. {{\bf{r}}'} \right|{\bf{s}}'} \right) \cdot {\bf{\hat f}}_{\omega '\alpha '}^\dag  \left( {{\bf{s}}'} \right)} \right], \nonumber\\ 
&& \left[ {{\bf{\hat F}}_{\omega \nu } \left( {\bf{r}} \right),{\bf{\hat G}}_{\omega 's}^\dag  \left( {\bf{n}} \right)} \right] = \int {do_{\bf{m}} } do_{{\bf{m}}'} \left[ {{\cal M}_{\omega \nu s} \left( {{\bf{r}}\left| {\bf{m}} \right.} \right) \cdot {\bf{\hat g}}_{\omega s} \left( {\bf{m}} \right),{\cal T}_{\omega 'ss}^* \left( {\left. {\bf{n}} \right|{\bf{m}}'} \right) \cdot {\bf{\hat g}}_{\omega 's}^\dag  \left( {{\bf{m}}'} \right)} \right] \nonumber \\ 
&&  + \sum\limits_{\alpha ,\alpha ' = e,m} {\int {d^3 {\bf{s}}d^3 {\bf{s}}'} } \left[ {{\cal V}_{\omega \nu \alpha } \left( {\left. {\bf{r}} \right|{\bf{s}}} \right) \cdot {\bf{\hat f}}_{\omega \alpha } \left( {\bf{s}} \right),{\cal E}_{\omega 's\alpha '}^* \left( {\left. {\bf{n}} \right|{\bf{s}}'} \right) \cdot {\bf{\hat f}}_{\omega '\alpha '}^\dag  \left( {{\bf{s}}'} \right)} \right],
\end{eqnarray}
where we exploited all the vanishing communtation relations among the ingoing polariton operators ${\bf{\hat g}}_{\omega s}$, ${\bf{\hat f}}_{\omega e}$ and ${\bf{\hat f}}_{\omega m}$. By using Eq.(\ref{FielComm2}) of Appendix A together with the commutation relations of Eqs.(\ref{gfComRel}) of the main text we get from Eqs.(\ref{COMM1}) 
\begin{eqnarray} \label{COMM2}
&& \left[ {{\bf{\hat G}}_{\omega s} \left( {\bf{n}} \right),{\bf{\hat G}}_{\omega 's}^\dag  \left( {{\bf{n}}'} \right)} \right] = \delta \left( {\omega  - \omega '} \right)\left[ {\int {do_{\bf{m}} } {\cal T}_{\omega ss} \left( {\left. {\bf{n}} \right|{\bf{m}}} \right) \cdot {\cal T}_{\omega ss}^{T*} \left( {\left. {{\bf{n}}'} \right|{\bf{m}}} \right) + \sum\limits_{\alpha =e,m}  {\int {d^3 {\bf{s}}} } {\cal E}_{\omega s\alpha } \left( {\left. {\bf{n}} \right|{\bf{s}}} \right) \cdot {\cal E}_{\omega s\alpha }^{T*} \left( {\left. {{\bf{n}}'} \right|{\bf{s}}} \right)} \right], \nonumber \\ 
&& \left[ {{\bf{\hat F}}_{\omega \nu } \left( {\bf{r}} \right),{\bf{\hat F}}_{\omega '\nu '}^\dag  \left( {{\bf{r}}'} \right)} \right] = \delta \left( {\omega  - \omega '} \right)\left[ {\int {do_{\bf{m}} } {\cal M}_{\omega \nu s} \left( {{\bf{r}}\left| {\bf{m}} \right.} \right) \cdot {\cal M}_{\omega \nu 's}^{T*} \left( {{\bf{r}}'\left| {\bf{m}} \right.} \right) + \sum\limits_{\alpha =e,m}  {\int {d^3 {\bf{s}}} } {\cal V}_{\omega \nu \alpha } \left( {\left. {\bf{r}} \right|{\bf{s}}} \right) \cdot {\cal V}_{\omega \nu '\alpha }^{T*} \left( {\left. {{\bf{r}}'} \right|{\bf{s}}} \right)} \right], \nonumber \\ 
&& \left[ {{\bf{\hat F}}_{\omega \nu } \left( {\bf{r}} \right),{\bf{\hat G}}_{\omega 's}^\dag  \left( {\bf{n}} \right)} \right] = \delta \left( {\omega  - \omega '} \right)\left[ {\int {do_{\bf{m}} } {\cal M}_{\omega \nu s} \left( {{\bf{r}}\left| {\bf{m}} \right.} \right) \cdot {\cal T}_{\omega ss}^{T*} \left( {\left. {\bf{n}} \right|{\bf{m}}} \right) + \sum\limits_{\alpha =e,m}  {\int {d^3 {\bf{s}}} {\cal V}_{\omega \nu \alpha } \left( {\left. {\bf{r}} \right|{\bf{s}}} \right) \cdot {\cal E}_{\omega s\alpha }^{T*} \left( {\left. {\bf{n}} \right|{\bf{s}}} \right)} } \right],
\end{eqnarray}
whose comparison with the commutation relations of Eqs.(\ref{GFComRel}) and (\ref{FGComRel}) directly requires to set
\begin{eqnarray} \label{COMM3}
&& \int {do_{\bf{m}} } {\cal T}_{\omega ss} \left( {\left. {\bf{n}} \right|{\bf{m}}} \right) \cdot {\cal T}_{\omega ss}^{T*} \left( {\left. {{\bf{n}}'} \right|{\bf{m}}} \right) + \sum\limits_{\alpha =e,m}  {\int {d^3 {\bf{s}}} } {\cal E}_{\omega s\alpha } \left( {\left. {\bf{n}} \right|{\bf{s}}} \right) \cdot {\cal E}_{\omega s\alpha }^{T*} \left( {\left. {{\bf{n}}'} \right|{\bf{s}}} \right) = \delta \left( {o_{\bf{n}}  - o_{{\bf{n}}'} } \right){\cal I}_{\bf{n}}, \nonumber  \\ 
&& \int {do_{\bf{m}} } {\cal M}_{\omega \nu s} \left( {{\bf{r}}\left| {\bf{m}} \right.} \right) \cdot {\cal M}_{\omega \nu 's}^{T*} \left( {{\bf{r}}'\left| {\bf{m}} \right.} \right) + \sum\limits_{\alpha =e,m}  {\int {d^3 {\bf{s}}} } {\cal V}_{\omega \nu \alpha } \left( {\left. {\bf{r}} \right|{\bf{s}}} \right) \cdot {\cal V}_{\omega \nu '\alpha }^{T*} \left( {\left. {{\bf{r}}'} \right|{\bf{s}}} \right) = \delta _{\nu \nu '} \delta \left( {{\bf{r}} - {\bf{r}}'} \right){\cal I}, \nonumber \\ 
&& \int {do_{\bf{m}} } {\cal M}_{\omega \nu s} \left( {{\bf{r}}\left| {\bf{m}} \right.} \right) \cdot {\cal T}_{\omega ss}^{T*} \left( {\left. {\bf{n}} \right|{\bf{m}}} \right) + \sum\limits_{\alpha =e,m}  {\int {d^3 {\bf{s}}} {\cal V}_{\omega \nu \alpha } \left( {\left. {\bf{r}} \right|{\bf{s}}} \right) \cdot {\cal E}_{\omega s\alpha }^{T*} \left( {\left. {\bf{n}} \right|{\bf{s}}} \right)}  = 0. 
\end{eqnarray}
At this point it becomes clear that it is convenient to adopt the compact notation introduced at the beginning of Section IV since, after noting that $\delta \left( {o_{\bf{n}}  - o_{{\bf{n}}'} } \right){\cal I}_{\bf{n}}$ and $
\delta \left( {{\bf{r}} - {\bf{r}}'} \right){\cal I}$ are the kernels of the identity operators ${\sf I}_{ss}$ and ${\sf I}_{ee} = {\sf I}_{mm}$, Eqs.(\ref{COMM3}) can be written as
\begin{eqnarray} \label{COMM4}
&& {\sf T}_{\omega ss} {\sf T}_{\omega ss}^ +   
+ \begin{pmatrix}
   {{\sf E}_{\omega se} } & {{\sf E}_{\omega sm} }  \\
\end{pmatrix}
\begin{pmatrix}
   {{\sf E}_{\omega se}^ +  }  \\
   {{\sf E}_{\omega sm}^ +  }  \\
\end{pmatrix} = {\sf I}_{ss}, \nonumber  \\ 
&& \begin{pmatrix}
   {{\sf M}_{\omega es} }  \\
   {{\sf M}_{\omega ms} }  \\
\end{pmatrix}
\begin{pmatrix}
   {{\sf M}_{\omega es}^ +  } & {{\sf M}_{\omega ms}^ +  }  \\
\end{pmatrix}
 + \begin{pmatrix}
   {{\sf V}_{\omega ee} } & {{\sf V}_{\omega em} }  \\
   {{\sf V}_{\omega me} } & {{\sf V}_{\omega mm} }  \\
\end{pmatrix}
\begin{pmatrix}
   {{\sf V}_{\omega ee}^ +  } & {{\sf V}_{\omega me}^ +  }  \\
   {{\sf V}_{\omega em}^ +  } & {{\sf V}_{\omega mm}^ +  }  \\
\end{pmatrix}
 = \begin{pmatrix}
   {{\sf I}_{ee} } & 0  \\
   0 & {{\sf I}_{mm} }  \\
\end{pmatrix}, \nonumber \\ 
&& \begin{pmatrix}
   {{\sf M}_{\omega es} }  \\
   {{\sf M}_{\omega ms} }  \\
\end{pmatrix}
{\sf T}_{\omega ss}^ +  
+ \begin{pmatrix}
   {{\sf V}_{\omega ee} } & {{\sf V}_{\omega em} }  \\
   {{\sf V}_{\omega me} } & {{\sf V}_{\omega mm} }  \\
\end{pmatrix}
\begin{pmatrix}
   {{\sf E}_{\omega se}^ +  }  \\
   {{\sf E}_{\omega sm}^ +  }  \\
\end{pmatrix} = 
\begin{pmatrix}
   0  \\
   0  \\
\end{pmatrix}.
\end{eqnarray}
These equations are easily seen to be equivalent to the matrix equation in Eq.(\ref{MainScEq}) we here rewrite for convenience
\begin{equation} \label{COMM5}
\begin{pmatrix}
   {{\sf T}_{\omega ss} } & {{\sf E}_{\omega se} } & {{\sf E}_{\omega sm} }  \\
   {{\sf M}_{\omega es} } & {{\sf V}_{\omega ee} } & {{\sf V}_{\omega em} }  \\
   {{\sf M}_{\omega ms} } & {{\sf V}_{\omega me} } & {{\sf V}_{\omega mm} }  \\
\end{pmatrix}
\begin{pmatrix}
   {{\sf T}_{\omega ss}^ +  } & {{\sf M}_{\omega es}^ +  } & {{\sf M}_{\omega ms}^ +  }  \\
   {{\sf E}_{\omega se}^ +  } & {{\sf V}_{\omega ee}^ +  } & {{\sf V}_{\omega me}^ +  }  \\
   {{\sf E}_{\omega sm}^ +  } & {{\sf V}_{\omega em}^ +  } & {{\sf V}_{\omega mm}^ +  }  \\
\end{pmatrix}
 = \begin{pmatrix}
   {{\sf I}_{ss} } & 0 & 0  \\
   0 & {{\sf I}_{ee} } & 0  \\
   0 & 0 & {{\sf I}_{mm} }  \\
\end{pmatrix},
\end{equation}
since the first two of Eqs.(\ref{COMM4}) reproduce the upper $1 \times 1$ and lower $2 \times 2$ submatrices on the diagonal of Eq.(\ref{COMM5}), respectively, whereas the third of Eqs.(\ref{COMM4}) and its  Hermitian conjugate 
\begin{equation}
{\sf T}_{\omega ss} \begin{pmatrix}
   {{\sf M}_{\omega es}^ +  } & {{\sf M}_{\omega ms}^ +  }  \\
\end{pmatrix}
+ \begin{pmatrix}
   {{\sf E}_{\omega se} } & {{\sf E}_{\omega sm} }  \\
\end{pmatrix}
\begin{pmatrix}
   {{\sf V}_{\omega ee}^ +  } & {{\sf V}_{\omega me}^ +  }  \\
   {{\sf V}_{\omega em}^ +  } & {{\sf V}_{\omega mm}^ +  }  \\
\end{pmatrix} = 
\begin{pmatrix}
   0 & 0  \\
\end{pmatrix}
\end{equation}
coincide with the lower left $2 \times 1$ and upper right $1 \times 2$ submatrices of Eq.(\ref{COMM5}).

\section{Relations connecting the transmission and emission/absorption dyadics to the dyadic Green's function}
As a prelude to the derivation of Eqs.(\ref{RelTAG}) and (\ref{RelTAGConj}) from the fundamental integral relation of Eq.(\ref{GreFunInt}), it is first convenient introducing the two operators ${{\sf \Gamma} _{\omega l e} }$ and ${{\sf \Gamma} _{\omega l m} }$ whose left-actions on the vector fields ${\bf{v}}_e$ and ${\bf{v}}_m$  are given by
\begin{eqnarray}
&& \left( {{\sf \Gamma} _{\omega le} {\bf{v}}_e } \right)\left( {\bf{r}} \right) =  - \sqrt {4k_\omega ^3 {\mathop{\rm Im}\nolimits} \left[ {\varepsilon _\omega  \left( {\bf{r}} \right)} \right]} {\bf{v}}_e \left( {\bf{r}} \right), \nonumber \\
&& \left( { {\sf \Gamma} _{\omega lm} {\bf{v}}_m } \right)\left( {\bf{r}} \right) = \nabla _{\bf{r}}  \times \left\{ { - \sqrt {4k_\omega  {\mathop{\rm Im}\nolimits} \left[ {\frac{{ - 1}}{{\mu _\omega  \left( {\bf{r}} \right)}}} \right]} {\bf{v}}_m \left( {\bf{r}} \right)} \right\}.
\end{eqnarray}
These operators are real (i.e. ${\sf \Gamma} _{\omega l e}^*  = {\sf \Gamma} _{\omega l e}$, ${\sf \Gamma} _{\omega l m}^*  = {\sf \Gamma} _{\omega l m}$) and their domain spaces are $L^2_{e}$ and $L^2_{m}$ which we remark to be copies of their image space $L^2_{l}$ (the distinction among the subscripts $e,m,l$ having a later notational convenience). We denote with ${\bf{v}}_l^ + $ the dual vector field (linear functional) associated to the vector field ${\bf{v}}_l$ by the $L^2_{l}$ scalar product, i.e. ${\bf{v}}_l^ +  {\bf{w}}_l  = \int {d^3 {\bf{r}}} \;{\bf{v}}_l^* \left( {\bf{r}} \right) \cdot {\bf{w}}_l \left( {\bf{r}} \right)$. The right-action of ${\sf \Gamma} _{\omega le}$ on the dual vector field ${\bf{v}}_l^ +$ is easily seen to be given by
\begin{equation} \label{vlGole1}
\left( {{\bf{v}}_l^ +  {\sf \Gamma} _{\omega le} } \right){\bf{v}}_e  = \int {d^3 {\bf{r}}} \;\left\{ { - {\bf{v}}_l \left( {\bf{r}} \right)\sqrt {4k_\omega ^3 {\mathop{\rm Im}\nolimits} \left[ {\varepsilon _\omega  \left( {\bf{r}} \right)} \right]} } \right\}^*  \cdot {\bf{v}}_e \left( {\bf{r}} \right)
\end{equation}
whereas, due to the presence of the curl, the right-action of ${\sf \Gamma} _{\omega l m}$ on the dual vector field ${\bf{v}}_l^ +$ is a little more involved and has to be deduced from the relation
\begin{equation} \label{vlGolm1}
\left( {{\bf{v}}_l^ +  {\sf \Gamma} _{\omega lm} } \right){\bf{v}}_m  = \int {d^3 {\bf{r}}} \;{\bf{v}}_l^* \left( {\bf{r}} \right) \cdot \nabla _{\bf{r}}  \times \left\{ { - \sqrt {4k_\omega  {\mathop{\rm Im}\nolimits} \left[ {\frac{{ - 1}}{{\mu _\omega  \left( {\bf{r}} \right)}}} \right]} {\bf{v}}_m \left( {\bf{r}} \right)} \right\}.
\end{equation}
Using Eq.(\ref{FrotF}) to integrate by parts the right hand side of Eq.(\ref{vlGolm1}), we get
\begin{eqnarray} \label{vlGolm2}
\left( {{\bf{v}}_l^ +  {\sf \Gamma} _{\omega lm} } \right){\bf{v}}_m  = \int {d^3 {\bf{r}}} \left\{ { - \sqrt {4k_\omega  {\mathop{\rm Im}\nolimits} \left[ {\frac{{ - 1}}{{\mu _\omega  \left( {\bf{r}} \right)}}} \right]} \nabla _{\bf{r}}  \times {\bf{v}}_l^* \left( {\bf{r}} \right)} \right\} \cdot {\bf{v}}_m \left( {\bf{r}} \right) + \int\limits_{S_\infty  } {dS} \;\sqrt {4k_\omega  {\mathop{\rm Im}\nolimits} \left[ {\frac{{ - 1}}{{\mu _\omega  \left( {\bf{r}} \right)}}} \right]} {\bf{u}}_r  \cdot {\bf{v}}_l^* \left( {\bf{r}} \right) \times {\bf{v}}_m \left( {\bf{r}} \right) \nonumber \\
\end{eqnarray}
which, after noting that the $\mu_\omega = 1$ outside the object (and hence on $S_\infty$) and exploiting the vector relation $\nabla  \times {\bf{V}} =  - {\bf{V}} \times \mathord{\buildrel{\lower3pt\hbox{$\scriptscriptstyle\leftarrow$}} 
\over \nabla }$, eventually yields
\begin{equation} \label{vlGolm3}
\left( {{\bf{v}}_l^ +  {\sf \Gamma} _{\omega lm} } \right){\bf{v}}_m  = \int {d^3 {\bf{r}}} \left\{ {{\bf{v}}_l \left( {\bf{r}} \right) \times \mathord{\buildrel{\lower3pt\hbox{$\scriptscriptstyle\leftarrow$}} 
\over \nabla } _{\bf{r}} \sqrt {4k_\omega  {\mathop{\rm Im}\nolimits} \left[ {\frac{{ - 1}}{{\mu _\omega  \left( {\bf{r}} \right)}}} \right]} } \right\}^*  \cdot {\bf{v}}_m \left( {\bf{r}} \right).
\end{equation}
The relations ${\bf{v}}_e^ +  {\sf \Gamma} _{\omega le}^ +  {\bf{v}}_l  = \left( {{\bf{v}}_l^ +  {\sf \Gamma} _{\omega le} {\bf{v}}_e } \right)^*$ and ${\bf{v}}_m^ +  {\sf \Gamma} _{\omega lm}^ +  {\bf{v}}_l  = \left( {{\bf{v}}_l^ +  {\sf \Gamma} _{\omega lm} {\bf{v}}_m } \right)^*$ enable to show that the Hermitian conjugate operators ${{\sf \Gamma} _{\omega le}^ +  }$ and ${{\sf \Gamma} _{\omega lm}^ +  }$ have left-actions on the vector field ${\bf{v}}_l$ 
given by 
\begin{eqnarray} \label{vlGolm4}
&& \left( {{\sf \Gamma} _{\omega le}^ +  {\bf{v}}_l } \right)\left( {\bf{r}} \right) =  - \sqrt {4k_\omega ^3 {\mathop{\rm Im}\nolimits} \left[ {\varepsilon _\omega  \left( {\bf{r}} \right)} \right]} {\bf{v}}_l \left( {\bf{r}} \right),\nonumber \\
&& \left( {{\sf \Gamma} _{\omega lm}^ +  {\bf{v}}_l } \right)\left( {\bf{r}} \right) =  - \sqrt {4k_\omega  {\mathop{\rm Im}\nolimits} \left[ {\frac{{ - 1}}{{\mu _\omega  \left( {\bf{r}} \right)}}} \right]} \nabla _{\bf{r}}  \times {\bf{v}}_l \left( {\bf{r}} \right),
\end{eqnarray}
and right-actions on dual vector fields ${{\bf{v}}_e^ +  }$ and ${{\bf{v}}_m^ +  }$ given by
\begin{eqnarray}
&& \left( {{\bf{v}}_e^ +  {\sf \Gamma} _{\omega le}^ +  } \right){\bf{v}}_l  = \int {d^3 {\bf{r}}} \;\left\{ { - {\bf{v}}_e \left( {\bf{r}} \right)\sqrt {4k_\omega ^3 {\mathop{\rm Im}\nolimits} \left[ {\varepsilon _\omega  \left( {\bf{r}} \right)} \right]} } \right\}^*  \cdot {\bf{v}}_l \left( {\bf{r}} \right), \nonumber \\ 
&& \left( {{\bf{v}}_m^ +  {\sf \Gamma} _{\omega lm}^ +  } \right){\bf{v}}_l  = \int {d^3 {\bf{r}}} \;\left\{ {\left\{ {{\bf{v}}_m \left( {\bf{r}} \right)\sqrt {4k_\omega  {\mathop{\rm Im}\nolimits} \left[ {\frac{{ - 1}}{{\mu _\omega  \left( {\bf{r}} \right)}}} \right]} } \right\} \times \mathord{\buildrel{\lower3pt\hbox{$\scriptscriptstyle\leftarrow$}} 
\over \nabla } _{\bf{r}} } \right\}^*  \cdot {\bf{v}}_l \left( {\bf{r}} \right). 
\end{eqnarray}
The main reason of the introduction of the operators ${\sf \Gamma} _{\omega le}$ and ${\sf \Gamma} _{\omega lm}$ lies in the fact that, by using Eqs.(\ref{vlGole1}) and (\ref{vlGolm3}), Eqs.(\ref{GoeGom}) and  the last two of Eqs.(\ref{TwAweAwm}) can be written as
\begin{eqnarray} \label{GoeGomAoseAosm}
&& {\cal G}_{\omega e} \left( {\left. {\bf{r}} \right|{\bf{r}}'} \right) = \frac{1}{{2i}}\sqrt {\frac{{\hbar k_\omega  }}{{\pi \varepsilon _0 }}} \left[\kern-0.15em\left[ {{\sf G}_{\omega ll} {\sf \Gamma} _{\omega le} } 
 \right]\kern-0.15em\right]\left( {{\bf{r}}\left| {{\bf{r}}'} \right.} \right),\quad \quad {\cal E}_{\omega se} \left( {\left. {\bf{n}} \right|{\bf{r}}} \right) = \left[\kern-0.15em\left[ {{\sf W}_{\omega sl} {\sf \Gamma} _{\omega le} } 
 \right]\kern-0.15em\right]\left( {{\bf{n}}\left| {\bf{r}} \right.} \right), \nonumber \\ 
&& {\cal G}_{\omega m} \left( {\left. {\bf{r}} \right|{\bf{r}}'} \right) = \frac{1}{{2i}}\sqrt {\frac{{\hbar k_\omega  }}{{\pi \varepsilon _0 }}} \left[\kern-0.15em\left[ {{\sf G}_{\omega ll} {\sf \Gamma} _{\omega lm} } 
 \right]\kern-0.15em\right]\left( {{\bf{r}}\left| {{\bf{r}}'} \right.} \right),\quad \quad {\cal E}_{\omega sm} \left( {\left. {\bf{n}} \right|{\bf{r}}} \right) = \left[\kern-0.15em\left[ {{\sf W}_{\omega sl} {\sf \Gamma} _{\omega lm} } 
 \right]\kern-0.15em\right]\left( {{\bf{n}}\left| {\bf{r}} \right.} \right),  
\end{eqnarray}
where ${\sf G}_{\omega ll}$ and ${\sf W}_{\omega sl}$ are the Green and asymptotic integral operators, respectively, such that $\left[\kern-0.15em\left[ {{\sf G}_{\omega ll} }  \right]\kern-0.15em\right]\left( {\left. {\bf{r}} \right|{\bf{r}}'} \right) = {\cal G}_\omega  \left( {{\bf{r}}\left| {{\bf{r}}'} \right.} \right)$ and $\left[\kern-0.15em\left[ {{\sf W}_{\omega sl} } \right]\kern-0.15em\right]\left( {\left. {\bf{n}} \right|{\bf{r}}} \right) = {\cal W}_\omega  \left( {{\bf{n}}\left| {\bf{r}} \right.} \right)$. As a final preliminar remark, we note that the reciprocity relations of the dyadic Green's function in Eq.(\ref{GomRec}) and of the transmission dyadic in Eq.(\ref{TosRec}) can be expressed in operatorial form as 
\begin{eqnarray} \label{OpeRecip}
&& {\sf G}_{\omega ll}^ +   = {\sf G}_{\omega ll}^*, \nonumber  \\ 
&& {\sf T}_{\omega ss}^ +   = {\sf J}_{ss} {\sf T}_{\omega ss}^* {\sf J}_{ss},
\end{eqnarray}
where ${\sf J}_{ss}$ is the inversion operator in the space $L^2_s$, i.e. $\left( {{\sf J}_{ss} {\bf{v}}_s } \right)\left( {\bf{n}} \right) = {\bf{v}}_s \left( { - {\bf{n}}} \right)$, which is evidently an Hermitian ${\sf J}_{ss}^ +   = {\sf J}_{ss}$ and involutive ${\sf J}_{ss} {\sf J}_{ss}  = {\sf I}_{ss}$ operator.

We are now prepared for the derivation of Eq.(\ref{RelTAG}). Using the left column of Eqs.(\ref{GoeGomAoseAosm}), the fundamental integral relation of Eq.(\ref{GreFunInt}) is easily seen to get the operatorial form
\begin{equation} \label{OpeFunGo}
\left( {{\sf W}_{\omega sl}^ +  {\sf W}_{\omega sl} } \right)^*  + \frac{1}{{4k_\omega ^2 }}{\sf G}_{\omega ll} \left( {{\sf \Gamma} _{\omega le} {\sf \Gamma} _{\omega le}^ +   + {\sf \Gamma} _{\omega lm} {\sf \Gamma} _{\omega lm}^ +  } \right){\sf G}_{\omega ll}^ +   = \frac{1}{{2ik_\omega  }}\left( {{\sf G}_{\omega ll}  - {\sf G}_{\omega ll}^* } \right).
\end{equation}
After left multiplication by the column $\begin{pmatrix} {{\sf \Gamma} _{\omega le}^ +  } \\ {{\sf \Gamma} _{\omega lm}^ +  } \end{pmatrix}$ and right multiplication by the row $\begin{pmatrix} {{\sf \Gamma} _{\omega le} } & {{\sf \Gamma} _{\omega lm} } \end{pmatrix}$, Eq.(\ref{OpeFunGo}) yields the operator matrix equation
\begin{eqnarray} \label{A+ARel}
&& \begin{pmatrix}
   {{\sf E}_{\omega se}^{ + *} }  \\
   {{\sf E}_{\omega sm}^{ + *} }  \\
\end{pmatrix}
\begin{pmatrix}
   {{\sf E}_{\omega se}^* } & {{\sf E}_{\omega sm}^* }  \\
\end{pmatrix} + \frac{1}{{4k_\omega ^2 }}\begin{pmatrix}
   {{\sf \Gamma} _{\omega le}^ +  }  \\
   {{\sf \Gamma} _{\omega lm}^ +  }  \\
\end{pmatrix}
{\sf G}_{\omega ll}
\begin{pmatrix}
   {{\sf \Gamma} _{\omega le} } & {{\sf \Gamma} _{\omega lm} }  \\
\end{pmatrix}
\begin{pmatrix}
   {{\sf \Gamma} _{\omega le}^ +  }  \\
   {{\sf \Gamma} _{\omega lm}^ +  }  \\
\end{pmatrix}
{\sf G}_{\omega ll}^ +  
\begin{pmatrix}
   {{\sf \Gamma} _{\omega le} } & {{\sf \Gamma} _{\omega lm} }  \\
\end{pmatrix} \nonumber \\ 
&&  = -\frac{i}{{2k_\omega  }}
\begin{pmatrix}
   {{\sf \Gamma} _{\omega le}^ +  }  \\
   {{\sf \Gamma} _{\omega lm}^ +  }  \\
\end{pmatrix}
{\sf G}_{\omega ll} 
\begin{pmatrix}
   {{\sf \Gamma} _{\omega le} } & {{\sf \Gamma} _{\omega lm} }  \\
\end{pmatrix} + \frac{i}{{2k_\omega  }}
\begin{pmatrix}
   {{\sf \Gamma} _{\omega le}^ +  }  \\
   {{\sf \Gamma} _{\omega lm}^ +  }  \\
\end{pmatrix}
{\sf G}_{\omega ll}^+ 
\begin{pmatrix}
   {{\sf \Gamma} _{\omega le} } & {{\sf \Gamma} _{\omega lm} }  \\
\end{pmatrix},
\end{eqnarray}
where, in the first term of the left hand side, the emission operators ${\sf E}_{\omega se} $ and ${\sf E}_{\omega sm} $ have shown up due to the right column of Eqs.(\ref{GoeGomAoseAosm}) whereas, in the last term of the right hand side, ${\sf G}_{\omega ll}^+ $ has replaced ${\sf G}_{\omega ll}^* $ as a consequence of the first of Eqs.(\ref{OpeRecip}) (reciprocity of the Green's operator). Now Eq.(\ref{A+ARel}) turns, after some algebra, into
\begin{eqnarray}
&& \begin{pmatrix}
   {{\sf E}_{\omega se}^{ + *} }  \\
   {{\sf E}_{\omega sm}^{ + *} }  \\
\end{pmatrix}
\begin{pmatrix}
   {{\sf E}_{\omega se}^* } & {{\sf E}_{\omega sm}^* }  \\
\end{pmatrix} + \left[ {
\begin{pmatrix}
   {{\sf I}_{ee} } & 0  \\
   0 & {{\sf I}_{mm} }  \\
\end{pmatrix} + \frac{i}{{2k_\omega  }}
\begin{pmatrix}
   {{\sf \Gamma} _{\omega le}^ +  }  \\
   {{\sf \Gamma} _{\omega lm}^ +  }  \\
\end{pmatrix}
{\sf G}_{\omega ll}
\begin{pmatrix}
   {{\sf \Gamma} _{\omega le} } & {{\sf \Gamma} _{\omega lm} }  \\
\end{pmatrix}} \right] \nonumber \\ 
&&  \cdot \left[ { 
\begin{pmatrix}
   {{\sf I}_{ee} } & 0  \\
   0 & {{\sf I}_{mm} }  \\
\end{pmatrix}  - \frac{i}{{2k_\omega  }}
\begin{pmatrix}
   {{\sf \Gamma} _{\omega le}^ +  }  \\
   {{\sf \Gamma} _{\omega lm}^ +  }  \\
\end{pmatrix} 
{\sf G}_{\omega ll}^ +
\begin{pmatrix}
   {{\sf \Gamma} _{\omega le} } & {{\sf \Gamma} _{\omega lm} }  \\
\end{pmatrix}} \right]   = 
\begin{pmatrix}
   {{\sf I}_{ee} } & 0  \\
   0 & {{\sf I}_{mm} }  \\
\end{pmatrix} 
\end{eqnarray}
which can also be written as
\begin{equation} \label{AAQQII}
\begin{pmatrix}
   {{\sf E}_{\omega se}^{ + *} }  \\
   {{\sf E}_{\omega sm}^{ + *} }  \\
\end{pmatrix}
\begin{pmatrix}
   {{\sf E}_{\omega se}^* } & {{\sf E}_{\omega sm}^* }  \\
\end{pmatrix} + 
\begin{pmatrix}
   {{\sf Q}_{\omega ee} } & {{\sf Q}_{\omega em} }  \\
   {{\sf Q}_{\omega me} } & {{\sf Q}_{\omega mm} }  \\
\end{pmatrix}
\begin{pmatrix}
   {{\sf Q}_{\omega ee}^ +  } & {{\sf Q}_{\omega em}^ +  }  \\
   {{\sf Q}_{\omega me}^ +  } & {{\sf Q}_{\omega mm}^ +  }  \\
\end{pmatrix} =
\begin{pmatrix}
   {{\sf I}_{ee} } & 0  \\
   0 & {{\sf I}_{mm} }  \\
\end{pmatrix},
\end{equation}
where we have introduced the operator matrix
\begin{equation} \label{OpeQ}
\begin{pmatrix}
   {{\sf Q}_{\omega ee} } & {{\sf Q}_{\omega em} }  \\
   {{\sf Q}_{\omega me} } & {{\sf Q}_{\omega mm} }  \\
\end{pmatrix} = - \left[
\begin{pmatrix}
   {{\sf I}_{ee} } & 0  \\
   0 & {{\sf I}_{mm} }  \\
\end{pmatrix} + \frac{i}{{2k_\omega  }}\begin{pmatrix}
   {{\sf \Gamma} _{\omega le}^ +  }  \\
   {{\sf \Gamma} _{\omega lm}^ +  }  \\
\end{pmatrix}
{\sf G}_{\omega ll} 
\begin{pmatrix}
   {{\sf \Gamma} _{\omega le} } & {{\sf \Gamma} _{\omega lm} }  \\
\end{pmatrix} \right] 
\end{equation}
(the minus sign has been introduced for later convenience). Note that, as a consequence of the reciprocity of Green's operator in the first of Eq.(\ref{OpeRecip}) and of the reality of the operators ${\sf \Gamma} _{\omega le}$ and ${\sf \Gamma} _{\omega lm}$, the operators ${\sf Q}_{\omega \nu \nu'}$ satisfy the matrix relation
\begin{equation} \label{Q+Q*}
\begin{pmatrix}
   {{\sf Q}_{\omega ee}^ +  } & {{\sf Q}_{\omega me}^ +  }  \\
   {{\sf Q}_{\omega em}^ +  } & {{\sf Q}_{\omega mm}^ +  }  \\
\end{pmatrix} = \begin{pmatrix}
   {{\sf Q}_{\omega ee}^* } & {{\sf Q}_{\omega em}^* }  \\
   {{\sf Q}_{\omega me}^* } & {{\sf Q}_{\omega mm}^* }  \\
\end{pmatrix}.
\end{equation}
Besides, the kernels of the integral operators ${\sf Q}_{\omega \nu \nu'}$ can be obtained from Eq.(\ref{OpeQ}) by using the left column of Eqs.(\ref{GoeGomAoseAosm}) together with Eqs.(\ref{vlGolm4}) and the result is
\begin{equation}
\begin{pmatrix}
   {{\cal Q}_{\omega ee} \left( {\left. {\bf{r}} \right|{\bf{r}}'} \right)} & {{\cal Q}_{\omega em} \left( {\left. {\bf{r}} \right|{\bf{r}}'} \right)}  \\
   {{\cal Q}_{\omega me} \left( {\left. {\bf{r}} \right|{\bf{r}}'} \right)} & {{\cal Q}_{\omega mm} \left( {\left. {\bf{r}} \right|{\bf{r}}'} \right)}  \\
\end{pmatrix} = -\delta \left( {{\bf{r}} - {\bf{r}}'} \right)
 \begin{pmatrix}
   {{\cal I}} & 0  \\
   0 & {{\cal I}}  \\
\end{pmatrix} - 
\begin{pmatrix}
   {\sqrt {\frac{{4\pi \varepsilon _0 }}{\hbar }{\mathop{\rm Im}\nolimits} \left[ {\varepsilon _\omega  \left( {\bf{r}} \right)} \right]} }  \\
   {\sqrt {\frac{{4\pi \varepsilon _0 }}{{\hbar k_\omega ^2 }}{\mathop{\rm Im}\nolimits} \left[ {\frac{{ - 1}}{{\mu _\omega  \left( {\bf{r}} \right)}}} \right]} \nabla _{\bf{r}}  \times }  \\
\end{pmatrix}
\begin{pmatrix}
   {{\cal G}_{\omega e} \left( {\left. {\bf{r}} \right|{\bf{r}}'} \right)} & {{\cal G}_{\omega m} \left( {\left. {\bf{r}} \right|{\bf{r}}'} \right)}  \\
\end{pmatrix} .
\end{equation}
which coincides with Eq.(\ref{KerQQQQ}) of the main text and which, in view of Eqs.(\ref{GoeGom}), has the suggestive explicit expression
\begin{eqnarray}
&&  \begin{pmatrix}
   {{\cal Q}_{\omega ee} \left( {\left. {\bf{r}} \right|{\bf{r}}'} \right)} & {{\cal Q}_{\omega em} \left( {\left. {\bf{r}} \right|{\bf{r}}'} \right)}  \\
   {{\cal Q}_{\omega me} \left( {\left. {\bf{r}} \right|{\bf{r}}'} \right)} & {{\cal Q}_{\omega mm} \left( {\left. {\bf{r}} \right|{\bf{r}}'} \right)}  \\
\end{pmatrix} = -\delta \left( {{\bf{r}} - {\bf{r}}'} \right)
\begin{pmatrix}
   {{\cal I}} & 0  \\
   0 & {{\cal I}}  \\
\end{pmatrix} \nonumber \\ 
&&  + 2i \begin{pmatrix}
   {- \sqrt {{\mathop{\rm Im}\nolimits} \left[ {\varepsilon _\omega  \left( {\bf{r}} \right)} \right]} \left[ {k_\omega ^2 {\cal G}_\omega  \left( {\left. {\bf{r}} \right|{\bf{r}}'} \right)} \right]\sqrt {{\mathop{\rm Im}\nolimits} \left[ {\varepsilon _\omega  \left( {{\bf{r}}'} \right)} \right]} } & {  \sqrt {{\mathop{\rm Im}\nolimits} \left[ {\varepsilon _\omega  \left( {\bf{r}} \right)} \right]} \left[ {k_\omega  {\cal G}_\omega  \left( {\left. {\bf{r}} \right|{\bf{r}}'} \right) \times \mathord{\buildrel{\lower3pt\hbox{$\scriptscriptstyle\leftarrow$}} 
\over \nabla } _{{\bf{r}}'} } \right]\sqrt {{\mathop{\rm Im}\nolimits} \left[ {\frac{{ - 1}}{{\mu _\omega  \left( {{\bf{r}}'} \right)}}} \right]} }  \\
   {-\sqrt {{\mathop{\rm Im}\nolimits} \left[ {\frac{{ - 1}}{{\mu _\omega  \left( {\bf{r}} \right)}}} \right]} \left[ {k_\omega  \nabla _{\bf{r}}  \times {\cal G}_\omega  \left( {\left. {\bf{r}} \right|{\bf{r}}'} \right)} \right]\sqrt {{\mathop{\rm Im}\nolimits} \left[ {\varepsilon _\omega  \left( {{\bf{r}}'} \right)} \right]} } & {  \sqrt {{\mathop{\rm Im}\nolimits} \left[ {\frac{{ - 1}}{{\mu _\omega  \left( {\bf{r}} \right)}}} \right]} \left[ {\nabla _{\bf{r}}  \times {\cal G}_\omega  \left( {\left. {\bf{r}} \right|{\bf{r}}'} \right) \times \mathord{\buildrel{\lower3pt\hbox{$\scriptscriptstyle\leftarrow$}} 
\over \nabla } _{{\bf{r}}'} } \right]\sqrt {{\mathop{\rm Im}\nolimits} \left[ {\frac{{ - 1}}{{\mu _\omega  \left( {{\bf{r}}'} \right)}}} \right]} }  \\
\end{pmatrix}. \nonumber \\ 
\end{eqnarray}

The fundamental relation in Eq.(\ref{OpeFunGo}) for the Green's integral operator has other important consequences which are essential for our purposes and which we now derive. Such consequences arise in examining the far-field behavior of Eq.(\ref{OpeFunGo}) so that our analysis starts from its dyadic counterpart
\begin{eqnarray} \label{FunAsy1}
&& \int {do_{\bf{m}} } {\cal W}_\omega ^T \left( {\left. {\bf{m}} \right|r{\bf{n}}} \right) \cdot {\cal W}_\omega ^* \left( {\left. {\bf{m}} \right|r'{\bf{n}}'} \right) \nonumber \\ 
&& + \frac{1}{{4k_\omega ^2 }}\int {d^3 {\bf{s}}d^3 {\bf{s}}'} {\cal G}_\omega  \left( {r{\bf{n}}\left| {\bf{s}} \right.} \right) \cdot \left[\kern-0.15em\left[ {{\sf \Gamma} _{\omega le} {\sf \Gamma} _{\omega le}^ +   + {\sf \Gamma} _{\omega lm} {\sf \Gamma} _{\omega lm}^ +  } 
 \right]\kern-0.15em\right]\left( {\left. {\bf{s}} \right|{\bf{s}}'} \right) \cdot {\cal G}_\omega ^{T*} \left( {\left. {r'{\bf{n}}'} \right|{\bf{s}}'} \right) = \frac{1}{{2ik_\omega  }}\left[ {{\cal G}_\omega  \left( {\left. {r{\bf{n}}} \right|r'{\bf{n}}'} \right) - {\cal G}_\omega ^ *  \left( {\left. {r{\bf{n}}} \right|r'{\bf{n}}'} \right)} \right], \nonumber \\ 
\end{eqnarray}
where we have set ${\bf{r}} = r{\bf{n}}$ and ${\bf{r}'} = r'{\bf{n}}'$ for the observation and source point, respectively, and $\bf n$ and ${\bf n}'$ are unit vectors. The idea is that of taking the limit $r \rightarrow +\infty$ of Eq.(\ref{FunAsy1}) and, in order to do so, we preliminary note that the basic Eq.(\ref{FwolProp}) can be written as ${\cal W}_\omega  \left( {{\bf{m}}\left| r{\bf{n}} \right.} \right) = \sqrt {\frac{{\pi \varepsilon _0 }}{{\hbar k_\omega ^3 }}} {\cal F}_{\omega s}^T \left( {\left. r{\bf{n}} \right| - {\bf{m}}} \right)$, a representation of the asympotic amplitude ${\cal W}_\omega$ in terms of the modal dyadic ${\cal F}_{\omega s}$ which is here very useful since its  combination with the fourth of Eq.(\ref{ESSBouVal}) directly yields the asymptotic behavior
\begin{equation}
{\cal W}_\omega  \left( {{\bf{m}}\left| {r{\bf{n}}} \right.} \right)\mathop  \approx \limits_{r \to \infty } \frac{1}{{4\pi }}\left[ {e^{i\left( {k_\omega  r} \right){\bf{n}} \cdot \left( { - {\bf{m}}} \right)} {\cal I}_{\bf{m}}  + \frac{{e^{ik_\omega  r} }}{r}{\cal S}_\omega ^T \left( {{\bf{n}}\left| { - {\bf{m}}} \right.} \right)} \right]
\end{equation}
which, by using Eq.(\ref{DistJonLem}) to handle the plane wave and the first of Eqs.(\ref{TwAweAwm}) to incorporate the transmission dyadic, turns into

\begin{equation} \label{WWWasy}
{\cal W}_\omega  \left( {{\bf{m}}\left| {r{\bf{n}}} \right.} \right)\mathop  \approx \limits_{r \to \infty } \frac{1}{{2k_\omega  }}\left[ {\frac{{e^{ - ik_\omega  r} }}{{ - ir}}\delta \left( {o_{\bf{n}}  - o_{\bf{m}} } \right){\cal I}_{\bf{m}}  + \frac{{e^{ik_\omega  r} }}{{ir}}{\cal T}_{\omega ss}^T \left( {\left. {\bf{n}} \right| - {\bf{m}}} \right)} \right].
\end{equation}
Now using Eq.(\ref{WWWasy}) and the fourth on Eq.(\ref{GreBouVal}), it is easily seen that in the leading order term of the asymptotic expansion of Eq.(\ref{FunAsy1}) the ingoing spherical wave $e^{ - ik_\omega  r} /r$ contribution vanishes so that, after multiplying both of its sides by $re^{ - ik_\omega  r}$ to remove the outgoing spherical wave $e^{ - ik_\omega  r} /r$ carrier and taking the limit $r \rightarrow +\infty$, we eventually get
\begin{eqnarray} \label{TWWGW}
&& \int {do_{\bf{m}} } {\cal T}_{\omega ss} \left( {\left. {\bf{n}} \right| - {\bf{m}}} \right) \cdot {\cal W}_\omega ^* \left( {\left. {\bf{m}} \right|r'{\bf{n}}'} \right) \nonumber\\ 
&&  + \frac{i}{{2k_\omega  }}\int {d^3 {\bf{s}}d^3 {\bf{s}}'} {\cal W}_\omega  \left( {\left. {\bf{n}} \right|{\bf{s}}} \right) \cdot \left[\kern-0.15em\left[ {{\sf \Gamma} _{\omega le} {\sf \Gamma} _{\omega le}^ +   + {\sf \Gamma} _{\omega lm} {\sf \Gamma} _{\omega lm}^ +  } 
 \right]\kern-0.15em\right]\left( {\left. {\bf{s}} \right|{\bf{s}}'} \right) \cdot {\cal G}_\omega ^{T*} \left( {\left. {r'{\bf{n}}'} \right|{\bf{s}}'} \right) = {\cal W}_\omega  \left( {\left. {\bf{n}} \right|r'{\bf{n}}'} \right).
\end{eqnarray}
We also need to take the limit $r' \rightarrow +\infty$ of this equation and this can again be done by using  Eq.(\ref{WWWasy}) and the fourth on Eq.(\ref{GreBouVal}) (with suitable variables relabelling) so that, after some algebra we get the asymptotic expansion of Eq.(\ref{TWWGW})
\begin{eqnarray} \label{TTWW1}
&& \frac{{e^{ - ik_\omega  r'} }}{{ - ir'}}\left[ {\int {do_{\bf{m}} } {\cal T}_{\omega ss} \left( {\left. {\bf{n}} \right|  {\bf{m}}} \right) \cdot {\cal T}_{\omega ss}^{T*} \left( {\left. {{\bf{n}}'} \right|  {\bf{m}}} \right) + \int {d^3 {\bf{s}}d^3 {\bf{s}}'} {\cal W}_\omega  \left( {\left. {\bf{n}} \right|{\bf{s}}} \right) \cdot \left[\kern-0.15em\left[ {{\sf \Gamma} _{\omega le} {\sf \Gamma} _{\omega le}^ +   + {\sf \Gamma} _{\omega lm} {\sf \Gamma} _{\omega lm}^ +  } 
 \right]\kern-0.15em\right]\left( {\left. {\bf{s}} \right|{\bf{s}}'} \right) \cdot {\cal W}_\omega ^{T*} \left( {\left. {{\bf{n}}'} \right|{\bf{s}}'} \right)} \right] \nonumber \\ 
&&  + \frac{{e^{ik_\omega  r'} }}{{ir'}}{\cal T}_{\omega ss} \left( {\left. {\bf{n}} \right| - {\bf{n}}'} \right) = \frac{{e^{ - ik_\omega  r'} }}{{ - ir'}}\delta \left( {o_{{\bf{n}}'}  - o_{\bf{n}} } \right){\cal I}_{\bf{n}}  + \frac{{e^{ik_\omega  r'} }}{{ir'}}{\cal T}_{\omega ss}^T \left( {\left. {{\bf{n}}'} \right| - {\bf{n}}} \right) + O\left( {\frac{1}{{r'^2 }}} \right).
\end{eqnarray}
It is now essential using the reciprocity of the transmission matrix (see Eq.(\ref{TosRec})) which implies that ${\cal T}_\omega  \left( {\left. {\bf{n}} \right| - {\bf{n}}'} \right) = {\cal T}_{\omega ss}^T \left( {\left. {{\bf{n}}'} \right| - {\bf{n}}} \right)$ so that the outgoing spherical wave
 $e^{  ik_\omega  r'} /r'$ contribution disappear from the leading order term in Eq.(\ref{TTWW1}) and, after multiplying both of its sides by $r'e^{ ik_\omega  r'}$ to remove the ingoing spherical wave $e^{ - ik_\omega  r'} /r'$ carrier and taking the limit $r' \rightarrow +\infty$, we obtain
\begin{eqnarray} \label{TTWW2}
&& \int {do_{\bf{m}} } {\cal T}_{\omega ss} \left( {\left. {\bf{n}} \right|  {\bf{m}}} \right) \cdot {\cal T}_{\omega ss}^{T*} \left( {\left. {{\bf{n}}'} \right|  {\bf{m}}} \right) + \int {d^3 {\bf{s}}d^3 {\bf{s}}'} {\cal W}_\omega  \left( {\left. {\bf{n}} \right|{\bf{s}}} \right) \cdot \left[\kern-0.15em\left[ {{\sf \Gamma} _{\omega le} {\sf \Gamma} _{\omega le}^ +   + {\sf \Gamma} _{\omega lm} {\sf \Gamma} _{\omega lm}^ +  } 
 \right]\kern-0.15em\right]\left( {\left. {\bf{s}} \right|{\bf{s}}'} \right) \cdot {\cal W}_\omega ^{T*} \left( {\left. {{\bf{n}}'} \right|{\bf{s}}'} \right) \nonumber \\ 
&&  = \delta \left( {o_{{\bf{n}}'}  - o_{\bf{n}} } \right){\cal I}_{\bf{n}}.
\end{eqnarray}
Equations Eq.(\ref{TWWGW}) and Eq.(\ref{TTWW2}) can be conveniently written in operatorial form as
\begin{eqnarray}
&& {\sf T}_{\omega ss} {\sf J}_{ss} {\sf W}_{\omega sl}^*  + \frac{i}{{2k_\omega  }}{\sf W}_{\omega sl} 
\begin{pmatrix}
   {{\sf \Gamma} _{\omega le} } & {{\sf \Gamma} _{\omega lm} }  \\
\end{pmatrix}
\begin{pmatrix}
   {{\sf \Gamma} _{\omega le}^ +  }  \\
   {{\sf \Gamma} _{\omega lm}^ +  }  \\
\end{pmatrix}
{\sf G}_{\omega ll}^ +   = {\sf W}_{\omega sl}, \nonumber \\ 
&& {\sf T}_{\omega ss} {\sf T}_{\omega ss}^ +   + {\sf W}_{\omega sl} \begin{pmatrix}
   {{\sf \Gamma} _{\omega le} } & {{\sf \Gamma} _{\omega lm} }  \\
\end{pmatrix}
\begin{pmatrix}
   {{\sf \Gamma} _{\omega le}^ +  }  \\
   {{\sf \Gamma} _{\omega lm}^ +  }  \\
\end{pmatrix}
{\sf W}_{\omega sl}^ +   = {\sf I}_{ss} 
\end{eqnarray}
so that, after right multiplying the first of these equations by the row $\begin{pmatrix} {{\sf \Gamma} _{\omega le} } & {{\sf \Gamma} _{\omega lm} } \end{pmatrix}$, using the defintion of the operator matrix in Eq.(\ref{OpeQ}) and the right column of Eqs.(\ref{GoeGomAoseAosm}), we get
\begin{eqnarray} \label{final0}
&& {\sf T}_{\omega ss} {\sf J}_{ss} 
\begin{pmatrix}
   {{\sf E}_{\omega se}^* } & {{\sf E}_{\omega sm}^* }  \\
\end{pmatrix} 
+ \begin{pmatrix}
   {{\sf E}_{\omega se} } & {{\sf E}_{\omega sm} }  \\
\end{pmatrix}
\begin{pmatrix}
   {{\sf Q}_{\omega ee}^ +  } & {{\sf Q}_{\omega me}^ +  }  \\
   {{\sf Q}_{\omega em}^ +  } & {{\sf Q}_{\omega mm}^ +  }  \\
\end{pmatrix} = 
\begin{pmatrix}
   0 & 0  \\
\end{pmatrix}
, \nonumber \\ 
&& {\sf T}_{\omega ss} {\sf T}_{\omega ss}^ +  
+ \begin{pmatrix}
   {{\sf E}_{\omega se} } & {{\sf E}_{\omega sm} }  \\
\end{pmatrix}
\begin{pmatrix}
   {{\sf E}_{\omega se}^ +  }  \\
   {{\sf E}_{\omega sm}^ +  }  \\
\end{pmatrix} = {\sf I}_{ss}  
\end{eqnarray}
Putting all together, Eq.(\ref{AAQQII}) and Eqs.(\ref{final0}) are easily seen to be equivalent to the matrix relation in Eq.(\ref{RelTAG}) we here rewrite for convenience
\begin{equation} \label{final1}
\begin{pmatrix}
   {{\sf T}_{\omega ss} } & {{\sf E}_{\omega se} } & {{\sf E}_{\omega sm} }  \\
   {{\sf E}_{\omega se}^{ +  * } {\sf J}_{ss} } & {  {\sf Q}_{\omega ee} } & {  {\sf Q}_{\omega em} }  \\
   {{\sf E}_{\omega sm}^{ +  * } {\sf J}_{ss} } & {  {\sf Q}_{\omega me} } & {  {\sf Q}_{\omega mm} }  \\
\end{pmatrix}
\begin{pmatrix}
   {{\sf T}_{\omega ss}^ +  } & {{\sf J}_{ss} {\sf E}_{\omega se}^ *  } & {{\sf J}_{ss} {\sf E}_{\omega sm}^ *  }  \\
   {{\sf E}_{\omega se}^ +  } & {  {\sf Q}_{\omega ee}^ +  } & {  {\sf Q}_{\omega me}^ +  }  \\
   {{\sf E}_{\omega sm}^ +  } & {  {\sf Q}_{\omega em}^ +  } & {  {\sf Q}_{\omega mm}^ +  }  \\
\end{pmatrix} 
= \begin{pmatrix}
   {{\sf I}_{ss} } & 0 & 0  \\
   0 & {{\sf I}_{ee} } & 0  \\
   0 & 0 & {{\sf I}_{mm} }  \\
\end{pmatrix}
\end{equation}
since the second of Eqs.(\ref{final0}) and Eq.(\ref{AAQQII}) reproduce the upper $1 \times 1$ and lower $2 \times 2$ submatrices on the diagonal of Eq.(\ref{final1}), respectively, whereas the first of Eqs.(\ref{final0}) and its Hermitian conjugate 
\begin{equation} \label{final2}
\begin{pmatrix}
   {{\sf E}_{\omega se}^{ +  * } }  \\
   {{\sf E}_{\omega sm}^{ +  * } }  \\
\end{pmatrix}
{\sf J}_{ss} {\sf T}_{\omega ss}^ +   
+ \begin{pmatrix}
   {{\sf Q}_{\omega ee} } & {{\sf Q}_{\omega em} }  \\
   {{\sf Q}_{\omega me} } & {{\sf Q}_{\omega mm} }  \\
\end{pmatrix}
\begin{pmatrix}
   {{\sf E}_{\omega se}^ +  }  \\
   {{\sf E}_{\omega sm}^ +  }  \\
\end{pmatrix}
 = \begin{pmatrix}
   0  \\
   0  \\
\end{pmatrix}
\end{equation}
coincide with the upper right $1 \times 2$ and  lower left $2 \times 1$ submatrices of Eq.(\ref{final1}).

We now multiply in reverse order the two operator matrices in the left hand side of Eq.(\ref{final1}) and, after a short manipulation, we get
\begin{eqnarray} \label{unita0}
&& \begin{pmatrix}
   {{\sf T}_{\omega ss}^ +  } & {{\sf J}_{ss} {\sf E}_{\omega se}^ *  } & {{\sf J}_{ss} {\sf E}_{\omega sm}^ *  }  \\
   {{\sf E}_{\omega se}^ +  } & { {\sf Q}_{\omega ee}^ +  } & { {\sf Q}_{\omega me}^ +  }  \\
   {{\sf E}_{\omega sm}^ +  } & { {\sf Q}_{\omega em}^ +  } & { {\sf Q}_{\omega mm}^ +  }  \\
\end{pmatrix}\begin{pmatrix}
   {{\sf T}_{\omega ss} } & {{\sf E}_{\omega se} } & {{\sf E}_{\omega sm} }  \\
   {{\sf E}_{\omega se}^{ +  * } {\sf J}_{ss} } & { {\sf Q}_{\omega ee} } & { {\sf Q}_{\omega em} }  \\
   {{\sf E}_{\omega sm}^{ +  * } {\sf J}_{ss} } & { {\sf Q}_{\omega me} } & { {\sf Q}_{\omega mm} } \nonumber \\
\end{pmatrix} \\ 
&&  = \begin{pmatrix}
   {{\sf J}_{ss} } & 0 & 0  \\
   0 & 1 & 0  \\
   0 & 0 & 1  \\
\end{pmatrix}\begin{pmatrix}
   {{\sf J}_{ss} {\sf T}_{\omega ss}^ +  {\sf J}_{ss} } & {{\sf E}_{\omega se}^ *  } & {{\sf E}_{\omega sm}^ *  }  \\
   {{\sf E}_{\omega se}^ +  {\sf J}_{ss} } & { {\sf Q}_{\omega ee}^ +  } & { {\sf Q}_{\omega me}^ +  }  \\
   {{\sf E}_{\omega sm}^ +  {\sf J}_{ss} } & { {\sf Q}_{\omega em}^ +  } & { {\sf Q}_{\omega mm}^ +  }  \\
\end{pmatrix}\begin{pmatrix}
   {{\sf J}_{ss} {\sf T}_{\omega ss} {\sf J}_{ss} } & {{\sf J}_{ss} {\sf E}_{\omega se} } & {{\sf J}_{ss} {\sf E}_{\omega sm} }  \\
   {{\sf E}_{\omega se}^{ +  * } } & { {\sf Q}_{\omega ee} } & { {\sf Q}_{\omega em} }  \\
   {{\sf E}_{\omega sm}^{ +  * } } & { {\sf Q}_{\omega me} } & { {\sf Q}_{\omega mm} }  \\
\end{pmatrix}\begin{pmatrix}
   {{\sf J}_{ss} } & 0 & 0  \\
   0 & 1 & 0  \\
   0 & 0 & 1  \\
\end{pmatrix} ,
\end{eqnarray}
as a consequence of the involutive property ${\sf J}_{ss} {\sf J}_{ss}  = {\sf I}_{ss}$. From the second of Eqs.(\ref{OpeRecip}) and from Eq.(\ref{Q+Q*}) we get the relations
\begin{eqnarray}
&& {\sf J}_{ss} {\sf T}_{\omega ss}^ +  {\sf J}_{ss}  = {\sf T}_{\omega ss}^* ,\quad \begin{pmatrix}
   {{\sf Q}_{\omega ee}^ +  } & {{\sf Q}_{\omega me}^ +  }  \\
   {{\sf Q}_{\omega em}^ +  } & {{\sf Q}_{\omega mm}^ +  }  \\
\end{pmatrix} = \begin{pmatrix}
   {{\sf Q}_{\omega ee}^* } & {{\sf Q}_{\omega em}^* }  \\
   {{\sf Q}_{\omega me}^* } & {{\sf Q}_{\omega mm}^* }  \\
\end{pmatrix}, \nonumber \\ 
&& {\sf J}_{ss} {\sf T}_{\omega ss} {\sf J}_{ss}  = {\sf T}_{\omega ss}^{ + *} ,\quad \begin{pmatrix}
   {{\sf Q}_{\omega ee} } & {{\sf Q}_{\omega em} }  \\
   {{\sf Q}_{\omega me} } & {{\sf Q}_{\omega mm} }  \\
\end{pmatrix} = \begin{pmatrix}
   {{\sf Q}_{\omega ee}^{ + *} } & {{\sf Q}_{\omega me}^{ + *} }  \\
   {{\sf Q}_{\omega em}^{ + *} } & {{\sf Q}_{\omega mm}^{ + *} }  \\
\end{pmatrix}, 
 \end{eqnarray}
which can be used to rewrite the matrix elements $(1,1)$ and $(2,2)$, $(2,3)$, $(3,2)$ and $(3,3)$ of the inner left and right matrices in right hand side of Eq.(\ref{unita0}), thus getting
\begin{eqnarray} \label{unita1}
&& \begin{pmatrix}
   {{\sf T}_{\omega ss}^ +  } & {{\sf J}_{ss} {\sf E}_{\omega se}^ *  } & {{\sf J}_{ss} {\sf E}_{\omega sm}^ *  }  \\
   {{\sf E}_{\omega se}^ +  } & { {\sf Q}_{\omega ee}^ +  } & { {\sf Q}_{\omega me}^ +  }  \\
   {{\sf E}_{\omega sm}^ +  } & { {\sf Q}_{\omega em}^ +  } & { {\sf Q}_{\omega mm}^ +  }  \\
\end{pmatrix}\begin{pmatrix}
   {{\sf T}_{\omega ss} } & {{\sf E}_{\omega se} } & {{\sf E}_{\omega sm} }  \\
   {{\sf E}_{\omega se}^{ +  * } {\sf J}_{ss} } & { {\sf Q}_{\omega ee} } & { {\sf Q}_{\omega em} }  \\
   {{\sf E}_{\omega sm}^{ +  * } {\sf J}_{ss} } & { {\sf Q}_{\omega me} } & { {\sf Q}_{\omega mm} }  \\
\end{pmatrix} \nonumber \\ 
&&  = \begin{pmatrix}
   {{\sf J}_{ss} } & 0 & 0  \\
   0 & 1 & 0  \\
   0 & 0 & 1  \\
\end{pmatrix}\begin{pmatrix} 
   {{\sf T}_{\omega ss}^* } & {{\sf E}_{\omega se}^ *  } & {{\sf E}_{\omega sm}^ *  }  \\
   {{\sf E}_{\omega se}^ +  {\sf J}_{ss} } & { {\sf Q}_{\omega ee}^* } & { {\sf Q}_{\omega em}^* }  \\
   {{\sf E}_{\omega sm}^ +  {\sf J}_{ss} } & { {\sf Q}_{\omega me}^* } & { {\sf Q}_{\omega mm}^* }  \\
\end{pmatrix}\begin{pmatrix}
   {{\sf T}_{\omega ss}^{ + *} } & {{\sf J}_{ss} {\sf E}_{\omega se} } & {{\sf J}_{ss} {\sf E}_{\omega sm} }  \\
   {{\sf E}_{\omega se}^{ +  * } } & { {\sf Q}_{\omega ee}^{ + *} } & { {\sf Q}_{\omega me}^{ + *} }  \\
   {{\sf E}_{\omega sm}^{ +  * } } & { {\sf Q}_{\omega em}^{ + *} } & { {\sf Q}_{\omega mm}^* }  \\
\end{pmatrix}\begin{pmatrix}
   {{\sf J}_{ss} } & 0 & 0  \\
   0 & 1 & 0  \\
   0 & 0 & 1  \\
\end{pmatrix}.
\end{eqnarray}
Now the product of the two inner matrices in right hand side of Eq.(\ref{unita1}) coincides with the complex conjugate of the left hand side of Eq.(\ref{final1}) so that, in view of the relation ${\sf J}_{ss} {\sf J}_{ss}  = {\sf I}_{ss}$, we straighforwardly obtain
\begin{equation}
\begin{pmatrix}
   {{\sf T}_{\omega ss}^ +  } & {{\sf J}_{ss} {\sf E}_{\omega se}^ *  } & {{\sf J}_{ss} {\sf E}_{\omega sm}^ *  }  \\
   {{\sf E}_{\omega se}^ +  } & { {\sf Q}_{\omega ee}^ +  } & { {\sf Q}_{\omega me}^ +  }  \\
   {{\sf E}_{\omega sm}^ +  } & { {\sf Q}_{\omega em}^ +  } & { {\sf Q}_{\omega mm}^ +  }  \\
\end{pmatrix}\begin{pmatrix}
   {{\sf T}_{\omega ss} } & {{\sf E}_{\omega se} } & {{\sf E}_{\omega sm} }  \\
   {{\sf E}_{\omega se}^{ +  * } {\sf J}_{ss} } & { {\sf Q}_{\omega ee} } & { {\sf Q}_{\omega em} }  \\
   {{\sf E}_{\omega sm}^{ +  * } {\sf J}_{ss} } & { {\sf Q}_{\omega me} } & { {\sf Q}_{\omega mm} }  \\
\end{pmatrix} = \begin{pmatrix}
   {{\sf I}_{ss} } & 0 & 0  \\
   0 & {{\sf I}_{ee} } & 0  \\
   0 & 0 & {{\sf I}_{mm} }  \\
\end{pmatrix}.
\end{equation}

\section{Change of basis matrix}
In order to derive the change of basis matrix in Eq.(\ref{ChaBasMat}), we start by expressing the ingoing polariton operators $\hat a_\mu^\dagger  \left( {\xi _\mu  } \right)$ in terms of the outgoing ones $\hat A_\mu^\dagger  \left( {\Xi _\mu  } \right)$. To this end we consider the inverse input-output relation in Eq.(\ref{InversInOutRel}) written for creation operators in explicit form as
\begin{eqnarray}
&& {\bf{\hat g}}_{\omega s}^\dag  \left( {\bf{n}} \right) = \int {do_{\bf{N}} } {\cal T}_{\omega ss}^T \left( {\left. {\bf{N}} \right|{\bf{n}}} \right) \cdot {\bf{\hat G}}_{\omega s}^\dag  \left( {\bf{N}} \right) + \int {d^3 {\bf{R}}} \left[{{\cal A}_{\omega es}^T \left( {{\bf{R}}\left| {\bf{n}} \right.} \right)  \cdot {\bf{\hat F}}_{\omega e}^\dag  \left( {\bf{R}} \right) + {\cal A}_{\omega ms}^T \left( {{\bf{R}}\left| {\bf{n}} \right.} \right) \cdot {\bf{\hat F}}_{\omega m}^\dag  \left( {\bf{R}} \right)} \right], \nonumber \\ 
&& {\bf{\hat f}}_{\omega e}^\dag  \left( {\bf{r}} \right) = \int {do_{\bf{N}} } {\cal E}_{\omega se}^T \left( {\left. {\bf{N}} \right|{\bf{r}}} \right) \cdot {\bf{\hat G}}_{\omega s}^\dag  \left( {\bf{N}} \right) + \int {d^3 {\bf{R}}} \left[ {{\cal Q}_{\omega ee}^T \left( {{\bf{R}}\left| {\bf{r}} \right.} \right) \cdot {\bf{\hat F}}_{\omega e}^\dag  \left( {\bf{R}} \right) + {\cal Q}_{\omega me}^T \left( {{\bf{R}}\left| {\bf{r}} \right.} \right) \cdot {\bf{\hat F}}_{\omega m}^\dag  \left( {\bf{R}} \right)} \right], \nonumber \\ 
&& {\bf{\hat f}}_{\omega m}^\dag  \left( {\bf{r}} \right) = \int {do_{\bf{N}} } {\cal E}_{\omega sm}^T \left( {\left. {\bf{N}} \right|{\bf{r}}} \right) \cdot {\bf{\hat G}}_{\omega s}^\dag  \left( {\bf{N}} \right) + \int {d^3 {\bf{R}}} \left[ {{\cal Q}_{\omega em}^T \left( {{\bf{R}}\left| {\bf{r}} \right.} \right) \cdot {\bf{\hat F}}_{\omega e}^\dag  \left( {\bf{R}} \right) + {\cal Q}_{\omega mm}^T \left( {{\bf{R}}\left| {\bf{r}} \right.} \right) \cdot {\bf{\hat F}}_{\omega m}^\dag  \left( {\bf{R}} \right)} \right],
\end{eqnarray}
from which, by using Eq.(\ref{Opegosfos}) and (\ref{OpeGosFos}) and setting 
\begin{eqnarray}
&& \xi _s  = \left( {\omega ,{\bf{n}},\lambda } \right),\quad \xi _e  = \xi _m  = \left( {\omega ,{\bf{r}},\lambda } \right), \nonumber \\
&& \Xi _s  = \left( {\Omega ,{\bf{N}},\Lambda } \right),\quad \Xi _e  = \Xi _m  = \left( {\Omega ,{\bf{R}},\Lambda } \right),
\end{eqnarray}
we readily get, in the notation introduced at the beginning of Sec.V,
\begin{equation} \label{amuAph}
\hat a_\mu^\dag  \left( {\xi _\mu  } \right) = \sum\limits_\tau  {\int {d\Xi _\tau  } Z_{\tau \mu } \left( {\Xi _\tau  \left| {\xi _\mu  } \right.} \right)\hat A_\tau ^\dag  \left( {\Xi _\tau  } \right)},
\end{equation}
(note the reverse order of the index $\mu$ and $\tau$) where the sum is evidently over $\tau  = s,e,m$ and the $3 \times 3$ matrix $Z_{\tau \mu } \left( {\Xi _\tau  \left| {\xi _\mu  } \right.} \right)$ is given by
\begin{eqnarray}
&& \begin{pmatrix}
   {Z_{ss} \left( {\Xi _s \left| {\xi _s } \right.} \right)} & {Z_{es} \left( {\Xi _e \left| {\xi _s } \right.} \right)} & {Z_{ms} \left( {\Xi _m \left| {\xi _s } \right.} \right)}  \\
   {Z_{se} \left( {\Xi _s \left| {\xi _e } \right.} \right)} & {Z_{ee} \left( {\Xi _e \left| {\xi _e } \right.} \right)} & {Z_{me} \left( {\Xi _m \left| {\xi _e } \right.} \right)}  \\
   {Z_{sm} \left( {\Xi _s \left| {\xi _m } \right.} \right)} & {Z_{em} \left( {\Xi _e \left| {\xi _m } \right.} \right)} & {Z_{mm} \left( {\Xi _m \left| {\xi _m } \right.} \right)}  \\
\end{pmatrix} \nonumber \\ 
&&  = \delta \left( {\Omega  - \omega } \right)\begin{pmatrix}
   {{\bf{e}}_{{\bf{N}}\Lambda }  \cdot {\cal T}_{\omega ss} \left( {\left. {\bf{N}} \right|{\bf{n}}} \right) \cdot {\bf{e}}_{{\bf{n}}\lambda } } & {{\bf{u}}_\Lambda   \cdot {\cal A}_{\omega es} \left( {{\bf{R}}\left| {\bf{n}} \right.} \right) \cdot {\bf{e}}_{{\bf{n}}\lambda } } &  {{\bf{u}}_\Lambda   \cdot {\cal A}_{\omega ms} \left( {{\bf{R}}\left| {\bf{n}} \right.} \right) \cdot {\bf{e}}_{{\bf{n}}\lambda } } \\
   {{\bf{e}}_{{\bf{N}}\Lambda }  \cdot {\cal E}_{\omega se} \left( {\left. {\bf{N}} \right|{\bf{r}}} \right) \cdot {\bf{u}}_\lambda  }
 & {{\bf{u}}_\Lambda   \cdot {\cal Q}_{\omega ee} \left( {{\bf{R}}\left| {\bf{r}} \right.} \right) \cdot {\bf{u}}_\lambda  } & {{\bf{u}}_\Lambda   \cdot {\cal Q}_{\omega me} \left( {{\bf{R}}\left| {\bf{r}} \right.} \right) \cdot {\bf{u}}_\lambda  }  \\
  {{\bf{e}}_{{\bf{N}}\Lambda }  \cdot {\cal E}_{\omega sm} \left( {\left. {\bf{N}} \right|{\bf{r}}} \right) \cdot {\bf{u}}_\lambda  } & {{\bf{u}}_\Lambda   \cdot {\cal Q}_{\omega em} \left( {{\bf{R}}\left| {\bf{r}} \right.} \right) \cdot {\bf{u}}_\lambda  } & {{\bf{u}}_\Lambda   \cdot {\cal Q}_{\omega mm} \left( {{\bf{R}}\left| {\bf{r}} \right.} \right) \cdot {\bf{u}}_\lambda  }  \\
\end{pmatrix}.
\end{eqnarray}
Equation (\ref{amuAph}) represents a very compact statement of the input-output relation whose unitarity reflects into the unitarity of the matrix $Z_{\tau \mu } \left( {\Xi _\tau  \left| {\xi _\mu  } \right.} \right)$, in turn vividly expressed by the relations
\begin{eqnarray}
&& \sum\limits_\tau  {\int {d\Xi _\tau  } } Z_{\tau \mu }^* \left( {\Xi _\tau  \left| {\xi _\mu  } \right.} \right)Z_{\tau \mu '} \left( {\Xi _\tau  \left| {\xi '_{\mu '} } \right.} \right) = \delta _{\mu \mu '} \delta _\mu  \left( {\xi _\mu   - \xi '_{\mu '} } \right), \nonumber \\ 
&& \sum\limits_\mu  {\int {d\xi _\mu  } } Z_{\tau \mu } \left( {\Xi _\tau  \left| {\xi _\mu  } \right.} \right)Z_{\tau '\mu }^* \left( {\Xi '_{\tau '} \left| {\xi _\mu  } \right.} \right) = \delta _{\tau \tau '} \delta _\tau  \left( {\Xi _\tau   - \Xi '_{\tau '} } \right),
\end{eqnarray}
which are easily derived from Eqs.(\ref{RelTAG}) and (\ref{RelTAGConj}). 

We are now prepared for the evaluation of the change of basis matrix which, after using Eq.(\ref{IngBas}), we write in the notationally convenient form
\begin{equation} \label{matr1}
\left\langle {{\Xi _s^{\left( P \right)} \Xi _e^{\left( Q \right)} \Xi _m^{\left( R \right)} }}
 \mathrel{\left | {\vphantom {{\Xi _s^{\left( P \right)} \Xi _e^{\left( Q \right)} \Xi _m^{\left( R \right)} } {\xi _s^{\left( p \right)} \xi _e^{\left( q \right)} \xi _m^{\left( r \right)} }}}
 \right. \kern-\nulldelimiterspace}
 {{\xi _s^{\left( p \right)} \xi _e^{\left( q \right)} \xi _m^{\left( r \right)} }} \right\rangle  = \frac{1}{{\sqrt {p!q!r!} }}\left\langle {\Xi _s^{\left( P \right)} \Xi _e^{\left( Q \right)} \Xi _m^{\left( R \right)} } \right|\prod\limits_{j = 1}^n {\hat a_{\chi \left( {\bar \xi ^i } \right)}^\dag  \left( {\bar \xi ^i } \right)\left| 0 \right\rangle } ,
\end{equation}
where $n = p+q+r$ is the total number of ingoing polaritons, 
\begin{equation}
\left( {\bar \xi ^1  \ldots \bar \xi ^n } \right) = \left( {\xi _s^1 , \ldots ,\xi _s^p ,\xi _e^1 , \ldots ,\xi _e^q ,\xi _m^1 , \ldots ,\xi _m^r } \right)
\end{equation}
is the $n$-tuple containing all ingoing polariton variables and $\chi \left( {\xi _\mu  } \right) = \mu$ detones the $\mu=s,e,m$ character of the variable ${\xi _\mu  }$. From Eq.(\ref{amuAph}) we get the creation operator for the $i$-th ingoing polariton variable
\begin{equation}
\hat a_{\chi \left( {\bar \xi ^i } \right)}^\dag  \left( {\bar \xi ^i } \right) = \sum\limits_{\tau _i } {\int {d\tilde \Xi _{\tau _i }^i } Z_{\tau _i \chi \left( {\bar \xi ^i } \right)} \left( {\tilde \Xi _{\tau _i }^i \left| {\bar \xi ^i } \right.} \right)\hat A_{\tau _i }^\dag  \left( {\tilde \Xi _{\tau _i }^i } \right)} 
\end{equation}
which inserted into Eq.(\ref{matr1}) yields
\begin{eqnarray}  \label{matr2}
&& \left\langle {{\Xi _s^{\left( P \right)} \Xi _e^{\left( Q \right)} \Xi _m^{\left( R \right)} }}
 \mathrel{\left | {\vphantom {{\Xi _s^{\left( P \right)} \Xi _e^{\left( Q \right)} \Xi _m^{\left( R \right)} } {\xi _s^{\left( p \right)} \xi _e^{\left( q \right)} \xi _m^{\left( r \right)} }}}
 \right. \kern-\nulldelimiterspace}
 {{\xi _s^{\left( p \right)} \xi _e^{\left( q \right)} \xi _m^{\left( r \right)} }} \right\rangle  = \frac{1}{{\sqrt {p!q!r!} }}\sum\limits_{\left( {\tau _1  \ldots \tau _n } \right)} {\int {d\tilde \Xi _{\tau _1 }^1  \ldots d\tilde \Xi _{\tau _n }^n } \left[ {\prod\limits_{i = 1}^n {Z_{\tau _i \chi \left( {\bar \xi ^i } \right)} \left( {\tilde \Xi _{\tau _i }^i \left| {\bar \xi ^i } \right.} \right)} } \right]} \nonumber \\ 
&& \cdot \left\langle {\Xi _s^{\left( P \right)} \Xi _e^{\left( Q \right)} \Xi _m^{\left( R \right)} } \right|\prod\limits_{i = 1}^n {\hat A_{\tau _i }^\dag  \left( {\tilde \Xi _{\tau _i }^i } \right)\left| 0 \right\rangle } .
\end{eqnarray}
As a consequence of Eq.(\ref{OutgBas}), it is evident that the only surviving matrix elements in the right hand side of Eq.(\ref{matr2}) are those ones with $n = N$, where $N = P+Q+R$ is the total number of outgoing polaritons, and such that the $n$-string $\left( {\tau _1  \ldots \tau _n } \right)$ precisely contains $P$ times the $s$ character, $Q$ times the $e$ character and $R$ times the $m$ character, so that Eq.(\ref{matr2}) can be written as
\begin{eqnarray}  \label{matr3}
&& \left\langle {{\Xi _s^{\left( P \right)} \Xi _e^{\left( Q \right)} \Xi _m^{\left( R \right)} }}
 \mathrel{\left | {\vphantom {{\Xi _s^{\left( P \right)} \Xi _e^{\left( Q \right)} \Xi _m^{\left( R \right)} } {\xi _s^{\left( p \right)} \xi _e^{\left( q \right)} \xi _m^{\left( r \right)} }}}
 \right. \kern-\nulldelimiterspace}
 {{\xi _s^{\left( p \right)} \xi _e^{\left( q \right)} \xi _m^{\left( r \right)} }} \right\rangle  = \frac{{\delta _{N,n} }}{{\sqrt {p!q!r!} }}\sum\limits_{\scriptstyle \left( {\tau _1  \ldots \tau _N } \right) \hfill \atop 
  \scriptstyle Ps,Qe,Rm \hfill} {\int {d\tilde \Xi _{\tau _1 }^1  \ldots d\tilde \Xi _{\tau _N }^N } \left[ {\prod\limits_{i = 1}^N {Z_{\tau _i \chi \left( {\bar \xi ^i } \right)} \left( {\tilde \Xi _{\tau _i }^i \left| {\bar \xi ^i } \right.} \right)} } \right]} \nonumber \\ 
&& \cdot \left\langle {\Xi _s^{\left( P \right)} \Xi _e^{\left( Q \right)} \Xi _m^{\left( R \right)} } \right|\prod\limits_{i = 1}^N {\hat A_{\tau _i }^\dag  \left( {\tilde \Xi _{\tau _i }^i } \right)\left| 0 \right\rangle }.
\end{eqnarray}
Now a quick combinatorial analysis reveals that the set of the $N$-strings $\tau = \left( {\tau _1  \ldots \tau _N } \right)$ with $Ps,Qe,Rm$ coincides with the set, we label as $\left( {sem} \right)_{PQR}$,  of all the $N!/\left( {P!Q!R!} \right)$ distinct anagrams of the fundamental $N$-string
\begin{equation} \label{FunString}
\left( {\Theta _1  \ldots \Theta _N } \right) = (\underbrace {s \ldots s}_P\underbrace {e \ldots e}_Q\underbrace {m \ldots m}_R).
\end{equation}
For example, setting $P=1$, $Q=2$ and $R=1$, we have 
\begin{eqnarray} \label{Example1}
&& \left( {\Theta _1 \Theta _2 \Theta _3 \Theta _4 } \right) = \left( {seem} \right), \nonumber \\ 
&& \left( {sem} \right)_{121}  = \left\{ {seem,mees,emes,eems,mese,emse,msee,smee,esme,seme,eesm,esem} \right\}.
\end{eqnarray}
Note that to each $N$-string $\tau \in \left( {sem} \right)_{PQR}$ is associated a permutation $\pi _\tau   \in S_N$ such that $\tau _i  = \Theta _{\pi _\tau  \left( i \right)} $ and that such permutation is not uniquely defined since,
if $\sigma _{PQR}  \in S_N$ is any permutation that solely shuffles the elements inside each of the three groups of $P$,$Q$ and $R$ elements, the permutation $\sigma _{PQR}  \circ \pi _\tau$ is associated to $\tau$ as well, i.e.
\begin{equation} \label{ThetInv}
\Theta _{\left( {\sigma _{PQR}  \circ \pi _\tau  } \right)\left( i \right)}  = \Theta _{\pi _\tau  \left( i \right)}.
\end{equation}
For example, in the above specific situation of Eq.(\ref{Example1}), we have
\begin{eqnarray} \label{Example2}
&& \tau  = \left( {emes} \right) = \left( {\Theta _2 \Theta _4 \Theta _3 \Theta _1 } \right) \quad   \Rightarrow \quad  \pi _\tau   = \begin{pmatrix}
   {1234}  \\
   {2431}  \\
\end{pmatrix}, \nonumber \\ 
&& \sigma _{121}  = \begin{pmatrix}
   {1234}  \\
   {1324}  \\
\end{pmatrix},\quad \pi ' = \sigma _{121}  \circ \pi _\tau   = \begin{pmatrix}
   {1234}  \\
   {3421}  \\
\end{pmatrix}\quad  \Rightarrow \quad  \left( {\Theta _3 \Theta _4 \Theta _2 \Theta _1 } \right) = \left( {emes} \right) = \tau.
\end{eqnarray}
Using such combinatorial observation, Eq.(\ref{matr3}) can be written as
\begin{eqnarray} \label{matr4}
&& \left\langle {{\Xi _s^{\left( P \right)} \Xi _e^{\left( Q \right)} \Xi _m^{\left( R \right)} }}
 \mathrel{\left | {\vphantom {{\Xi _s^{\left( P \right)} \Xi _e^{\left( Q \right)} \Xi _m^{\left( R \right)} } {\xi _s^{\left( p \right)} \xi _e^{\left( q \right)} \xi _m^{\left( r \right)} }}}
 \right. \kern-\nulldelimiterspace}
 {{\xi _s^{\left( p \right)} \xi _e^{\left( q \right)} \xi _m^{\left( r \right)} }} \right\rangle  = \frac{{\delta _{N,n} }}{{\sqrt {p!q!r!} }}\sum\limits_{\tau  \in \left( {sem} \right)_{PQR} } {\int {d\tilde \Xi _{\Theta _{\pi _\tau  \left( 1 \right)} }^1  \ldots d\tilde \Xi _{\Theta _{\pi _\tau  \left( n \right)} }^N } \left[ {\prod\limits_{i = 1}^N {Z_{\Theta _{\pi _\tau  \left( i \right)} \chi \left( {\bar \xi ^i } \right)} \left( {\tilde \Xi _{\Theta _{\pi _\tau  \left( i \right)} }^i \left| {\bar \xi ^i } \right.} \right)} } \right]} \nonumber \\ 
&& \cdot \left\langle {\Xi _s^{\left( P \right)} \Xi _e^{\left( Q \right)} \Xi _m^{\left( R \right)} } \right|\prod\limits_{i = 1}^N {\hat A_{\Theta _{\pi _\tau  \left( i \right)} }^\dag  \left( {\tilde \Xi _{\Theta _{\pi _\tau  \left( i \right)} }^i } \right)\left| 0 \right\rangle }.
\end{eqnarray}
To evaluate the matrix element inside the integral in Eq.(\ref{matr4}), we perform the change of index $i = \pi _\tau ^{ - 1} \left( j \right)$ and we resort to Eq.(\ref{FunString}), thus getting
\begin{eqnarray}
&& \left\langle {\Xi _s^{\left( P \right)} \Xi _e^{\left( Q \right)} \Xi _m^{\left( R \right)} } \right|\prod\limits_{i = 1}^N {\hat A_{\Theta _{\pi _\tau  \left( i \right)} }^\dag  \left( {\tilde \Xi _{\Theta _{\pi _\tau  \left( i \right)} }^i } \right)\left| 0 \right\rangle } \nonumber \\ 
&&  = \left\langle {\Xi _s^{\left( P \right)} \Xi _e^{\left( Q \right)} \Xi _m^{\left( R \right)} } \right|\prod\limits_{I = 1}^P {\prod\limits_{J = 1}^Q {\prod\limits_{K = 1}^R {\hat A_s^\dag  \left( {\tilde \Xi _s^{\pi _\tau ^{ - 1} \left( I \right)} } \right)\hat A_e^\dag  \left( {\tilde \Xi _e^{\pi _\tau ^{ - 1} \left( {P + J} \right)} } \right)\hat A_m^\dag  \left( {\tilde \Xi _m^{\pi _\tau ^{ - 1} \left( {P + Q + K} \right)} } \right)} } } \left| 0 \right\rangle,
\end{eqnarray}
which, using the expression of the basis states in Eq.(\ref {OutgBas}) and to their orthonormality (see the first of Eqs.(\ref{OrtCom}) with substituted capital letters),  yields
\begin{eqnarray} \label{matrinter1}
&& \left\langle {\Xi _s^{\left( P \right)} \Xi _e^{\left( Q \right)} \Xi _m^{\left( R \right)} } \right|\prod\limits_{i = 1}^N {\hat A_{\Theta _{\pi _\tau  \left( i \right)} }^\dag  \left( {\tilde \Xi _{\Theta _{\pi _\tau  \left( i \right)} }^i } \right)\left| 0 \right\rangle } = \sqrt {P!Q!R!} \nonumber \\ 
&& \cdot  \Delta _s \left( {\Xi _s^{\left( P \right)} \left| {\left( {\tilde \Xi _s^{\pi _\tau ^{ - 1} \left( 1 \right)} , \ldots ,\tilde \Xi _s^{\pi _\tau ^{ - 1} \left( P \right)} } \right)} \right.} \right)\Delta _e \left( {\Xi _e^{\left( Q \right)} \left| {\left( {\tilde \Xi _e^{\pi _\tau ^{ - 1} \left( {P + 1} \right)} , \ldots ,\tilde \Xi _e^{\pi _\tau ^{ - 1} \left( {P + Q} \right)} } \right)} \right.} \right)\Delta _m \left( {\Xi _m^{\left( R \right)} \left| {\left( {\tilde \Xi _m^{\pi _\tau ^{ - 1} \left( {P + Q + 1} \right)} , \ldots ,\tilde \Xi _m^{\pi _\tau ^{ - 1} \left( N \right)} } \right)} \right.} \right). \nonumber \\ 
\end{eqnarray}
The expression for the symmetrized delta functions in Eq.(\ref{SimmDelta}) enable to write Eq.(\ref{matrinter1}) as
\begin{equation} \label{matrinter2}
\left\langle {\Xi _s^{\left( P \right)} \Xi _e^{\left( Q \right)} \Xi _m^{\left( R \right)} } \right|\prod\limits_{i = 1}^N {\hat A_{\Theta _{\pi _\tau  \left( i \right)} }^\dag  \left( {\tilde \Xi _{\Theta _{\pi _\tau  \left( i \right)} }^i } \right)\left| 0 \right\rangle }  = \frac{1}{{\sqrt {P!Q!R!} }}\sum\limits_{\sigma _{PQR} } {\prod\limits_{j = 1}^N {\delta _{\Theta _j } \left( {\tilde \Xi _{\Theta _j }^{\pi _\tau ^{ - 1} \left( j \right)}  - \bar \Xi ^{\sigma _{PQR} \left( j \right)} } \right)} } ,
\end{equation}
where $\sigma _{PQR}$ spans the permutations in $S_n$ that solely shuffles the elements inside each of the three groups of $P$,$Q$ and $R$ elements and we have introduced the $N$-tuple containing all the outgoing polariton variables
\begin{equation}
\left( {\bar \Xi ^1  \ldots \bar \Xi ^N } \right) = \left( {\Xi _s^1 , \ldots ,\Xi _s^P ,\Xi _e^1 , \ldots ,\Xi _e^Q ,\Xi _m^1 , \ldots ,\Xi _m^R } \right).
\end{equation}
Setting $j = \pi _\tau  \left( i \right)$ in the right hand side of Eq.(\ref{matrinter2}) we get
\begin{equation}
\left\langle {\Xi _s^{\left( P \right)} \Xi _e^{\left( Q \right)} \Xi _m^{\left( R \right)} } \right|\prod\limits_{i = 1}^N {\hat A_{\Theta _{\pi _\tau  \left( i \right)} }^\dag  \left( {\tilde \Xi _{\Theta _{\pi _\tau  \left( i \right)} }^i } \right)\left| 0 \right\rangle }  = \frac{1}{{\sqrt {P!Q!R!} }}\sum\limits_{\sigma _{PQR}  } {\prod\limits_{i = 1}^N {\delta _{\Theta _{\pi _\tau  \left( i \right)} } \left( {\tilde \Xi _{\Theta _{\pi _\tau  \left( i \right)} }^i  - \bar \Xi ^{\sigma _{PQR} \left( {\pi _\tau  \left( i \right)} \right)} } \right)} }, 
\end{equation}
which, inserted into Eq.(\ref{matr4}), yields
\begin{equation} \label{matr5}
\left\langle {{\Xi _s^{\left( P \right)} \Xi _e^{\left( Q \right)} \Xi _m^{\left( R \right)} }}
 \mathrel{\left | {\vphantom {{\Xi _s^{\left( P \right)} \Xi _e^{\left( Q \right)} \Xi _m^{\left( R \right)} } {\xi _s^{\left( p \right)} \xi _e^{\left( q \right)} \xi _m^{\left( r \right)} }}}
 \right. \kern-\nulldelimiterspace}
 {{\xi _s^{\left( p \right)} \xi _e^{\left( q \right)} \xi _m^{\left( r \right)} }} \right\rangle  = \frac{{\delta _{N,n} }}{{\sqrt{P!Q!R!} \sqrt {p!q!r!} }}\sum\limits_{\tau  \in \left( {sem} \right)_{PQR} } {\sum\limits_{\sigma _{PQR}  \in S_n } {\prod\limits_{i = 1}^N {Z_{\chi \left( {\bar \Xi ^{\left( {\sigma _{PQR}  \circ \pi _\tau  } \right)\left( i \right)} } \right)\chi \left( {\bar \xi ^i } \right)} \left( {\bar \Xi ^{\left( {\sigma _{PQR}  \circ \pi _\tau  } \right)\left( i \right)} \left| {\bar \xi ^i } \right.} \right)} } }.
\end{equation}
This expression can be further simplified by noting that here the permutation $\pi _N  = \sigma _{PQR}  \circ \pi _\tau$ first separately shuffles the three groups of $P,Q,R$ variables of $\left( {\bar \Xi ^1 , \ldots ,\bar \Xi ^N } \right)$ and then produces the anagram so that, due to the double summation over $\tau$ and $\sigma _{PQR}$, $\pi _N$ explores the whole permutation group $S_n$. Accordingly, Eq.(\ref{matr5}) can be written as
\begin{equation} 
\left\langle {{\Xi _s^{\left( P \right)} \Xi _e^{\left( Q \right)} \Xi _m^{\left( R \right)} }}
 \mathrel{\left | {\vphantom {{\Xi _s^{\left( P \right)} \Xi _e^{\left( Q \right)} \Xi _m^{\left( R \right)} } {\xi _s^{\left( p \right)} \xi _e^{\left( q \right)} \xi _m^{\left( r \right)} }}}
 \right. \kern-\nulldelimiterspace}
 {{\xi _s^{\left( p \right)} \xi _e^{\left( q \right)} \xi _m^{\left( r \right)} }} \right\rangle  = \frac{{\delta _{N,n} }}{{\sqrt{P!Q!R!}\sqrt {p!q!r!} }}\sum\limits_{\pi _N  \in S_N } {\prod\limits_{i = 1}^N {Z_{\chi \left( {\bar \Xi ^{\pi _N \left( i \right)} } \right)\chi \left( {\bar \xi ^i } \right)} \left( {\bar \Xi ^{\pi _N \left( i \right)} \left| {\bar \xi ^i } \right.} \right)} }, 
\end{equation}
or, since $\pi_N$ spans all the permutation of $S_N$, eventually as
\begin{equation} \label{matr6}
\left\langle {{\Xi _s^{\left( P \right)} \Xi _e^{\left( Q \right)} \Xi _m^{\left( R \right)} }}
 \mathrel{\left | {\vphantom {{\Xi _s^{\left( P \right)} \Xi _e^{\left( Q \right)} \Xi _m^{\left( R \right)} } {\xi _s^{\left( p \right)} \xi _e^{\left( q \right)} \xi _m^{\left( r \right)} }}}
 \right. \kern-\nulldelimiterspace}
 {{\xi _s^{\left( p \right)} \xi _e^{\left( q \right)} \xi _m^{\left( r \right)} }} \right\rangle  = \frac{{\delta _{N,n} }}{{\sqrt {P!Q!R!} \sqrt {p!q!r!} }}\sum\limits_{\pi _N  \in S_N } {\prod\limits_{i = 1}^N {Z_{\chi \left( {\bar \Xi ^i } \right)\chi \left( {\bar \xi ^{\pi _N \left( i \right)} } \right)} \left( {\bar \Xi ^i \left| {\bar \xi ^{\pi _N \left( i \right)} } \right.} \right)} } 
\end{equation}
which is Eq.(\ref{ChaBasMat}) of the main text. Note that an equivalent expression of Eq.(\ref{matr6}) is
\begin{equation}
\left\langle {{\Xi _s^{\left( P \right)} \Xi _e^{\left( Q \right)} \Xi _m^{\left( R \right)} }}
 \mathrel{\left | {\vphantom {{\Xi _s^{\left( P \right)} \Xi _e^{\left( Q \right)} \Xi _m^{\left( R \right)} } {\xi _s^{\left( p \right)} \xi _e^{\left( q \right)} \xi _m^{\left( r \right)} }}}
 \right. \kern-\nulldelimiterspace}
 {{\xi _s^{\left( p \right)} \xi _e^{\left( q \right)} \xi _m^{\left( r \right)} }} \right\rangle  = \frac{{\delta _{N,n} }}{{\sqrt {P!Q!R!} \sqrt{p!q!r!}}}{\rm Perm} \left( {\bar Z_{ij} } \right)
\end{equation}
where we have introduced the $N \times N$ matrix
\begin{equation}
\bar Z_{ij}  = Z_{\chi \left( {\bar \Xi ^i } \right)\chi \left( {\bar \xi ^j } \right)} \left( {\bar \Xi ^i \left| {\bar \xi ^j } \right.} \right)
\end{equation}
and ${\rm Perm} (A_{ij})$ is the permanent of the $N \times N$ matrix $A_{ij}$, the symmetric analog of the determinant,  defined as
\begin{equation}
{\rm Perm} \left( {A_{ij} } \right) = \sum\limits_{\pi _N  \in S_N } {\prod\limits_{i = 1}^N {A_{i \pi _N \left( i \right)} } } .
\end{equation}

\section{Outgoing wavefunction and probability $p_{PQR}$}
Inserting Eq.(\ref{IniRadStat}) into Eq.(\ref{StatIngOut}) and using Eq.(\ref{ChaBasMat}), we get for the outgoing wavefunction
\begin{equation} \label{PsiOut1}
\Psi _{PQR}^{\left( {out} \right)} \left( {\Xi _s^{\left( P \right)} ,\Xi _e^{\left( Q \right)} ,\Xi _m^{\left( R \right)} } \right) = \frac{1}{{\sqrt {P!Q!R!} \sqrt {N!} }}\int {d\xi _s^{\left( N \right)} } \left[ {\sum\limits_{\pi _N  \in S_N } {\prod\limits_{i = 1}^N {Z_{\chi \left( {\bar \Xi ^i } \right)s} \left( {\bar \Xi ^i \left| {\xi _s^{\pi _N \left( i \right)} } \right.} \right)} } } \right]\psi _N^{\left( {in} \right)} \left( {\xi _s^{\left( N \right)} } \right)
\end{equation}
which, by exploiting the exchange symmetry of the ingoing wavefunction and of the differential, $\psi _N^{\left( {in} \right)} ( {\xi _s^{\left( N \right)} } ) = \psi _N^{\left( {in} \right)} ( {\pi _N \xi _s^{\left( N \right)} } )$ and $d\xi _s^{\left( N \right)}  = d( {\pi _N \xi _s^{\left( N \right)} } )$, where $\pi _N \xi _s^{\left( N \right)}  = ( {\xi _s^{\pi _N \left( 1 \right)} , \ldots ,\xi _s^{\pi _N \left( N \right)} } )$, readily yields
\begin{equation} \label{PsiOut2}
\Psi _{PQR}^{\left( {out} \right)} \left( {\Xi _s^{\left( P \right)} ,\Xi _e^{\left( Q \right)} ,\Xi _m^{\left( R \right)} } \right) = \frac{1}{{\sqrt {P!Q!R!} \sqrt {N!} }}\int {d\left( {\pi _N \xi _s^{\left( N \right)} } \right)} \left[ {\sum\limits_{\pi _N  \in S_N } {\prod\limits_{i = 1}^N {Z_{\chi \left( {\bar \Xi ^i } \right)s} \left( {\bar \Xi ^i \left| {\left( {\pi _N \xi _s^{\left( N \right)} } \right)^i } \right.} \right)} } } \right]\psi _N^{\left( {in} \right)} \left( {\pi _N \xi _s^{\left( N \right)} } \right).
\end{equation}
We now perform the change of integration variable $\pi _N \xi _s^{\left( N \right)}  \to \xi _s^{\left( N \right)}$ and, using the fact that the resulting integral does not depend on the permutation $\pi_N$, we get
\begin{equation} \label{PsiOut3}
\Psi _{PQR}^{\left( {out} \right)} \left( {\Xi _s^{\left( P \right)} ,\Xi _e^{\left( Q \right)} ,\Xi _m^{\left( R \right)} } \right) = \sqrt {\frac{{N!}}{{P!Q!R!}}} \int {d\xi _s^{\left( N \right)} } \left[ {\prod\limits_{i = 1}^N {Z_{\chi \left( {\bar \Xi ^i } \right)s} \left( {\bar \Xi ^i \left| {\xi _s^i } \right.} \right)} } \right]\psi _N^{\left( {in} \right)} \left( {\xi _s^{\left( N \right)} } \right).
\end{equation}
which is Eq.(\ref{IniRadOutStat}). By using Eqs.(\ref{PsiOut3}), it is easy to show that the probability 
\begin{equation}
p_{PQR}  = \int {d\Xi _s^{\left( P \right)} d\Xi _e^{\left( Q \right)} d\Xi _m^{\left( R \right)} } \left| {\Psi _{PQR}^{\left( {out} \right)} \left( {\Xi _s^{\left( P \right)} ,\Xi _e^{\left( Q \right)} ,\Xi _m^{\left( R \right)} } \right)} \right|^2 
\end{equation}
can be written as

\begin{eqnarray}
&& p_{PQR}  = \int {d\xi _s^{\left( N \right)} d{\xi '}_s^{\left( N \right)} } \psi _N^{\left( {in} \right)*} \left( {\xi _s^{\left( N \right)} } \right)\psi _N^{\left( {in} \right)} \left( {{\xi '}_s^{\left( N \right)} } \right) \nonumber \\
&& \cdot \left[ { \frac{{N!}}{{P!Q!R!}}\prod\limits_{I = 1}^P {J_s \left( {\xi _s^I \left| {{\xi '}_s^I } \right.} \right)} \prod\limits_{J = 1}^Q {J_e \left( {\xi _s^{P + J} \left| {{\xi '}_s^{P + J} } \right.} \right)} \prod\limits_{K = 1}^R {J_m \left( {\xi _s^{P + Q + K} \left| {{\xi '}_s^{P + Q + K} } \right.} \right)} } \right],
\end{eqnarray}
which is Eq.(\ref{pPQR}) and where
\begin{equation} \label{Jtau}
J_\tau  \left( {\xi _s \left| {\xi '_s } \right.} \right) = \int {d\Xi _\tau  } Z_{\tau s}^* \left( {\Xi _\tau  \left| {\xi _s } \right.} \right)Z_{\tau s} \left( {\Xi _\tau  \left| {\xi '_s } \right.} \right).
\end{equation}
To manipulate Eq.(\ref{Jtau}), we use Eqs.(\ref{PolInteg}) and the first three of Eqs.(\ref{ZmatrEntr}) to make explicit the integral and the  matrix entries $Z_{s s}$ and $Z_{\nu s}$ with $\nu = e,m$, respectively, thus getting
\begin{eqnarray}
&& J_s \left( {\xi _s \left| {\xi '_s } \right.} \right) = \int_0^{ + \infty } {d\Omega } \int {do_{\bf{N}} } \sum\limits_{\Lambda  = 1}^2 {\left[ {\delta \left( {\Omega  - \omega } \right){\bf{e}}_{{\bf{N}}\Lambda }  \cdot {\cal T}_{\omega ss}^* \left( {\left. {\bf{N}} \right|{\bf{n}}} \right) \cdot {\bf{e}}_{{\bf{n}}\lambda } } \right]} \left[ {\delta \left( {\Omega  - \omega '} \right){\bf{e}}_{{\bf{N}}\Lambda }  \cdot {\cal T}_{\omega 'ss} \left( {\left. {\bf{N}} \right|{\bf{n}}'} \right) \cdot {\bf{e}}_{{\bf{n}}'\lambda '} } \right], \nonumber \\ 
&& J_\nu  \left( {\xi _s \left| {\xi '_s } \right.} \right) = \int_0^{ + \infty } {d\Omega } \int {d^3 {\bf{R}}} \sum\limits_{\Lambda  = 1}^3 {\left[ {\delta \left( {\Omega  - \omega } \right){\bf{u}}_\Lambda   \cdot {\cal A}_{\omega \nu s}^* \left( {{\bf{R}}\left| {\bf{n}} \right.} \right) \cdot {\bf{e}}_{{\bf{n}}\lambda } } \right]\left[ {\delta \left( {\Omega  - \omega '} \right){\bf{u}}_\Lambda   \cdot {\cal A}_{\omega '\nu s} \left( {{\bf{R}}\left| {{\bf{n}}'} \right.} \right) \cdot {\bf{e}}_{{\bf{n}}'\lambda '} } \right]} , \nonumber  \\ 
\end{eqnarray}
which, by using the completeness of the polarization unit vectors, $\sum\nolimits_{\Lambda  = 1}^2 {{\bf{e}}_{{\bf{N}}\Lambda } {\bf{e}}_{{\bf{N}}\Lambda } }  = {\cal I}_{\bf{N}}$ and $\sum\nolimits_{\Lambda  = 1}^3 {{\bf{u}}_\Lambda  {\bf{u}}_\Lambda  }  = {\cal I}$, turn into
\begin{eqnarray} \label{JsJnu}
&& J_s \left( {\xi _s \left| {\xi '_s } \right.} \right) = \delta \left( {\omega  - \omega '} \right){\bf{e}}_{{\bf{n}}\lambda }  \cdot \left[ {\int {do_{\bf{N}} } {\cal T}_{\omega ss}^{T*} \left( {\left. {\bf{N}} \right|{\bf{n}}} \right) \cdot {\cal T}_{\omega ss} \left( {\left. {\bf{N}} \right|{\bf{n}}'} \right)} \right] \cdot {\bf{e}}_{{\bf{n}}'\lambda '}, \nonumber  \\ 
&& J_\nu  \left( {\xi _s \left| {\xi '_s } \right.} \right) = \delta \left( {\omega  - \omega '} \right){\bf{e}}_{{\bf{n}}\lambda }  \cdot \left[ {\int {d^3 {\bf{R}}} {\cal A}_{\omega \nu s}^{T*} \left( {{\bf{R}}\left| {\bf{n}} \right.} \right) \cdot {\cal A}_{\omega \nu s} \left( {{\bf{R}}\left| {{\bf{n}}'} \right.} \right)} \right] \cdot {\bf{e}}_{{\bf{n}}'\lambda '} .
\end{eqnarray}
Now the integrals appearing in the right hand sides of Eqs.(\ref{JsJnu}) can be conveniently rewritten by resorting to the transmission operator ${\sf T}_{\omega ss}$ and the absorption operators ${\sf A}_{\omega \nu s}$, thus getting
\begin{eqnarray} \label{JsJnuAPPENDIX}
&&  J_s \left( {\xi _s \left| {\xi '_s } \right.} \right) = \delta \left( {\omega  - \omega '} \right){\bf{e}}_{{\bf{n}}\lambda }  \cdot \left[\kern-0.15em\left[  {{\sf T}_{\omega ss}^ +  {\sf T}_{\omega ss} } 
 \right]\kern-0.15em\right]\left( {{\bf{n}},{\bf{n}}'} \right) \cdot {\bf{e}}_{{\bf{n}}'\lambda '}, \nonumber  \\ 
&& J_\nu  \left( {\xi _s \left| {\xi '_s } \right.} \right) = \delta \left( {\omega  - \omega '} \right){\bf{e}}_{{\bf{n}}\lambda }  \cdot \left[\kern-0.15em\left[  {{\sf A}_{\omega \nu s}^ +  {\sf A}_{\omega \nu s} }   \right]\kern-0.15em\right]\left( {{\bf{n}},{\bf{n}}'} \right) \cdot {\bf{e}}_{{\bf{n}}'\lambda '} .
\end{eqnarray} 
which can be concisely written as Eq.(\ref{Jtauxx'}), i.e.
\begin{equation} \label{JtauAPPENDIX}
J_\tau  \left( {\xi _s \left| {\xi '_s } \right.} \right) = \delta \left( {\omega  - \omega '} \right){\bf{e}}_{{\bf{n}}\lambda }  \cdot  {\left[\kern-0.15em\left[ {{\sf K}_{\omega s s }^{\tau} } 
 \right]\kern-0.15em\right]\left( {{\bf{n}},{\bf{n}}'} \right)}  \cdot {\bf{e}}_{{\bf{n}}'\lambda '},
\end{equation}
where
\begin{eqnarray}
&& {\sf K}_{\omega ss}^s  = {\sf T}_{\omega ss}^ +  {\sf T}_{\omega ss} , \nonumber  \\ 
&& {\sf K}_{\omega ss}^\nu  =  {\sf A}_{\omega \nu s}^ +  {\sf A}_{\omega \nu s} .
\end{eqnarray}
Note that the operators ${{\sf K}_{\omega ss}^\tau  }$ are Hermitian so that from Eq.(\ref{JtauAPPENDIX}) it is straightforward showing that 
\begin{equation} \label{JtauStar}
J_\tau ^* \left( {\xi _s \left| {\xi '_s } \right.} \right) = J_\tau  \left( {\xi '_s \left| {\xi _s } \right.} \right).
\end{equation}

\section{Reduced density operator of the scattered radiation}
We here evaluate the reduced density operator $\hat \rho _s^{\left( {out} \right)}  = {\rm Tr}_{\Xi _e \Xi _m } \left( {\left| {\Psi ^{\left( {out} \right)} } \right\rangle \left\langle {\Psi ^{\left( {out} \right)} } \right|} \right)$ of the scattered radiation. The partial trace over the outgoing $e$- and $m$-polaritons space $F_e  \otimes F_m$ of an operator $\hat O$ acting on $H = F_s  \otimes (F_e  \otimes F_m)$ is given by
\begin{equation}
{\rm Tr}_{\Xi _e \Xi _m } \left( {\hat O} \right) = \sum\limits_{Q,R = 0}^\infty  {\int {d\Xi _e^{\left( Q \right)} d\Xi _m^{\left( R \right)} } } \left[ {\hat I_s  \otimes \left\langle {\Xi _e^{\left( Q \right)} \Xi _m^{\left( R \right)} } \right|} \right]\hat O\left[ {\hat I_s  \otimes \left| {\Xi _e^{\left( Q \right)} \Xi _m^{\left( R \right)} } \right\rangle } \right],
\end{equation}
where $\hat I_s$ is the identity operator of the outgoing $s$-polarion Fock space $F_s$ so that the operator ${\hat I_s  \otimes  \langle {\Xi _e^{\left( Q \right)} \Xi _m^{\left( R \right)} }  |}$ acts on the Hilbert space $H$ and produces states in the Fock space $F_s$. Accordingly the reduced density operator is explicitly written as
\begin{equation} \label{RhoOut1}
\hat \rho _s^{\left( {out} \right)}  = \sum\limits_{Q,R = 0}^\infty  {\int {d\Xi _e^{\left( Q \right)} d\Xi _m^{\left( R \right)} } } \left[ {\hat I_s  \otimes \left\langle {\Xi _e^{\left( Q \right)} \Xi _m^{\left( R \right)} } \right|} \right]\left| {\Psi ^{\left( {out} \right)} } \right\rangle \left\langle {\Psi ^{\left( {out} \right)} } \right|\left[ {\hat I_s  \otimes \left| {\Xi _e^{\left( Q \right)} \Xi _m^{\left( R \right)} } \right\rangle } \right],
\end{equation}
so that, after noting that Eq.(\ref{PsiOutgo}) yields
\begin{equation}
\left[ {\hat I_s  \otimes \left\langle {\Xi _e^{\left( Q \right)} \Xi _m^{\left( R \right)} } \right|} \right]\left| {\Psi ^{\left( {out} \right)} } \right\rangle  = \sum\limits_{P = 0}^\infty  {\int {d\Xi _s^{\left( P \right)} } } \Psi _{PQR}^{\left( {out} \right)} \left( {\Xi _s^{\left( P \right)} ,\Xi _e^{\left( Q \right)} ,\Xi _m^{\left( R \right)} } \right)\left| {\Xi _s^{\left( P \right)} } \right\rangle,
\end{equation}
Eq.(\ref{RhoOut1}) can be written as
\begin{equation} \label{RhoOut2}
\hat \rho _s^{\left( {out} \right)}  = \sum\limits_{P,P' = 0}^\infty  {\int {d\Xi _s^{\left( P \right)} d{\Xi '}_s^{\left( {P'} \right)} } } \rho _s^{\left( {out} \right)} \left( {\Xi _s^{\left( P \right)} \left| {{\Xi '}_s^{\left( {P'} \right)} } \right.} \right)\left| {\Xi _s^{\left( P \right)} } \right\rangle \left\langle {{\Xi '}_s^{\left( {P'} \right)} } \right|,
\end{equation}
where the reduced density matrix $\rho _s^{\left( {out} \right)} ( {\Xi _s^{\left( P \right)} | {{\Xi '}_s^{\left( {P'} \right)} } } ) = \langle {\Xi _s^{\left( P \right)} } |\hat \rho _s^{\left( {out} \right)} | {{\Xi '}_s^{\left( {P'} \right)} } \rangle$ is
\begin{eqnarray} \label{RhoMatr1}
&&\rho _s^{\left( {out} \right)} \left( {\Xi _s^{\left( P \right)} \left| {{\Xi '}_s^{\left( {P'} \right)} } \right.} \right) = \sum\limits_{Q,R = 0}^\infty  {\int {d\Xi _e^{\left( Q \right)} d\Xi _m^{\left( R \right)} } } \nonumber \\ 
&& \cdot \left[ {\frac{1}{{P!}}\sum\limits_{\pi _P  \in S_P } {\Psi _{PQR}^{\left( {out} \right)} \left( {\pi _P \Xi _s^{\left( P \right)} ,\Xi _e^{\left( Q \right)} ,\Xi _m^{\left( R \right)} } \right)} } \right]\left[ {\frac{1}{{P'!}}\sum\limits_{\pi _{P'}  \in S_{P'} } {\Psi _{P'QR}^{\left( {out} \right)*} \left( {\pi _{P'} {\Xi '}_s^{\left( {P'} \right)} ,\Xi _e^{\left( Q \right)} ,\Xi _m^{\left( R \right)} } \right)} } \right].
\end{eqnarray}
We now use Eq.(\ref{PsiOut3}) of Appendix G for the outgoing wavefunction which inserted  Eq.(\ref{RhoMatr1}) yields
\begin{eqnarray} \label{PsioutH2}
&& \rho _s^{\left( {out} \right)} \left( {\Xi _s^{\left( P \right)} \left| {{\Xi '}_s^{\left( {P'} \right)} } \right.} \right) = \sum\limits_{Q,R = 0}^\infty  {\int {d\xi _s^{\left( N \right)} }  {d{\xi '}_s^{\left( {N'} \right)} } {\phi _{Q + R}^P } \left( {\Xi _s^{\left( P \right)} \left| {\xi _s^{\left( N \right)} } \right.} \right) {\phi _{Q + R}^{P'*} } \left( {{\Xi '}_s^{\left( {P'} \right)} \left| {{\xi '}_s^{\left( {N'} \right)} } \right.} \right)} \sqrt {\frac{{N!N'!}}{{P!P'!}}} 
 \frac{1}{{Q!R!}} \nonumber \\ 
&&  \cdot \left[ {\prod\limits_{J = 1}^Q {\int {d\Xi _e } Z_{es} \left( {\Xi _e \left| {\xi _s^{P + J} } \right.} \right)Z_{es}^* \left( {\Xi _e \left| {{\xi '}_s^{P' + J} } \right.} \right)} } \right]\left[ {\prod\limits_{K = 1}^R {\int {d\Xi _m } Z_{ms} \left( {\Xi _m \left| {\xi _s^{P + Q + K} } \right.} \right)Z_{ms}^* \left( {\Xi _m \left| {{\xi '}_s^{P' + Q + K} } \right.} \right)} } \right], \nonumber \\ 
\end{eqnarray}
where
\begin{equation}
\phi _{Q + R}^P \left( {\Xi _s^{\left( P \right)} \left| {\xi _s^{\left( N \right)} } \right.} \right) =  \left[ {\frac{1}{{P!}}\sum\limits_{\pi _P  \in S_P } {\prod\limits_{I = 1}^P {Z_{ss} \left( {\Xi _s^{\pi _P \left( I \right)} \left| {\xi _s^I } \right.} \right)} } } \right]\psi _N^{\left( {in} \right)} \left( {\xi _s^{\left( N \right)} } \right),
\end{equation}
so that, by using Eq.(\ref{Jtau}) of Appendix G, Eq.(\ref{PsioutH2}) turns into
\begin{eqnarray} \label{PsioutH3}
&& \rho _s^{\left( {out} \right)} \left( {\Xi _s^{\left( P \right)} \left| {{\Xi '}_s^{\left( {P'} \right)} } \right.} \right) = \sum\limits_{Q,R = 0}^\infty  {\int {d\xi _s^{\left( N \right)} d{\xi '}_s^{\left( {N'} \right)} } {\phi _{Q + R}^P } \left( {\Xi _s^{\left( P \right)} \left| {\xi _s^{\left( N \right)} } \right.} \right) {\phi _{Q + R}^{P'*} } \left( {{\Xi '}_s^{\left( {P'} \right)} \left| {{\xi '}_s^{\left( {N'} \right)} } \right.} \right)} \nonumber \\
&& \cdot \left[ {   \sqrt {\frac{{N!N'!}}{{P!P'!}}}  \frac{1}{{Q!R!}}  \prod\limits_{J = 1}^Q {J_e^* \left( {\xi _s^{P + J} \left| {{\xi '}_s^{P' + J} } \right.} \right)\prod\limits_{K = 1}^R {J_m^* \left( {\xi _s^{P + Q + K} \left| {{\xi '}_s^{P' + Q + K} } \right.} \right)} } } \right].
\end{eqnarray}
We repeat, for clarity purposes, that $N= P + Q +R$ and $N' = P' + Q +R$ so that the evident summation relation
\begin{equation}
\sum\limits_{Q = 0}^\infty  {\sum\limits_{R = 0}^\infty  {f\left( {Q,R} \right)} }  = \sum\limits_{S = 0}^\infty  {\left[ {\sum\limits_{Q = 0}^S {f\left( {Q,S - Q} \right)} } \right]}
\end{equation}
enable Eq.(\ref{PsioutH3}) to be conveniently rewritten as
\begin{equation}
\rho _s^{\left( {out} \right)} \left( {\Xi _s^{\left( P \right)} \left| {{\Xi '}_s^{\left( {P'} \right)} } \right.} \right) = \sum\limits_{S = 0}^\infty  {\int {d\xi _s^{\left( {P + S} \right)} d{{\xi '}}_s^{\left( {P' + S} \right)} } {\phi _S^P } \left( {\Xi _s^{\left( P \right)} \left| {\xi _s^{\left( {P + S} \right)} } \right.} \right) {\phi _S^{P'*} } \left( {{\Xi '}_s^{\left( {P'} \right)} \left| {{\xi '}_s^{\left( {P' + S} \right)} } \right.} \right)D_S^{PP'} \left( {\xi _s^{\left( {P + S} \right)} \left| {{\xi '}_s^{\left( {P' + S} \right)} } \right.} \right)},
\end{equation}
where
\begin{equation}
D_S^{PP'} \left( {\xi _s^{\left( {P + S} \right)} \left| {{\xi '}_s^{\left( {P' + S} \right)} } \right.} \right) = 
\sqrt {\frac{{\left( {P + S} \right)!\left( {P' + S} \right)!}}{{P!P'!}}}  \sum\limits_{Q = 0}^S {\frac{1}{{Q!\left( {S - Q} \right)!}}\prod\limits_{J = 1}^Q {J_e^* \left( {\xi _s^{P + J} \left| {{\xi '}_s^{P' + J} } \right.} \right)\prod\limits_{K = 1}^{S - Q} {J_m^* \left( {\xi _s^{P + Q + K} \left| {{\xi '}_s^{P' + Q + K} } \right.} \right)} } }.
\end{equation}

\section{One-polariton scattering: reduced density matrix}
To evaluate the reduced density operator of Eq.(\ref{RedDenOpe}) we start by inserting the one-polariton ingoing wavefunction of Eq.(\ref{OnePolIngWav}) into Eq.(\ref{PhiPS}) so that, using the relation $\delta _{\left( {P + S} \right)1}  = \delta _{S0} \delta _{P1} +\delta _{S1} \delta _{P0}$ for the Kronecker delta (since $P \ge 0$ and $S \ge 0$), we get
\begin{equation}
\phi _S^P \left( {\Xi _s^{\left( P \right)} \left| {\xi _s^{\left( {P + S} \right)} } \right.} \right) = \left[ {\delta _{S0} \delta _{P1} Z_{ss} \left( {\Xi _s^1 \left| {\xi _s^1 } \right.} \right) + \delta _{S1} \delta _{P0} } \right]\psi _1^{\left( {in} \right)} \left( {\xi _s^{\left( 1 \right)} } \right)
\end{equation}
which in turn directly yields
\begin{eqnarray} \label{PhiPPhiPpri}
&& \phi _S^P \left( {\Xi _s^{\left( P \right)} \left| {\xi _s^{\left( {P + S} \right)} } \right.} \right)\phi _S^{P'*} \left( {{\Xi '}_s^{\left( {P'} \right)} \left| {{\xi '}_s^{\left( {P' + S} \right)} } \right.} \right) = \left[ { \delta _{S0} \delta _{P1} \delta _{P'1} Z_{ss} \left( {\Xi _s^1 \left| {\xi _s^1 } \right.} \right)Z_{ss}^* \left( {{\Xi '}_s^1 \left| {{\xi '}_s^1 } \right.} \right) + \delta _{S1} \delta _{P0} \delta _{P'0} } \right] \nonumber \\
&& \cdot \psi _1^{\left( {in} \right)} \left( {\xi _s^{\left( 1 \right)} } \right)\psi _1^{\left( {in} \right)*} \left( {{\xi '}_s^{\left( 1 \right)} } \right).
\end{eqnarray}
We now substitute Eq.(\ref{PhiPPhiPpri}) into the reduced density matrix of Eq.(\ref{RedDenMat}) thus getting 
\begin{eqnarray}
&& \rho _s^{\left( {out} \right)} \left( {\Xi _s^{\left( P \right)} \left| {{\Xi '}_s^{\left( {P'} \right)} } \right.} \right) = \int {d\xi _s^{\left( 1 \right)} d{\xi '}_s^{\left( 1 \right)} } \psi _1^{\left( {in} \right)} \left( {\xi _s^{\left( 1 \right)} } \right)\psi _1^{\left( {in} \right)*} \left( {{\xi '}_s^{\left( 1 \right)} } \right) \nonumber \\
&& \cdot \left[ {   \delta _{P1} \delta _{P'1} Z_{ss} \left( {\Xi _s^1 \left| {\xi _s^1 } \right.} \right)Z_{ss}^* \left( {{\Xi '}_s^1 \left| {{\xi '}_s^1 } \right.} \right)D_0^{11} \left( {\xi _s^{\left( 1 \right)} \left| {{\xi '}_s^{\left( 1 \right)} } \right.} \right) +\delta _{P0} \delta _{P'0} D_1^{00} \left( {\xi _s^{\left( 1 \right)} \left| {{\xi '}_s^{\left( 1 \right)} } \right.} \right)  } \right]
\end{eqnarray}
where the relevant decoherence factors, from Eq.(\ref{DecohFac}), are readily seen to be
\begin{eqnarray}
&& D_0^{11} \left( {\xi _s^{\left( 1 \right)} \left| {{\xi '}_s^{\left( 1 \right)} } \right.} \right) = 1, \nonumber \\  
&& D_1^{00} \left( {\xi _s^{\left( 1 \right)} \left| {{\xi '}_s^{\left( 1 \right)} } \right.} \right) = J_e^* \left( {\xi _s^1 \left| {{\xi '}_s^1 } \right.} \right) + J_m^* \left( {\xi _s^1 \left| {{\xi '}_s^1 } \right.} \right).
\end{eqnarray}
Using these results, the reduced density operator of Eq.(\ref{RedDenOpe}) can be conveniently written as
\begin{equation} \label{reddenopeG}
\hat \rho _s^{\left( {out} \right)}  = p_1 \left| {\Phi _{1s} } \right\rangle \left\langle {\Phi _{1s} } \right| + p_0 \left| {\Phi _{0s} } \right\rangle \left\langle {\Phi _{0s} } \right|
\end{equation}
where
\begin{eqnarray} \label{p0phi1}
&& p_1  = \int {d\xi _s^{\left( 1 \right)} d{\xi '}_s^{\left( 1 \right)} \psi _1^{\left( {in} \right)*} \left( {\xi _s^{\left( 1 \right)} } \right)} \psi _1^{\left( {in} \right)} \left( {{\xi '}_s^{\left( 1 \right)} } \right)J_s \left( {\xi _s^1 \left| {{\xi '}_s^1 } \right.} \right), \nonumber \\ 
&&  \left| {\Phi _{1s} } \right\rangle  = \frac{1}{{\sqrt {p_1 } }}\int {d\Xi _s^{\left( 1 \right)} } \left[ {\int {d\xi _s^{\left( 1 \right)} } Z_{ss} \left( {\Xi _s^1 \left| {\xi _s^1 } \right.} \right)\psi _1^{\left( {in} \right)} \left( {\xi _s^{\left( 1 \right)} } \right)} \right]\left| {\Xi _s^{\left( 1 \right)} } \right\rangle, \nonumber  \\ 
&&  p_0  = \int {d\xi _s^{\left( 1 \right)} d{\xi '}_s^{\left( 1 \right)} } \psi _1^{\left( {in} \right)} \left( {\xi _s^{\left( 1 \right)} } \right)\psi _1^{\left( {in} \right)*} \left( {{\xi '}_s^{\left( 1 \right)} } \right)\left[ {J_e^* \left( {\xi _s^1 \left| {{\xi '}_s^1 } \right.} \right) + J_m^* \left( {\xi _s^1 \left| {{\xi '}_s^1 } \right.} \right)} \right], \nonumber \\ 
&&  \left| {\Phi _{0s} } \right\rangle  = \left| {\Xi _s^{\left( 0 \right)} } \right\rangle,  
\end{eqnarray}
where the coefficient $p_1$ has been introduced in such a way that  $\left\langle {{\Phi _{1s} }}  \mathrel{\left | {\vphantom {{\Phi _{1s} } {\Phi _{1s} }}} \right. \kern-\nulldelimiterspace}  {{\Phi _{1s} }} \right\rangle  = 1$. As a matter of fact, the squared norm of $\left| {\Phi _{1s} } \right\rangle$ is
\begin{equation}
\left\langle {{\Phi _{1s} }}
 \mathrel{\left | {\vphantom {{\Phi _{1s} } {\Phi _{1s} }}}
 \right. \kern-\nulldelimiterspace}
 {{\Phi _{1s} }} \right\rangle  = \frac{1}{{p_1 }}\int {d\xi _s^{\left( 1 \right)} d{\xi '}_s^{\left( 1 \right)} } \psi _1^{\left( {in} \right)*} \left( {\xi _s^{\left( 1 \right)} } \right)\psi _1^{\left( {in} \right)} \left( {{\xi '}_s^{\left( 1 \right)} } \right)\int {d\Xi _s^{\left( 1 \right)} } Z_{ss}^* \left( {\Xi _s^1 \left| {\xi _s^1 } \right.} \right)Z_{ss} \left( {\Xi _s^1 \left| {{\xi '}_s^1 } \right.} \right)
\end{equation}
so that Eq.(\ref{Jtau}) proves the assertion. The comparison of the first and third of Eqs.(\ref{p0phi1}) with Eq.(\ref{OnePolProb}) directly yield $p_1  = P_s$ and $p_0  = P_e  + P_m$. Therefore, the reduced density operator of Eq.(\ref{reddenopeG}) is 
\begin{equation}
\hat \rho _s^{\left( {out} \right)}  = P_s \left| {\Phi _{1s} } \right\rangle \left\langle {\Phi _{1s} } \right| + \left( {P_e  + P_m } \right)\left| {\Phi _{0s} } \right\rangle \left\langle {\Phi _{0s} } \right|.
\end{equation}
In addition, considering the outgoing state $\left| {\Psi ^{\left( {out} \right)} } \right\rangle$ in Fig.5a, it is evident that
\begin{equation}
\left| {\Phi _{1s} } \right\rangle  = \frac{1}{{\sqrt {P_s } }}\left[ {\hat I_s  \otimes \left\langle {\Phi _{0em} } \right|} \right] \left| {\Psi _{out} } \right\rangle,
\end{equation}
where ${\hat I_s }$ is the identity operator of the Fock space $F_s$ and $| {\Phi _{0em} } \rangle  = | \Xi _e ^{\left( 0 \right)} \Xi _m  ^{\left( 0 \right)}  \rangle$ is the vacuum state of the space $F_e  \otimes F_m$.

\section{Two-polariton scattering: reduced density matrix}
We now evaluate the reduced density operator of Eq.(\ref{RedDenOpe}). We start by inserting the two-polariton ingoing wavefunction of Eq.(\ref{TwoPolIngWav}) into Eq.(\ref{PhiPS}) so that, using the relation $\delta _{\left( {P + S} \right)2}  = \delta _{S0} \delta _{P2}  + \delta _{S1} \delta _{P1}  + \delta _{S2} \delta _{P0}$ for the Kronecker delta (since $P \ge 0$ and $S \ge 0$), we get
\begin{eqnarray}
&&\phi _S^P \left( {\Xi _s^{\left( P \right)} \left| {\xi _s^{\left( {P + S} \right)} } \right.} \right) = \left\{ {\delta _{S0} \delta _{P2} \frac{1}{2}\left[ {Z_{ss} \left( {\Xi _s^1 \left| {\xi _s^1 } \right.} \right)Z_{ss} \left( {\Xi _s^2 \left| {\xi _s^2 } \right.} \right) + Z_{ss} \left( {\Xi _s^2 \left| {\xi _s^1 } \right.} \right)Z_{ss} \left( {\Xi _s^1 \left| {\xi _s^2 } \right.} \right)} \right] + \delta _{S1} \delta _{P1} Z_{ss} \left( {\Xi _s^1 \left| {\xi _s^1 } \right.} \right) + \delta _{S2} \delta _{P0} } \right\} \nonumber \\
&& \cdot \psi _2^{\left( {in} \right)} \left( {\xi _s^{\left( 2 \right)} } \right)
\end{eqnarray}
so that
\begin{eqnarray} \label{PhiPPhiPpri2}
&& \phi _S^P \left( {\Xi _s^{\left( P \right)} \left| {\xi _s^{\left( {P + S} \right)} } \right.} \right)\phi _S^{P'*} \left( {{\Xi '}_s^{\left( {P'} \right)} \left| {{\xi '}_s^{\left( {P' + S} \right)} } \right.} \right) = \left\{ {\delta _{S0} \delta _{P2} \delta _{P'2} \frac{1}{4}\left[ {Z_{ss} \left( {\Xi _s^1 \left| {\xi _s^1 } \right.} \right)Z_{ss} \left( {\Xi _s^2 \left| {\xi _s^2 } \right.} \right) + Z_{ss} \left( {\Xi _s^2 \left| {\xi _s^1 } \right.} \right)Z_{ss} \left( {\Xi _s^1 \left| {\xi _s^2 } \right.} \right)} \right]} \right. \nonumber \\ 
&&  \cdot \left[ {Z_{ss}^* \left( {{\Xi '}_s^1 \left| {{\xi '}_s^1 } \right.} \right)Z_{ss}^* \left( {{\Xi '}_s^2 \left| {{\xi '}_s^2 } \right.} \right) + Z_{ss}^* \left( {{\Xi '}_s^2 \left| {{\xi '}_s^1 } \right.} \right)Z_{ss}^* \left( {{\Xi '}_s^1 \left| {{\xi '}_s^2 } \right.} \right)} \right]\left. { + \delta _{S1} \delta _{P1} \delta _{P'1} Z_{ss} \left( {\Xi _s^1 \left| {\xi _s^1 } \right.} \right)Z_{ss}^* \left( {{\Xi '}_s^1 \left| {{\xi '}_s^1 } \right.} \right) + \delta _{S2} \delta _{P0} \delta _{P'0} } \right\} \nonumber\\ 
&&  \cdot \psi _2^{\left( {in} \right)} \left( {\xi _s^{\left( 2 \right)} } \right)\psi _2^{\left( {in} \right)*} \left( {{\xi '}_s^{\left( 2 \right)} } \right). 
\end{eqnarray}
After substituting Eq.(\ref{PhiPPhiPpri2}) into the reduced density matrix of Eq.(\ref{RedDenMat}) we get
\begin{eqnarray}
&& \rho _s^{\left( {out} \right)} \left( {\Xi _s^{\left( P \right)} \left| {{\Xi '}_s^{\left( {P'} \right)} } \right.} \right) = \int {d\xi _s^{\left( 2 \right)} d{\xi '}_s^{\left( 2 \right)} } \psi _2^{\left( {in} \right)} \left( {\xi _s^{\left( 2 \right)} } \right)\psi _2^{\left( {in} \right)*} \left( {{\xi '}_s^{\left( 2 \right)} } \right)\left\{ {\delta _{P2} \delta _{P'2} \frac{1}{4}} \right. \nonumber \\ 
&&  \cdot \left[ {Z_{ss} \left( {\Xi _s^1 \left| {\xi _s^1 } \right.} \right)Z_{ss} \left( {\Xi _s^2 \left| {\xi _s^2 } \right.} \right) + Z_{ss} \left( {\Xi _s^2 \left| {\xi _s^1 } \right.} \right)Z_{ss} \left( {\Xi _s^1 \left| {\xi _s^2 } \right.} \right)} \right] \cdot \left[ {Z_{ss}^* \left( {{\Xi '}_s^1 \left| {{\xi '}_s^1 } \right.} \right)Z_{ss}^* \left( {{\Xi '}_s^2 \left| {{\xi '}_s^2 } \right.} \right) + Z_{ss}^* \left( {{\Xi '}_s^2 \left| {{\xi '}_s^1 } \right.} \right)Z_{ss}^* \left( {{\Xi '}_s^1 \left| {{\xi '}_s^2 } \right.} \right)} \right] \nonumber \\ 
&& \cdot \left. {D_0^{22} \left( {\xi _s^{\left( 2 \right)} \left| {{\xi '}_s^{\left( 2 \right)} } \right.} \right) + \delta _{P1} \delta _{P'1} Z_{ss} \left( {\Xi _s^1 \left| {\xi _s^1 } \right.} \right)Z_{ss}^* \left( {{\Xi '}_s^1 \left| {{\xi '}_s^1 } \right.} \right)D_1^{11} \left( {\xi _s^{\left( 2 \right)} \left| {{\xi '}_s^{\left( 2 \right)} } \right.} \right) + \delta _{P0} \delta _{P'0} D_2^{00} \left( {\xi _s^{\left( 2 \right)} \left| {{\xi '}_s^{\left( 2 \right)} } \right.} \right)} \right\} 
\end{eqnarray}
whose relevant decoherence factors, from Eq.(\ref{DecohFac}), are readily seen to be
\begin{eqnarray}
&& D_0^{22} \left( {\xi _s^{\left( 2 \right)} \left| {{\xi '}_s^{\left( 2 \right)} } \right.} \right) = 1, \nonumber \\ 
&& D_1^{11} \left( {\xi _s^{\left( 2 \right)} \left| {{\xi '}_s^{\left( 2 \right)} } \right.} \right) = 2\left[ {J_e^* \left( {\xi _s^2 \left| {{\xi '}_s^2 } \right.} \right) + J_m^* \left( {\xi _s^2 \left| {{\xi '}_s^2 } \right.} \right)} \right], \nonumber \\ 
&& D_2^{00} \left( {\xi _s^{\left( 2 \right)} \left| {{\xi '}_s^{\left( 2 \right)} } \right.} \right) =  2J_e^* \left( {\xi _s^1 \left| {{\xi '}_s^1 } \right.} \right)J_m^* \left( {\xi _s^2 \left| {{\xi '}_s^2 } \right.} \right) + J_e^* \left( {\xi _s^1 \left| {{\xi '}_s^1 } \right.} \right)J_e^* \left( {\xi _s^2 \left| {{\xi '}_s^2 } \right.} \right) + J_m^* \left( {\xi _s^1 \left| {{\xi '}_s^1 } \right.} \right)J_m^* \left( {\xi _s^2 \left| {{\xi '}_s^2 } \right.} \right).
\end{eqnarray}
Using these results, the reduced density operator of Eq.(\ref{RedDenOpe}) can be conveniently written as
\begin{equation} \label{reddenopeGGGGGG}
\hat \rho _s^{\left( {out} \right)}  = p_2 \left| {\Phi _{2s} } \right\rangle \left\langle {\Phi _{2s} } \right| + p_1  \hat \rho _{1s}  + p_0 \left| {\Phi _{0s} } \right\rangle \left\langle {\Phi _{0s} } \right|
\end{equation}
where
\begin{eqnarray} \label{p1ph1rho2p3pihi3}
&& p_2  = \int {d\xi _s^{\left( 2 \right)} d{\xi '}_s^{\left( 2 \right)} } \psi _2^{\left( {in} \right)*} \left( {\xi _s^{\left( 2 \right)} } \right)\psi _2^{\left( {in} \right)} \left( {{\xi '}_s^{\left( 2 \right)} } \right)J_s \left( {\xi _s^1 \left| {{\xi '}_s^1 } \right.} \right)J_s \left( {\xi _s^2 \left| {{\xi '}_s^2 } \right.} \right), \nonumber \\ 
&& \left| {\Phi _{2s} } \right\rangle  = \frac{1}{{\sqrt {p_1 } }}\int {d\Xi _s^{\left( 2 \right)} } \left[ {\int {d\xi _s^{\left( 2 \right)} } Z_{ss} \left( {\Xi _s^1 \left| {\xi _s^1 } \right.} \right)Z_{ss} \left( {\Xi _s^2 \left| {\xi _s^2 } \right.} \right)\psi _2^{\left( {in} \right)} \left( {\xi _s^{\left( 2 \right)} } \right)} \right]\left| {\Xi _s^{\left( 2 \right)} } \right\rangle, \nonumber  \\ 
&& p_1  = 2\int {d\xi _s^{\left( 2 \right)} d{\xi '}_s^{\left( 2 \right)} } \psi _2^{\left( {in} \right)} \left( {\xi _s^{\left( 2 \right)} } \right)\psi _2^{\left( {in} \right)*} \left( {{\xi '}_s^{\left( 2 \right)} } \right)J_s^* \left( {\xi _s^2 \left| {{\xi '}_s^2 } \right.} \right)\left[ {J_e^* \left( {\xi _s^2 \left| {{\xi '}_s^2 } \right.} \right) + J_m^* \left( {\xi _s^2 \left| {{\xi '}_s^2 } \right.} \right)} \right], \nonumber \\
&& \hat \rho _{1s}  = \int {d\Xi _s^{\left( 1 \right)} d{\Xi '}_s^{\left( 1 \right)} } \left\{ {2\int {d\xi _s^{\left( 2 \right)} d{\xi '}_s^{\left( 2 \right)} } \psi _2^{\left( {in} \right)} \left( {\xi _s^{\left( 2 \right)} } \right)\psi _2^{\left( {in} \right)*} \left( {{\xi '}_s^{\left( 2 \right)} } \right)Z_{ss} \left( {\Xi _s^1 \left| {\xi _s^1 } \right.} \right)Z_{ss}^* \left( {{\Xi '}_s^1 \left| {{\xi '}_s^1 } \right.} \right)} \right. \nonumber \\ 
&&  \cdot \left. {\left[ {J_e^* \left( {\xi _s^2 \left| {{\xi '}_s^2 } \right.} \right) + J_m^* \left( {\xi _s^2 \left| {{\xi '}_s^2 } \right.} \right)} \right]} \right\}\left| {\Xi _s^{\left( 1 \right)} } \right\rangle \left\langle {{\Xi '}_s^{\left( 1 \right)} } \right|, \nonumber \\ 
&& p_0  = \int {d\xi _s^{\left( 2 \right)} d{\xi '}_s^{\left( 2 \right)} } \psi _2^{\left( {in} \right)} \left( {\xi _s^{\left( 2 \right)} } \right)\psi _2^{\left( {in} \right)*} \left( {{\xi '}_s^{\left( 2 \right)} } \right) \nonumber \\ 
&& \cdot \left[ 2J_e^* \left( {\xi _s^1 \left| {{\xi '}_s^1 } \right.} \right)J_m^* \left( {\xi _s^2 \left| {{\xi '}_s^2 } \right.} \right) + J_e^* \left( {\xi _s^1 \left| {{\xi '}_s^1 } \right.} \right)J_e^* \left( {\xi _s^2 \left| {{\xi '}_s^2 } \right.} \right) + J_m^* \left( {\xi _s^1 \left| {{\xi '}_s^1 } \right.} \right)J_m^* \left( {\xi _s^2 \left| {{\xi '}_s^2 } \right.} \right) \right], \nonumber \\ 
&& \left| {\Phi _{0s} } \right\rangle  = \left| {\Xi _s^{\left( 0 \right)} } \right\rangle.
\end{eqnarray}
where we have used the exchange symmetry of $\psi _2^{\left( {in} \right)} ( {\xi _s^{\left( 2 \right)} } )$ to simplify the expression of the state $\left| {\Phi _{2s} } \right\rangle$ whereas the coefficients $p_2$ and $p_1$ have been introduced in such a way that  $\left\langle {{\Phi _{2s} }}  \mathrel{\left | {\vphantom {{\Phi _{2s} } {\Phi _{2s} }}} \right. \kern-\nulldelimiterspace}  {{\Phi _{2s} }} \right\rangle  = 1$ and ${\rm Tr} \left( {\hat \rho _{1s} } \right) = 1$, respectively. As a matter of fact, the squared norm of $\left| {\Phi _{2s} } \right\rangle$ and the trace of $\hat \rho _{1s}$ are
\begin{eqnarray}
&& \left\langle {{\Phi _{2s} }} \mathrel{\left | {\vphantom {{\Phi _{2s} } {\Phi _{2s} }}}
 \right. \kern-\nulldelimiterspace} {{\Phi _{2s} }} \right\rangle   = \frac{1}{{p_1 }}\int {d\xi _s^{\left( 2 \right)} d{\xi '}_s^{\left( 2 \right)} } \psi _2^{\left( {in} \right)*} \left( {\xi _s^{\left( 2 \right)} } \right)\psi _2^{\left( {in} \right)} \left( {{\xi '}_s^{\left( 2 \right)} } \right)\int {d\Xi _s^{\left( 2 \right)} } Z_{ss}^* \left( {\Xi _s^1 \left| {\xi _s^1 } \right.} \right)Z_{ss} \left( {\Xi _s^1 \left| {{\xi '}_s^1 } \right.} \right)Z_{ss}^* \left( {\Xi _s^2 \left| {\xi _s^2 } \right.} \right)Z_{ss} \left( {\Xi _s^2 \left| {{\xi '}_s^2 } \right.} \right), \nonumber \\ 
&& {\rm Tr} \left( {\hat \rho _{1s} } \right) = \frac{2}{{p_2 }}\int {d\xi _s^{\left( 2 \right)} d{\xi '}_s^{\left( 2 \right)} } \psi _2^{\left( {in} \right)} \left( {\xi _s^{\left( 2 \right)} } \right)\psi _2^{\left( {in} \right)*} \left( {{\xi '}_s^{\left( 2 \right)} } \right)\int {d\Xi _s^{\left( 1 \right)} } Z_{ss} \left( {\Xi _s^1 \left| {\xi _s^1 } \right.} \right)Z_{ss}^* \left( {\Xi _s^1 \left| {{\xi '}_s^1 } \right.} \right)\left[ {J_e^* \left( {\xi _s^2 \left| {{\xi '}_s^2 } \right.} \right) + J_m^* \left( {\xi _s^2 \left| {{\xi '}_s^2 } \right.} \right)} \right], \nonumber  \\ 
\end{eqnarray}
so that Eq.(\ref{Jtau}) proves the assertions. The comparison of the first, third and fifth of Eqs.(\ref{p1ph1rho2p3pihi3}) with Eq.(\ref{TwoPolProb}) directly yield $p_2  = P_{ss}$, $p_1  = P_{se}  + P_{sm}$ and $p_0  = P_{em} + P_{ee} + P_{mm}$, and hence the reduced density operator of Eq.(\ref{reddenopeGGGGGG}) is 
\begin{equation}
\hat \rho _s^{\left( {out} \right)}  = P_{ss} \left| {\Phi _{2s} } \right\rangle \left\langle {\Phi _{2s} } \right| + 
\left( {P_{se}  + P_{sm} } \right) \hat \rho _{1s}  + \left( {P_{em}  + P_{ee}  +  P_{mm} } \right)\left| {\Phi _{0s} } \right\rangle \left\langle {\Phi _{0s} } \right|.
\end{equation}
In addition, considering the outgoing state $\left| {\Psi ^{\left( {out} \right)} } \right\rangle$ in Fig.5b, it is evident that
\begin{equation}
\left| {\Phi _{2s} } \right\rangle  = \frac{1}{{\sqrt {P_{ss} } }}\left[ {\hat I_s  \otimes \left\langle {\Phi _{0em} } \right|} \right]\left| {\Psi ^{\left( {out} \right)} } \right\rangle 
\end{equation}
where ${\hat I_s }$ is the identity operator of the Fock space $F_s$ and $| {\Phi _{0em} } \rangle  = | \Xi _e ^{\left( 0 \right)} \Xi _m  ^{\left( 0 \right)}  \rangle$ is the vacuum state of the space $F_e  \otimes F_m$.

\section{Two-polariton scattering: Role of ingoing polaritons entanglement}

We first analyze the operator $\hat \rho _{1s}$ (see the second of Eqs.(\ref{TwoPolDefin})) in the specific situation where the two ingoing $s$-polariton are not entangled, i.e. the ingoing wavefunction is factored as in Eq.(\ref{NotEntangled})  we here rewrite
\begin{equation} \label{NotEntangledK}
\psi _2^{\left( {in} \right)} \left( {\xi _s^{\left( 2 \right)} } \right) = \varphi \left( {\xi _s^1 } \right)\varphi \left( {\xi _s^2 } \right),
\end{equation}
expressing the fact that both polaritons are in the same quantum state $\left| {\varphi  } \right\rangle$ of $F_s$, normalized as $\int {d\xi _s^{(1)} } \left| {\varphi  \left( {\xi _s^1 } \right)} \right|^2  = 1$. Inserting Eq.(\ref{NotEntangledK}) into Eq.(\ref{TwoPolProb}) we get
\begin{equation}
P_{\tau \mu }  = \left( {2 - \delta _{\tau \mu } } \right)\left[ {\int {d\xi _s^1 d{\xi '}_s^1 } \varphi ^* \left( {\xi _s^1 } \right)\varphi \left( {{\xi '}_s^1 } \right)J_\tau  \left( {\xi _s^1 \left| {{\xi '}_s^1 } \right.} \right)} \right]\left[ {\int {d\xi _s^2 d{\xi '}_s^2 } \varphi ^* \left( {\xi _s^2 } \right)\varphi \left( {{\xi '}_s^2 } \right)J_\mu  \left( {\xi _s^2 \left| {{\xi '}_s^2 } \right.} \right)} \right]
\end{equation}
which directly yields
\begin{equation} \label{TwoPolPse+Psm}
P_{se}  + P_{sm}  = \left[ {\int {d\xi _s^1 d{\xi '}_s^1 } \varphi ^* \left( {\xi _s^1 } \right)\varphi  \left( {{\xi '}_s^1 } \right)J_s  \left( {\xi _s^1 \left| {{\xi '}_s^1 } \right.} \right)} \right]\left\{ {2\int {d\xi _s^2 d{\xi '}_s^2 } \varphi ^* \left( {\xi _s^2 } \right)\varphi  \left( {{\xi '}_s^2 } \right)\left[ {J_e \left( {\xi _s^2 \left| {{\xi '}_s^2 } \right.} \right) + J_m \left( {\xi _s^2 \left| {{\xi '}_s^2 } \right.} \right)} \right]} \right\}.
\end{equation}
Besides, substituting Eq.(\ref{NotEntangledK}) into the second of Eqs.(\ref{TwoPolDefin}), we get
\begin{eqnarray}
&& \hat \rho _{1s}  = \frac{2}{{P_{se}  + P_{sm} }} {\int {d\xi _s^2 d{\xi '}_s^2 } \varphi  \left( {\xi _s^2 } \right)\varphi ^* \left( {{\xi '}_s^2 } \right)\left[ {J_e^* \left( {\xi _s^2 \left| {{\xi '}_s^2 } \right.} \right) + J_m^* \left( {\xi _s^2 \left| {{\xi '}_s^2 } \right.} \right)} \right]} \nonumber \\ 
&& \cdot \left[ {\int {d\Xi _s^{\left( 1 \right)} } \int {d\xi _s^1 } Z_{ss} \left( {\Xi _s^1 \left| {\xi _s^1 } \right.} \right) \varphi  \left( {\xi _s^1 } \right) \left| {\Xi _s^{\left( 1 \right)} } \right\rangle } \right]\left[ {\int {d{\Xi '}_s^{\left( 1 \right)} } \int {d{\xi '}_s^1 } Z_{ss}^* \left( {{\Xi '}_s^1 \left| {{\xi '}_s^1 } \right.} \right) \varphi ^* \left( {{\xi '}_s^1 } \right) \left\langle {{\Xi '}_s^{\left( 1 \right)} } \right|} \right] 
\end{eqnarray}
which, with the help of Eq.(\ref{TwoPolPse+Psm}), can be rewritten as
\begin{equation}
\hat \rho _{1s}  = \left| {\Phi _{1s} } \right\rangle \left\langle {\Phi _{1s} } \right|
\end{equation}
where 
\begin{equation}
\left| {\Phi _{1s} } \right\rangle  = \frac{\displaystyle \int {d\Xi _s^{\left( 1 \right)} } \left[ {\int {d\xi _s^{\left( 1 \right)} } Z_{ss} \left( {\Xi _s^1 \left| {\xi _s^1 } \right.} \right)\varphi _{1s} \left( {\xi _s^1 } \right)} \right]\left| {\Xi _s^{\left( 1 \right)} } \right\rangle }{\displaystyle {\sqrt {\int {d\xi _s^{\left( 1 \right)} d{\xi '}_s^{\left( 1 \right)} } \varphi _{1s}^* \left( {\xi _s^1 } \right)\varphi  \left( {{\xi '}_s^1 } \right)J_s  \left( {\xi _s^1 \left| {{\xi '}_s^1 } \right.} \right)} }}
\end{equation}
which has precisely the structure of the one-polariton state in the second of Eqs.(\ref{p0phi1}) so that it is normalized as $\left\langle {{\Phi _{1s} }} \mathrel{\left | {\vphantom {{\Phi _{1s} } {\Phi _{1s} }}} \right. \kern-\nulldelimiterspace} {{\Phi _{1s} }} \right\rangle  = 1$

We now explore the more involved example situation where the ingoing wavefunction is
\begin{equation} \label{EntangledK}
\psi _2^{\left( {in} \right)} \left( {\xi _s^{\left( 2 \right)} } \right) = \frac{1}{{\sqrt 2 }}\left[ {\varphi ^1 \left( {\xi _s^1 } \right)\varphi ^2 \left( {\xi _s^2 } \right) + \varphi ^1 \left( {\xi _s^1 } \right)\varphi ^2 \left( {\xi _s^2 } \right)} \right]
\end{equation}
where $\varphi^i$ are one-polariton wavefunctions which are normalized and orthogonal, i.e.
\begin{equation}
\int {d\xi _s \varphi ^{i*} \left( {\xi _s } \right)\varphi ^j \left( {\xi _s } \right)}  = \delta _{ij}, 
\end{equation}
this assuring the overall normalization condition $\int {d\xi _s^{\left( 2 \right)} } | {\psi _2^{\left( {in} \right)} ( {\xi _s^{\left( 2 \right)} } )} |^2  = 1$. Here, in order to deal with the operator $\hat \rho _{1s}$ in the second of Eqs.(\ref{TwoPolDefin}), it is  convenient introducing the two $2\times2$ matrices $
X_s$ and $X_{em}$ whose $i,j=(1,2)$ entries are
\begin{eqnarray} \label{XsXem}
&& X_s^{ij}  = \int {d\xi _s d\xi '_s } \varphi ^{i*} \left( {\xi _s } \right)\varphi _s^j \left( {\xi '_s } \right)J_s \left( {\xi _s \left| {\xi '_s } \right.} \right), \nonumber \\ 
&& X_{em}^{ij}  = \int {d\xi _s d\xi '_s } \varphi ^{i*} \left( {\xi _s } \right)\varphi _s^j \left( {\xi '_s } \right)\left[ {J_e \left( {\xi _s \left| {\xi '_s } \right.} \right) + J_m \left( {\xi _s \left| {\xi '_s } \right.} \right)} \right]
\end{eqnarray}
which, as a consequence of Eq.(\ref{JtauStar}), are such that $X_s^{ij*}  = X_s^{ji}$ and $X_{em}^{ij*}  = X_{em}^{ji}$, i.e the matrices $X_s$ and $X_{em}$ are Hermitian. In addition, due to the expression of $J_\tau  \left( {\xi _s \left| {\xi '_s } \right.} \right)$ in Eqs.(\ref{JsJnuAPPENDIX}) it is evident that
\begin{equation}
\int {d\xi _s } \int {d\xi '_s } \chi ^* \left( {\xi _s } \right)\chi \left( {\xi '_s } \right)J_\tau  \left( {\xi _s \left| {\xi '_s } \right.} \right) \ge 0,
\end{equation}
for any $\chi \left( {\xi _s } \right)$, so that 
\begin{eqnarray}
&& {\rm tr} \left( {X_s } \right) = X_s^{11}  + X_s^{22}  \ge 0, \nonumber \\ 
&& {\rm tr} \left( {X_{em} } \right) = X_{em}^{11}  + X_{em}^{22}  \ge 0
\end{eqnarray}
and, besides, the Schwartz inequalities
\begin{eqnarray} \label{semiposdefinit}
&& X_s^{11} X_s^{22}  \ge \left| {X_s^{12} } \right|^2, \nonumber  \\ 
&& X_{em}^{11} X_{em}^{22}  \ge \left| {X_{em}^{12} } \right|^2 
\end{eqnarray}
hold, this implying that the eigenvalues of the matrices $X_s$ and $X_{em}$  are greater/equal than zero, i.e. the matrices are positive semi-definite. Inserting Eq.(\ref{EntangledK}) into Eq.(\ref{TwoPolProb}), with the help of the second of Eqs.(\ref{XsXem}), we get
\begin{equation}
P_{se}  + P_{sm}  = \int {d\xi _s^1 } \int {d{\xi '}_s^1 } \begin{pmatrix}
   {\varphi ^{1*} \left( {\xi _s^1 } \right)} & {\varphi ^{2*} \left( {\xi _s^1 } \right)}  \\
\end{pmatrix}\begin{pmatrix}
   {X_{em}^{22} } & {X_{em}^{12*} }  \\
   {X_{em}^{12} } & {X_{em}^{11} }  \\
\end{pmatrix}\begin{pmatrix}
   {\varphi _s^1 \left( {{\xi '}_s^1 } \right)}  \\
   {\varphi _s^2 \left( {{\xi '}_s^1 } \right)}  \\
\end{pmatrix}J_s \left( {\xi _s^1 \left| {{\xi '}_s^1 } \right.} \right)
\end{equation}
which, using Eq.(\ref{JtauStar}) and the evident relation $x^T y = {\rm tr} \left( {yx^T } \right)$ (valid for any pair of vectors $x,y \in \mathbb{C}^2$), can be conveniently rewritten as
\begin{equation} \label{Pse+PsmENTA1}
P_{se}  + P_{sm}  = {\rm tr} 
\left[ { \begin{pmatrix}
   {X_{em}^{22} } & {X_{em}^{12*} }  \\
   {X_{em}^{12} } & {X_{em}^{11} }  \\
\end{pmatrix} \int {d{\xi '}_s^1 } \int {d\xi _s^1 } \begin{pmatrix}
   {\varphi _s^1 \left( {{\xi '}_s^1 } \right)}  \\
   {\varphi _s^2 \left( {{\xi '}_s^1 } \right)}  \\
\end{pmatrix}\begin{pmatrix}
   {\varphi ^{1*} \left( {\xi _s^1 } \right)} & {\varphi ^{2*} \left( {\xi _s^1 } \right)}  \\
\end{pmatrix}J_s^* \left( {{\xi '} _s^1 \left| {\xi_s^1 } \right.} \right)} \right].
\end{equation}
Now using the first of Eqs.(\ref{XsXem}), Eq.(\ref{Pse+PsmENTA1}) becomes 
\begin{equation}
P_{se}  + P_{sm}  = {\rm tr} \left[ {\begin{pmatrix}
   {X_{em}^{22} } & {X_{em}^{12*} }  \\
   {X_{em}^{12} } & {X_{em}^{11} }  \\
\end{pmatrix}\begin{pmatrix}
   {X_s^{11} } & {X_s^{12*} }  \\
   {X_s^{12} } & {X_s^{22} }  \\
\end{pmatrix}} \right]
\end{equation}
which, for later convenience (see below) we write as
\begin{equation} \label{trace1}
{\rm tr} \left( {\frac{{\tilde X_{em} X_s }}{{P_{se}  + P_{sm} }}} \right) = 1
\end{equation}
where
\begin{equation}
\tilde X_{em}  = \begin{pmatrix}
   {X_{em}^{22} } & {X_{em}^{12} }  \\
   {X_{em}^{12*} } & {X_{em}^{11} }  \\
\end{pmatrix}.
\end{equation}
We now substitute Eq.(\ref{EntangledK}) into the second of Eqs.(\ref{TwoPolDefin}) so that, using the second of Eqs.(\ref{XsXem}), we get after some algebra
\begin{eqnarray} \label{rho1s1}
&& \hat \rho _{1s}  = \int {d\Xi _s^{\left( 1 \right)} d{\Xi '}_s^{\left( 1 \right)} }   \left[ {\int {d\xi _s^1 } \int {d{\xi '}_s^1 } \begin{pmatrix}
   {\varphi ^1 \left( {\xi _s^1 } \right)} & {\varphi ^2 \left( {\xi _s^1 } \right)}  \\
\end{pmatrix}  \frac{\tilde X_{em}}{{P_{se}  + P_{sm} }} \begin{pmatrix}
   {\varphi ^{1*} \left( {{\xi '}_s^1 } \right)}  \\
   {\varphi ^{2*} \left( {{\xi '}_s^1 } \right)}  \\
\end{pmatrix}Z_{ss} \left( {\Xi _s^1 \left| {\xi _s^1 } \right.} \right)Z_{ss}^* \left( {{\Xi '}_s^1 \left| {{\xi '}_s^1 } \right.} \right)} \right] \left| {\Xi _s^{\left( 1 \right)} } \right\rangle \left\langle {{\Xi '}_s^{\left( 1 \right)} } \right| \nonumber \\
\end{eqnarray}
which we now diagonalize, for simplicity, in the situation $\det \left( {X_s } \right) \ne 0$. To this end, a suitable decomposition of the matrix ${\tilde X_{em} /\left( {P_{se}  + P_{sm} } \right)}$ is in order but Eq.(\ref{trace1}) suggests to consider the eigenvalue problem
\begin{equation} \label{EigProb1}
\frac{{\tilde X_{em} X_s }}{{P_{se}  + P_{sm} }}\Phi _{1s}  = \rho _{1s} \Phi _{1s} 
\end{equation}
since its eigenvalues $\rho _{1s}^a$ and $\rho _{1s}^b$ are consistently such that 
\begin{equation} \label{eigsum=1}
\rho _{1s}^a  + \rho _{1s}^b  = 1.
\end{equation}
Now, since we are supposing that $\det \left( {X_s } \right) \ne 0$, the Hermitian matrix ${X_s }$ is invertible and positive definite (see the first of Eqs.(\ref{semiposdefinit})) so that an invertible  and positive definite Hermitian matrix ${\left( {X_s } \right)^{1/2} }$ can be found such that  $X_s  = \left( {X_s } \right)^{1/2} \left( {X_s } \right)^{1/2}$. Accordingly, Eq.(\ref{EigProb1}) is equivalent to
\begin{equation} \label{EigProb2}
\frac{{\left( {X_s } \right)^{1/2} \tilde X_{em} \left( {X_s } \right)^{1/2} }}{{P_{se}  + P_{sm} }}\left[ {\left( {X_s } \right)^{1/2} \Phi _{1s} } \right] = \rho _{1s} \left[ {\left( {X_s } \right)^{1/2} \Phi _{1s} } \right].
\end{equation}
since $\left( {X_s } \right)^{1/2}$ is invertible so that, since the matrix ${\left( {X_s } \right)^{1/2} \tilde X_{em} \left( {X_s } \right)^{1/2} }$ is Hermitian and semi-positive definite, we conclude that the eigenvalues $\rho _{1s}^a$ and $\rho _{1s}^b$ are nonnegative real numbers which, from Eq.(\ref{eigsum=1}), are not greater than one, i.e.   
\begin{eqnarray}
 0 \le \rho _{1s}^a  \le 1, \nonumber \\
 0 \le \rho _{1s}^b  \le 1.
\end{eqnarray}
In addition, due again to its hemiticity, the matrix ${\left( {X_s } \right)^{1/2} \tilde X_{em} \left( {X_s } \right)^{1/2} }$ has two orthonormal eigenvectors ${\left( {X_s } \right)^{1/2} \Phi _{1s}^a }$ and ${\left( {X_s } \right)^{1/2} \Phi _{1s}^b }$, i.e.
\begin{equation} \label{orthonormAPPP}
\left[ {\left( {X_s } \right)^{1/2} \Phi _{1s}^\alpha  } \right]^{T*} \left[ {\left( {X_s } \right)^{1/2} \Phi _{1s}^\beta  } \right] = \delta _{\alpha \beta }, 
\end{equation}
and it admits the spectral decomposition
\begin{equation} \label{spectdecAPPP}
\frac{{\left( {X_s } \right)^{1/2} \tilde X_{em} \left( {X_s } \right)^{1/2} }}{{P_{se}  + P_{sm} }} = \rho _{1s}^a \left[ {\left( {X_s } \right)^{1/2} \Phi _{1s}^a } \right]\left[ {\left( {X_s } \right)^{1/2} \Phi _{1s}^a } \right]^{T*}  + \rho _{1s}^b \left[ {\left( {X_s } \right)^{1/2} \Phi _{1s}^b } \right]\left[ {\left( {X_s } \right)^{1/2} \Phi _{1s}^b } \right]^{T*},
\end{equation}
so that, after a simple algebraic manipulation exploiting the fact that the matrix $\left( {X_s } \right)^{1/2}$ is invertible, Eqs.(\ref{orthonormAPPP}) and (\ref{spectdecAPPP}) readily yield
\begin{eqnarray} \label{Xemdecomp}
&& \left( {\Phi _{1s}^\alpha  } \right)^{T*} X_s \Phi _{1s}^\beta   = \delta _{\alpha \beta }, \nonumber  \\ 
&& \frac{{\tilde X_{em} }}{{P_{se}  + P_{sm} }} = \rho _{1s}^a \Phi _{1s}^a \left( {\Phi _{1s}^a } \right)^{T*}  + \rho _{1s}^b \Phi _{1s}^b \left( {\Phi _{1s}^b } \right)^{T*}. 
\end{eqnarray}
The second of Eqs.(\ref{Xemdecomp}) provides the sought after decomposition of the matrix ${\tilde X_{em} /\left( {P_{se}  + P_{sm} } \right)}$ in terms of the eingevectors ${\Phi _{1s}^\alpha  }$ of the matrix $\tilde X_{em} X_s /\left( {P_{se}  + P_{sm} } \right)$ satisfying the normalization condition in the first of Eqs.(\ref{Xemdecomp}). Inserting the  second of Eqs.(\ref{Xemdecomp}) into Eq.(\ref{rho1s1}) we get
\begin{equation} \label{rho1s2}
\hat \rho _{1s}  = \rho _{1s}^a \left| {\Phi _{1s}^a } \right\rangle \left\langle {\Phi _{1s}^a } \right| + \rho _{1s}^b \left| {\Phi _{1s}^b } \right\rangle \left\langle {\Phi _{1s}^b } \right|
\end{equation}
where
\begin{equation}
\left| {\Phi _{1s}^\alpha  } \right\rangle  = \int {d\Xi _s^{\left( 1 \right)} } \left[ {\int {d\xi _s^{\left( 1 \right)} Z_{ss} \left( {\Xi _s^1 \left| {\xi _s^1 } \right.} \right)} \begin{pmatrix}
   {\varphi ^1 \left( {\xi _s^1 } \right)} & {\varphi ^2 \left( {\xi _s^1 } \right)} 
\end{pmatrix} \Phi _{1s}^\alpha  } \right]\left| {\Xi _s^{\left( 1 \right)} } \right\rangle 
\end{equation}
for $\alpha = a,b$. The scalar product of two states $| {\Phi _{1s}^\alpha  } \rangle $ and $| {\Phi _{1s}^\beta  } \rangle $ is easily seen to be given by
\begin{equation}
\left\langle {{\Phi _{1s}^\alpha  }}
 \mathrel{\left | {\vphantom {{\Phi _{1s}^\alpha  } {\Phi _{1s}^\beta  }}}
 \right. \kern-\nulldelimiterspace}
 {{\Phi _{1s}^\beta  }} \right\rangle  = \left(\Phi _{1s}^{\alpha}\right)^{ *T} \left[ {\int {d\xi _s^{\left( 1 \right)} d{\xi '}_s^{\left( 1 \right)} } \begin{pmatrix}
   {\varphi ^{1*} \left( {\xi _s^1 } \right)}  \\
   {\varphi ^{2*} \left( {\xi _s^1 } \right)}  \\
\end{pmatrix} \begin{pmatrix}
   {\varphi ^1 \left( {{\xi '}_s^1 } \right)} & {\varphi ^2 \left( {{\xi '}_s^1 } \right)}  \\
\end{pmatrix} \int {d\Xi _s^{\left( 1 \right)} }  Z_{ss}^* \left( {\Xi _s^1 \left| {\xi _s^1 } \right.} \right)Z_{ss} \left( {\Xi _s^1 \left| {{\xi '}_s^1 } \right.} \right)} \right]\Phi _{1s}^\alpha  
\end{equation}
so that, since $\int {d\Xi _s^{\left( 1 \right)} } Z_{ss}^* ( {\Xi _s^1 | {\xi _s^1 } } )Z_{ss} ( {\Xi _s^1 | {{\xi '}_s^1 } } ) = J_s ( {\xi _s^1 | {{\xi '}_s^1 } } )$ (see Eq.(\ref{Jtau})), the matrix in square bracket is precisely $X_s$ (see the first of Eqs.(\ref{XsXem})), i.e.
\begin{equation}
\left\langle {{\Phi _{1s}^\alpha  }}
 \mathrel{\left | {\vphantom {{\Phi _{1s}^\alpha  } {\Phi _{1s}^\beta  }}}
 \right. \kern-\nulldelimiterspace}
 {{\Phi _{1s}^\beta  }} \right\rangle  = \left( {\Phi _{1s}^\alpha  } \right)^{T*} X_s \Phi _{1s}^\beta  
\end{equation}
or, by using the first of Eqs.(\ref{Xemdecomp})
\begin{equation}
\left\langle {{\Phi _{1s}^\alpha  }}
 \mathrel{\left | {\vphantom {{\Phi _{1s}^\alpha  } {\Phi _{1s}^\beta  }}}
 \right. \kern-\nulldelimiterspace}
 {{\Phi _{1s}^\beta  }} \right\rangle  = \delta _{\alpha \beta }.
\end{equation}
Therefore the states $| {\Phi _{1s}^a  } \rangle $ and $| {\Phi _{1s}^b  } \rangle $ are orthonormal and consequently, from Eq.(\ref{rho1s2}), they are the eigenvectors of the operator $\hat \rho _{1s}$ corresponding to the eigenvalues $\rho _{1s}^a$ and $\rho _{1s}^b$.

\end{widetext}

\end{document}